\documentclass[useAMS,usenatbib,a4paper]{mn2e}

\usepackage{times}
\usepackage{psfrag}
\usepackage[dvips]{graphicx}
\usepackage{amssymb}
\usepackage{bm}
\usepackage[french,english]{babel}
\usepackage{color}
\usepackage{amsmath}
\usepackage{algorithmic}
\usepackage{algorithm}
\usepackage{url}
\usepackage{hyperref}
\hypersetup{breaklinks,colorlinks,citecolor=blue}

\title[PURIFY]
{PURIFY: a new approach to radio-interferometric imaging}

\author[Carrillo et al.]
{R. E. Carrillo$^{1}$\thanks{E-mail: rafael.carrillo@epfl.ch}, J. D. McEwen$^{2,3}$ and Y. Wiaux$^{1,4,5,6}$\\
$^{1}$Institute of Electrical Engineering, Ecole Polytechnique F{\'e}d{\'e}rale de Lausanne (EPFL), CH-1015 Lausanne, Switzerland\\
$^{2}$Department of Physics and Astronomy, University College London (UCL), London WC1E 6BT, UK\\
$^{3}$Mullard Space Science Laboratory, University College London (UCL), Holmbury St Mary, Surrey RH5 6NT, UK\\
$^{4}$Department of Medical Radiology, University Hospital Center (CHUV) and University of Lausanne (UNIL), CH-1011 Lausanne, Switzerland\\
$^{5}$Department of Radiology and Medical Informatics, University of Geneva (UniGE), CH-1211 Geneva, Switzerland\\
$^{6}$Institute of Sensors, Signals \& Systems, Heriot-Watt University, Edinburgh EH14 4AS, UK\\}

\begin{document}

\date{Accepted ---. Received ---; in original form ---}

\pagerange{\pageref{firstpage}--\pageref{lastpage}} \pubyear{2014}

\maketitle

\label{firstpage}

\begin{abstract}
In a recent article series, the authors have promoted convex optimization algorithms for radio-interferometric imaging in the framework of compressed sensing, which leverages sparsity regularization priors for the associated inverse problem and defines a minimization problem for image reconstruction. This approach was shown, in theory and through simulations in a simple discrete visibility setting, to have the potential to outperform significantly CLEAN and its evolutions. In this work, we leverage the versatility of convex optimization in solving minimization problems to both handle realistic continuous visibilities and offer a highly parallelizable structure paving the way to significant acceleration of the reconstruction and high-dimensional data scalability. The new algorithmic structure promoted relies on the simultaneous-direction method of multipliers (SDMM), and contrasts with the current major-minor cycle structure of CLEAN and its evolutions, which in particular cannot handle the state-of-the-art minimization problems under consideration where neither the regularization term nor the data term are differentiable functions. We release a beta version of an SDMM-based imaging software written in C and dubbed PURIFY (\url{http://basp-group. github.io/purify/}) that handles various sparsity priors, including our recent average sparsity approach SARA. We evaluate the performance of different priors through simulations in the continuous visibility setting, confirming the superiority of SARA.
\end{abstract}


\section{Introduction}
\label{sec:Introduction}
Radio interferometry is a powerful technique that allows observation of the radio emission from the sky with high angular resolution and sensitivity, providing valuable information for astrophysics, astrometry and cosmology~\citep{ryle46,blythe57,ryle59,ryle60,thompson01}. The measurement equation for radio interferometry defines an ill-posed linear inverse problem in the perspective of signal reconstruction. Under restrictive assumptions of monochromatic non-polarized imaging on small fields of view (FOV), the measured visibilities relates to Fourier measurements of the observed signal. Next-generation radio telescopes, such as the new LOw Frequency ARray (LOFAR\footnote{\url{http://www.lofar.org/}}), or the recently upgraded Karl G. Jansky Very Large Array (VLA\footnote{\url{https://science.nrao.edu/facilities/vla}}), or the future Square Kilometer Array (SKA\footnote{\url{http://www.skatelescope.org/}}), will achieve much higher dynamic range than current instruments, also at higher angular resolution. Also, these telescopes will acquire a massive amount of data, thus posing large-scale problems. Classical imaging techniques developed in the field, such as the CLEAN algorithm and its multi-scale variants \citep{hogbom74,bhatnagar04,cornwell08b}, are known to be slow and to provide suboptimal imaging quality \citep{li11,carrillo12}. This state of things has triggered an intense research to reformulate imaging techniques for radio interferometry in the perspective of next-generation instruments.

The theory of compressed sensing (CS) introduces a signal acquisition and reconstruction framework that goes beyond the traditional Nyquist sampling paradigm~\citep{donoho06,candes06,baraniuk07a,fornasier11}. Recently, CS and convex optimization techniques have been applied to image deconvolution in radio interferometry~\citep{wiaux09,wiaux09b,wenger10,mcewen11a,li11,carrillo12} showing promising results. These techniques promise improved image fidelity, flexibility and computation speed over traditional approaches. This speed enhancement is crucial for the scalability of imaging techniques to very high dimensions in the perspective of next-generation telescopes. However, CS-based imaging techniques have only been studied for low dimensional discrete visibility coverages. The works in \cite{wiaux09,wiaux09b} and \cite{mcewen11a} consider idealised random and discrete visibility coverages in order to remain as close to the CS theory as possible. First steps towards more realistic visibility coverages have been taken by \cite{wenger10} and \cite{li11}, who consider coverages due to specific interferometer configurations but which remain discrete. \cite{carrillo12} consider variable density sampling patterns, which mimic common generic sampling patterns in radio-interferometric (RI) imaging but also remaining discrete. These preliminary works suggest that the performance of CS reconstructions is likely to hold for more realistic visibility coverages. Therefore, the extension of CS techniques to more realistic continuous interferometric measurements is of great importance.

In the present work, we extend the previously proposed imaging approaches in \cite{wiaux09}, \cite{wiaux09b} and \cite{carrillo12} to handle continuous visibilities and open the door to large-scale optimization problems. We propose a general algorithmic framework based on the simultaneous-direction method of multipliers (SDMM) to solve sparse imaging problems. The proposed framework offers a parallel implementation structure that decomposes the original problem into several small simple problems, hence allowing implementation in multicore architectures or in computer clusters, or on graphics processing units. These implementations provide both flexibility in memory requirements and a significant gain in terms of speed, thus enabling scalability to large-scale problems. SDMM stands in stark contrast with the current major-minor cycle structure of CLEAN and evolutions, which in particular cannot handle the state-of-the-art minimization problems under consideration \citep{carrillo12}, where neither the regularization term nor the data term are differentiable functions. We release a beta version of an SDMM-based imaging software written in C and dubbed PURIFY (\url{http://basp-group. github.io/purify/}) that handles various sparsity priors, including our recent average sparsity approach SARA \citep{carrillo12}, thus providing a new powerful framework for RI imaging. We evaluate the performance of different priors through simulations in the continuous visibility setting. Simulation results confirm the superiority of SARA for continuous Fourier measurements. Even though this beta version of PURIFY is not parallelized, we discuss in detail the extraordinary parallel and distributed optimization potential of SDMM, to be exploited in future versions.

The remainder of the paper is organized as follows. In Section~\ref{sec:CS}, we review the theory of CS briefly. In Section~\ref{sec:Radio interferometric imaging}, we recall the inverse problem for image reconstruction from RI data and describe the state-of-the-art image reconstruction techniques used in radio astronomy. Section~\ref{sec:OP} presents the SDMM-based algorithm for RI imaging, which enables the incorporation of any convex sparsity regularization prior. In Section~\ref{sec:purify} we describe the PURIFY package, including implementation details. Numerical results evaluating the different regularization priors included in PURIFY, in particular SARA, are presented in Section~\ref{sec:Simulations and results}. Finally we conclude in Section~\ref{sec:Conclusion}.

\section{Compressed sensing}
\label{sec:CS}
CS introduces a signal acquisition framework that goes beyond the traditional Nyquist sampling paradigm~\citep{donoho06,candes06,baraniuk07a,fornasier11}, demonstrating that sparse signals may be recovered accurately from incomplete data. Consider a complex-valued signal $\bm{x}\in\mathbb{C}^{N}$, assumed to be sparse in some orthonormal basis $\mathsf{\Psi}\in\mathbb{C}^{N\times N}$ with $K\ll N$ nonzero coefficients, and also consider the measurement model $\bm{y}=\mathsf{\Phi}\bm{x}+\bm{n}$, where $\bm{y}\in\mathbb{C}^{M}$ denotes the measurement vector, $\mathsf{\Phi}\in\mathbb{C}^{M\times N}$ is the sensing matrix and $\bm{n}\in\mathbb{C}^{M}$ represents the observation noise. The standard condition $M<N$ characterizes the incompleteness of the data. The most common approach to recover $\bm{x}$ from $\bm{y}$ is to solve the following convex problem \citep{fornasier11}:
\begin{equation}\label{cs1}
\min_{\bar{\bm{\alpha}}\in\mathbb{C}^{N}}\|\bar{\bm{\alpha}}\|_{1}
\textnormal{ subject to }\| \bm{y}-\mathsf{\Phi \Psi}\bar{\bm{\alpha}}\|_{2}\leq\epsilon,
\end{equation}
where $\epsilon$ is an upper bound on the $\ell_{2}$ norm of the noise and \mbox{$\|\cdot\|_1$} denotes the $\ell_{1}$ norm of a complex-valued vector. The signal is recovered as $\hat{\bm{x}}=\mathsf{\Psi}\hat{\bm{\alpha}}$, where $\hat{\bm{\alpha}}$ denotes the solution to the above problem. Such problems that solve for the representation of the signal in a sparsity basis are known as synthesis-based problems. 

The standard theory of CS provides results for the recovery of $\bm{x}$ from $\bm{y}$ if $\mathsf{\Phi}$ obeys a Restricted Isometry Property (RIP)~\citep{fornasier11}. A sufficient condition is that $M$ is larger than roughly the signal sparsity: $M > 2K \ll N$. Note that incomplete Fourier measurements, on discrete or continuous spatial frequencies, represent a good sampling approach in this context. In the continuous setting, the theory applies also for $M>N$. It is not strictly ``compressed'' sensing any more but the inverse problem remains ill-posed. The basic theory  also requires $\mathsf{\Psi}$ to be orthonormal. However, signals often exhibit better sparsity in an overcomplete dictionary  \citep{gribonval03,bobin07,starck10}. Therefore recent works have begun to address the case of CS with redundant dictionaries. In this setting the signal $\bm{x}$ is expressed in terms of a dictionary $\mathsf{\Psi}\in\mathbb{C}^{N\times D}$, $N<D$, as $\bm{x} = \mathsf{\Psi}\bm{\alpha}$, $\bm{\alpha}\in\mathbb{C}^{D}$. \cite{rauhut08} find conditions on the dictionary $\mathsf{\Psi}$ such that the compound matrix $\mathsf{\Phi \Psi}$ obeys the RIP to accurately recover $\bm{\alpha}$ by solving a synthesis-based problem. Note that the problem is now more severely undertermined since the dimensionality of the unknonw has increased from $N$ to $D$.

As opposed to synthesis-based problems, analysis-based problems recover the signal itself solving:
\begin{equation}\label{cs6}
\min_{\bar{\bm{x}}\in\mathbb{C}^{N}}\|\mathsf{\Psi}^{\dagger}\bar{\bm{x}}\|_{1}
\textnormal{ subject to }\| \bm{y}-\mathsf{\Phi}\bar{\bm{x}}\|_{2}\leq\epsilon,
\end{equation}
where $\mathsf{\Psi}^{\dagger}$ denotes the adjoint operator of $\mathsf{\Psi}$. In this paper the superscript $^{\dagger}$ is used to denote both operator adjoint or conjugate transpose. \cite{candes10} provide a theoretical analysis of the $\ell_1$ analysis-based problem, extending the standard CS theory to coherent and redundant dictionaries. They provide theoretical stability guarantees based on a general condition of the sensing matrix $\mathsf{\Phi}$, coined the Dictionary Restricted Isometry Property (D-RIP). Note that  in the case when redundant dictionaries are used, the analysis problem does not increase the dimensionality of the problem as it solves for the signal itself.  Empirical and theoretical studies have shown clear advantages of the analysis approach over the synthesis approach for imaging problems \citep{carrillo13}. See \cite{nam13} and references therein for further discussion of the analysis model.

\section{Radio-interferometric imaging}
\label{sec:Radio interferometric imaging}

\subsection{Interferometric inverse problem}
A radio interferometer is an array of spatially separated antennas that takes measurements of the radio emissions of the sky, the so-called visibilities. The visibility coordinates are given by the relative position between each pair of antennas. The baseline components $(u,v,w)$ are measured in units of the wavelength $\lambda$ of the incoming signal. The components $\bm{u}=(u,v)$ specify the planar baseline coordinates, while the third component $w$ is associated with the basis vector of the coordinate pointing towards the center of the FOV of the telescope. The sky brightness distribution $x$ can be described in the same coordinate system as the baseline, with components $(l,m,n)$ where $\bm{l}=(l,m)$ denotes the coordinates on the image plane and $n(\bm{l})=\sqrt{1-l^2-m^2}$. The general RI equation for monochromatic non-polarized imaging reads as:
\begin{equation}
y\left(\bm{u}\right) = \int A\left(\bm{l},\bm{u}\right)x\left(\bm{l}\right){\rm e}^{-2\pi{\rm i} \bm{u} \bm{\cdot} \bm{l} }\: {\rm d}^2\bm{l}, \label{ri1}
\end{equation}
where $A\left(\bm{l},\bm{u}\right)=A'\left(\bm{l},\bm{u}\right)n^{-1}(\bm{l})$ and $A'\left(\bm{l},\bm{u}\right)$ stands for all contributions of direction dependent effects (DDE). Examples of DDEs are the primary beam, which limits the observed FOV, and the $w$-term ${\rm e}^{-2\pi{\rm i} w(n(\bm{l})-1)}$. This general equation defines a linear inverse problem in the perspective of recovering the intensity signal $x$ from the measured visibilities \citep{rau09}. Under the assumptions of small FOV ($n\approx 1$) or when the array is coplanar ($w\approx 0$), each visibility  corresponds to the measurement of the Fourier transform of a planar signal at the spatial frequency $\bm{u}$. This result is known as the van Cittert-Zernike theorem~\citep{thompson01}. The total number of points $\bm{u}$ probed by all telescope pairs of the array during the observation provides some incomplete coverage in the Fourier plane, the so-called $u$-$v$ coverage, characterizing the interferometer. 

To recover the source image from incomplete visibility measurements, we pose the inverse problem \eqref{ri1} for a sampled version of the image. The band-limited functions considered are completely identified by their Nyquist-Shannon sampling on a discrete uniform grid of $N=N^{1/2}\times N^{1/2}$ points in real space. The sampled intensity signal is denoted by the vector $\bm{x}\in\mathbb{R}^{N}$. We take $M$ visibilities denoted by the vector $\bm{y}\in\mathbb{C}^{M}$, which are related to the discrete image by the following linear model:
\begin{equation}
\bm{y}=\mathsf{\Phi}\bm{x}+\bm{n},\label{ri4}
\end{equation}
where $\mathsf{\Phi}\in\mathbb{C}^{M\times N}$ represents the general linear map from the image space domain to the visibility domain, which defines an ill-posed inverse problem in the perspective of image reconstruction. In the particular case when the visibilities identify with Fourier samples the measurement essentially reduces to a Fourier matrix sampled on $M$ spatial frequencies (see eq. \eqref{opmodel} in Section \ref{sec:purify}). In a realistic continuous visibility setting, one usually has $M>N$ and sometimes $M \gg N$, which will be increasingly the case for next-generation telescopes.

\subsection{State-of-the-art of classic imaging algorithms}
The most standard image reconstruction algorithm from visibility measurements is called CLEAN, which is a non-linear deconvolution method based on local iterative beam removal \citep{hogbom74,schwarz78,thompson01}. A sparsity prior on the original signal in real space is implicitly introduced thus already taking advantage of CS theory guarantees. Furthermore, as discussed in \cite{cornwell08b} and \cite{wiaux09} the CLEAN algorithm and its variants are examples of the Matching Pursuit algorithm \citep{mallat93}, which is well known in the CS community. CLEAN can be considered as a steepest descend algorithm to minimize the objective function $\chi^2=\| \bm{y}-\mathsf{\Phi}\bm{x}\|_{2}^{2}$ subject to an image model regularization \citep{rau09}. Most variants operate iteratively in two steps called the major and minor cycles. The major cycle computes the residual image $\bm{r}^{(t)}=\mathsf{\Phi}^{\dagger}(\bm{y}-\mathsf{\Phi}\bm{x}^{(t)})$, which is the gradient of the $\chi^2$ objective function at iteration $t$. The minor cycle regularizes the image update by applying an operator $\mathsf{T}$, which represents a deconvolution of the operator $\mathsf{\Phi}$, to the residual image yielding updates of the form
\begin{equation}
\bm{x}^{(t+1)}=\bm{x}^{(t)}+\mathsf{T}(\bm{r}^{(t)}).
\end{equation}

A multi-scale version of CLEAN, MS-CLEAN, has also been developed \citep{cornwell08b}, where the sparsity model is improved by multi-scale decomposition, hence enabling better recovery of the signal. The MS-CLEAN method was shown to perform better than the standard CLEAN, but still suffers from an empirical choice of basis profiles and scales. An adaptive scale pixel decomposition method called ASP-CLEAN was also introduced to improve on multi-scale CLEAN by relying on an adaptive choice of scales \citep{bhatnagar04}. ASP-CLEAN models an image as a superposition of atoms in a redundant dictionary parametrized by amplitude, location and scale. Thus, this algorithm can be seen as a Matching Pursuit algorithm with an overcomplete dictionary. Note that these approaches are known to be slow, sometimes prohibitively so. Variants of CLEAN that addresses wide-band effects, or atmospheric effects have also been proposed in the literature (see \cite{rau09}, \cite{bhatnagar08}, \cite{bhatnagar13} and references therein).

Another approach to the reconstruction of images from visibility measurements is the Maximum Entropy Method  (MEM). In contrast to CLEAN, MEM solves a global optimization problem in which the inverse problem is regularized by the introduction of an entropic prior on the signal, but sparsity is not explicitly required \citep{cornwell85}. In practice, CLEAN and variants have found more widespread application than MEM.

\subsection{State-of-the-art of convex imaging algorithms}
\label{ssec:RICS}
Reconstruction techniques based on CS and convex optimization have also been proposed. The relationship between CLEAN and $\ell_1$ minimization coupled with a Dirac basis was first studied by \cite{marsh87}. The first application of CS and convex optimization to radio interferometry was performed by \citet{wiaux09}, where the versatility of the approach and its superiority relative to standard interferometric imaging techniques was demonstrated. It was reported that an $\ell_1$ minimization problem of the form of \eqref{cs1} coupled with a Dirac basis yields similar reconstruction quality to CLEAN, while including a positivity constraint in a convex formulation significantly enhances the reconstruction quality relative to CLEAN. The spread spectrum phenomenon associated with the $w$ component on wide FOV observations was shown in \cite{wiaux09b} to underpin a significant enhancement of the imaging quality independently of the sparsity basis chosen. These considerations pave the way to potential optimization strategies at the acquisition level in terms of antenna distribution design. A CS approach was developed and evaluated by \citet{wiaux10a} to recover the signal induced by cosmic strings in the cosmic microwave background. \citet{mcewen11a} generalise the previous CS imaging techniques to a wide FOV, recovering interferometric images defined directly on the sphere, rather than a tangent plane.  All of these works consider uniformly random and discrete visibility coverage in order to remain as close to the CS theory as possible. \cite{wiaux10a} and \cite{mcewen11a} exploited the fact that many signals in nature are also sparse or compressible in the magnitude of their gradient space, in which case the total variation (TV) minimization problem, 
\begin{equation}\label{tvproblem}
\min_{\bar{\bm{x}}\in\mathbb{C}^{N}} \|\bar{\bm{x}}\|_{\rm{TV}}
\textnormal{ subject to }\| \bm{y}-\mathsf{\Phi}\bar{\bm{x}}\|_{2}\leq\epsilon,
\end{equation} 
has been shown to yield superior reconstruction results. The TV norm is defined as $\|\bar{\bm{x}}\|_{\rm{TV}} = \| \nabla \bar{\bm{x}} \|_1$, where $\nabla \bar{\bm{x}}$ denotes the image gradient magnitude \citep{rudin92}. 

First steps towards more realistic visibility coverages have been taken by \citet{suksmono09} and \citet{wenger10}, who consider coverages due to specific interferometer configurations but which remain discrete. The aforementioned works use the following unconstrained synthesis problem:
\begin{equation}\label{bpdn}
\min_{\bar{\bm{\alpha}}\in\mathbb{C}^{N}}\frac{1}{2}\| \bm{y}-\mathsf{\Phi \Psi}\bar{\bm{\alpha}}\|_{2}^{2}+\lambda\|\bar{\bm{\alpha}}\|_{1},
\end{equation}
where $\lambda$ is a regularization parameter that balances the weight between the fidelity term and the regularization term. \citet{wenger10} reports superior reconstruction quality relative to an automatic CLEAN reconstruction and similar results relative to a user-guided CLEAN reconstruction. \cite{li11} studied a CS imaging approach based on \eqref{bpdn} and the isotropic undecimated wavelet transform, reporting results from discrete simulated coverages of ASKAP. The reconstruction quality of the isotropic undecimated wavelet transform method was reported to be superior to those of CLEAN and MS-CLEAN. Minimization of the problem \eqref{bpdn} is done iteratively by a projected gradient algorithm with updates of the form:
\begin{equation}
\bm{\alpha}^{(t+1)}=\mathsf{S}_{\lambda}\left (\bm{\alpha}^{(t)}+\mu\mathsf{\Psi}^{\dagger}\mathsf{\Phi}^{\dagger}(\bm{y}-\mathsf{\Phi \Psi}\bm{\alpha}^{(t)}) \right ),
\end{equation}
where $\mathsf{S}_{\lambda}(\cdot)$ is the soft-thresholding operator, which will be defined in Section \ref{ssec:opdet}. This algorithm can be seen as a major-minor cycle update where the major cycle computes the gradient of the $\chi^2$ data fidelity term and the minor cycle regularizes the solution by applying the soft-thresholding operator.

\citet{carrillo12} proposed an imaging algorithm dubbed sparsity averaging reweighted analysis (SARA) based on average sparsity over multiple bases, showing superior reconstruction qualities relative to state-of-the-art imaging methods in the field. A sparsity dictionary composed of a concatenation of $q$ bases, 
$\mathsf{\Psi}=[\mathsf{\Psi}_1, \mathsf{\Psi}_2, \ldots, \mathsf{\Psi}_q]$,
with $\mathsf{\Psi}\in\mathbb{C}^{N\times D}$, $N<D$, is used and average sparsity is promoted through the minimization of an analysis $\ell_0$ prior, $\|\mathsf{\Psi}^{\dagger}\bar{\bm{x}}\|_{0} $. The concatenation of the Dirac basis and the first eight orthonormal Daubechies wavelet bases (Db1-Db8) was proposed as an effective and simple candidate for a dictionary in the RI imaging context. See \cite{carrillo13} for further discussions on the average sparsity model, the dictionary selection and other applications to compressive imaging.

SARA adopts a reweighted $\ell_1$ minimization scheme to promote average sparsity through the prior $\|\mathsf{\Psi}^{\dagger}\bar{\bm{x}}\|_{0} $. The algorithm replaces the $\ell_0$ norm by a weighted $\ell_1$ norm and solves a sequence of weighted $\ell_1$ problems where the weights are essentially the inverse of the values of the solution of the previous problem \citep{candes08a}. The weighted $\ell_1$ problem is defined as:
\begin{equation}\label{delta}
\min_{\bar{\bm{x}}\in\mathbb{R}_{+}^{N}}\|\mathsf{W\Psi}^{\dagger}\bar{\bm{x}}\|_{1}
\textnormal{ subject to }\| \bm{y}-\mathsf{\Phi}\bar{\bm{x}}\|_{2}\leq\epsilon,
\end{equation}
where $\mathsf{W}\in\mathbb{R}^{D\times D}$ denotes the diagonal matrix with positive weights and $\mathbb{R}^{N}_{+}$ denotes the positive orthant in $\mathbb{R}^{N}$, which represents the positivity prior on $\bm{x}$. Note that problems of the form \eqref{tvproblem} and \eqref{delta} involve the minimization of a constrained problem with non-differentiable functions, which rules out smooth optimization techniques and do not fit in the major-minor cycle structure of CLEAN and the projected gradient algorithm. Therefore one must resort to more sophisticated optimization techniques to solve these non-smooth problems.

\section{A large-scale optimization algorithm}
\label{sec:OP}
\subsection{Proximal splitting methods}
Convex optimization problems have many attractive properties, in particular the essential property that any local minimum must be a global minimum and thus there exist efficient methods to solve them. Among convex optimization methods, proximal splitting methods offer great flexibility and are shown to capture and extend several well-known algorithms in a unifying framework. Examples of proximal splitting algorithms include Douglas-Rachford, iterative thresholding, projected Landweber, projected gradient, forward-backward, alternating projections, alternating direction method of multipliers and alternating split Bregman \citep{combettes11}. Proximal splitting methods solve optimization problems of the form
\begin{equation}\label{cvx1}
\min_{\bm{x}\in\mathbb{R}^{N}} f_1(\bm{x})+\ldots +f_S(\bm{x}),
\end{equation}
where $f_1(\bm{x}),\ldots,f_S(\bm{x})$ are convex lower semicontinuous functions from $\mathbb{R}^{N}$ to $\mathbb{R}$, not necessarily differentiable. Note that any convex constrained problem can be formulated as an unconstrained problem by using the indicator function of the convex constraint set as one of the functions in \eqref{cvx1}, i.e. $f_k(\bm{x})=i_{C}(\bm{x})$ where $C$ represents the convex constraint set. The indicator function, defined as $i_{C}(\bm{x})=0$ if $\bm{x}\in C$ or $i_{C}(\bm{x})=+\infty$ otherwise, belongs to the class of convex lower semicontinuous functions. Also, note that complex-valued vectors are treated as real-valued vectors with twice the dimension (accounting for real and imaginary parts). 

Proximal splitting methods proceed by splitting the contribution of the functions $f_1(\bm{x}),\ldots,f_S(\bm{x})$ individually so as to yield an easily implementable algorithm. They are called proximal because each non-smooth function in \eqref{cvx1} is incorporated in the minimization via its proximity operator. The proximity operator is an extension of the notion of the set projection operator to more general functions. Let $f$ be a convex lower semicontinuous function from $\mathbb{R}^{N}$ to $\mathbb{R}$, then the proximity operator of $f$ is defined as:
\begin{equation}
\mathrm{prox}_{f}(\bm{x}) \triangleq \arg\min_{\bm{z}\in\mathbb{R}^{N}} f(\bm{z})+\frac{1}{2}\| \bm{x}-\bm{z} \|_2^2.
\end{equation}
Typically, the solution to \eqref{cvx1} is reached iteratively by successive application of the proximity operator associated with each function. An important feature of proximal splitting methods is that they offer a powerful framework for solving convex problems in terms of speed and scalability of the techniques to very high dimensions. See \cite{combettes11} for a review of proximal splitting methods and their applications in signal and image processing. 

\subsection{Shortcomings of previously used algorithms}
The works in \cite{wiaux09b}, \cite{mcewen11a} and \cite{carrillo12} solved problems of the form in \eqref{delta}, whereas \citet{wenger10} and \citet{li11} solved the unconstrained problem \eqref{bpdn}. Unconstrained problems are easier to handle and there exist fast algorithms to solve them, such as the FISTA algorithm \citep{beck09}. However, there is no optimal strategy to fix the regularization parameter even if the noise level is known, therefore constrained problems, such as \eqref{delta}, offer a stronger fidelity term when the noise power is known, or can be estimated \emph{a priori}. Hence, we focus our attention on solving problem \eqref{delta} efficiently. \cite{wiaux09b}, \cite{mcewen11a} and \cite{carrillo12} used a Douglas-Rachford splitting algorithm \citep{combettes07} to solve \eqref{delta} in a simple discrete setting. However, in a realistic continuous setting this algorithm presents several shortcomings. In the following we discuss the main limitations of the Douglas-Rachford algorithm. 

The Douglas-Rachford splitting algorithm solves the problem by iteratively minimizing the $\ell_1$ norm and then projecting the result onto the constraint set $C'=\{ \bm{x}\in\mathbb{C}^{N} : \| \bm{y}-\mathsf{\Phi}\bm{x}\|_{2}\leq\epsilon \}\cap\mathbb{R}_{+}^{N}$ until some stopping criteria is achieved. The projection onto the set $C'$ is a hard optimization problem which in itself requires an iterative algorithm such as the generalized forward-backward algorithm. This iterative algorithm requires knowledge of the exact operator norm (maximum singular value) of $\mathsf{\Phi}$ or at least a closed upper bound to guarantee convergence. In the discrete case the exact operator norm can be computed and the algorithm achieves a fast convergence rate. However, in the continuous case the operator norm is unknown and its estimation poses a new problem. If the estimate of the operator norm is not precise enough, the algorithm takes many sub-iterations to converge. Hence, it would be advantageous to have an algorithm that does not need prior knowledge of the operator norm to achieve a fast convergence rate. Another tenet of the Douglas-Rachford algorithm is that it does not offer a parallel structure, which is a desirable property when solving large scale-problems such as those envisaged for the upcoming telescopes. For these reasons, we propose to use the simultaneous-direction method of multipliers (SDMM)~\citep{combettes11} which is also tailored to solve problems of the form of \eqref{cvx1} and circumvents the shortcomings of a Douglas-Rachford approach.

\subsection{Simultaneous Direction Method of Multipliers (SDMM)}
SDMM has two important properties: (i) it does not require differentiability of any of the functions, and (ii) it offers a parallel implementation structure where all the proximity operators can be computed in parallel rather than sequentially~\citep{combettes11}. Such a parallel structure is useful when implementing the algorithms on multicore architectures or on graphics processing units, thus providing a significant gain in terms of speed and scalability to large-scale problems. SDMM is a generalization of the alternating-direction method of multipliers \citep{boyd10} to a sum of more than two functions. As such, SDMM uses augmented Lagrangian techniques and duality arguments in its derivation. In the following we highlight the main steps in the derivation of SDMM tailored to solve \eqref{delta}. 

First, observe that the problem in \eqref{delta} can be reformulated as in \eqref{cvx1} in the following way:
\begin{equation}\label{cvx2}
\min_{\bm{x}\in\mathbb{C}^{N}} f_1(\mathsf{L}_1\bm{x})+f_2(\mathsf{L}_2\bm{x})+f_3(\mathsf{L}_3\bm{x}),
\end{equation}
where $\mathsf{L}_1=\mathsf{\Psi}^{\dagger}\in\mathbb{C}^{D\times N}$, $\mathsf{L}_2=\mathsf{\Phi}\in\mathbb{C}^{M\times N}$ and $\mathsf{L}_3=\mathsf{I}\in\mathbb{R}^{N\times N}$ is the identity matrix. In this formulation, $f_1(\bm{r}_1)=\| \mathsf{W}\bm{r}_1\|_1$ for $\bm{r}_1\in\mathbb{C}^D$, $f_2(\bm{r}_2)=i_{B}(\bm{r}_2)$ with $B=\{ \bm{r}_2\in\mathbb{C}^{M} : \| \bm{y}-\bm{r}_2\|_{2}\leq\epsilon \}$, and $f_3(\bm{r}_3)=i_{C}(\bm{r}_3)$ with $C=\mathbb{R}_{+}^{N} $. This problem is also equivalent to solving
\begin{align}\label{cvx3}
\min_{\substack{\bm{x}\in\mathbb{C}^{N}, \bm{r}_1\in\mathbb{C}^{D}, \\ \bm{r}_2\in\mathbb{C}^{M}, \bm{r}_3\in\mathbb{C}^{N}}} &f_1(\bm{r}_1)+f_2(\bm{r}_2)+f_3(\bm{r}_3)\\ \nonumber
\textnormal{subject to }&\mathsf{L}_i\bm{x}=\bm{r}_i, \textnormal{ for } i=1,2,3.
\end{align}
The augmented Lagrangian associated with \eqref{cvx3} is the saddle function
\begin{align}\label{cvx4}
\mathcal{L}_{\gamma}(\bm{x}, &\bm{r}_1, \bm{r}_2, \bm{r}_3, \bm{z}_1, \bm{z}_2, \bm{z}_3) = \\ \nonumber
&\sum_{i=1}^{3}f_i(\bm{r}_i)+\frac{1}{\gamma}\bm{z}_i^{\dagger}(\mathsf{L}_i\bm{x}-\bm{r}_i)+\frac{1}{2\gamma}\| \mathsf{L}_i\bm{x}-\bm{r}_i\|_2^2,
\end{align}
where $\gamma>0$ is a so-called penalty parameter and $\bm{z}_1\in\mathbb{C}^{D}$, $\bm{z}_2\in\mathbb{C}^{M}$ and $\bm{z}_3\in\mathbb{C}^{N}$ are the dual variables or the Langrange multipliers. SDMM is a primal dual algorithm that proceeds iteratively by first minimizing $\mathcal{L}_{\gamma}$ with respect to the primal variables, $\bm{x}$, $\bm{r}_1$, $\bm{r}_2$, $\bm{r}_3$, and as second step, solving the dual problem
\begin{equation}
\max_{\bm{z}_1\in\mathbb{C}^{D}, \bm{z}_2\in\mathbb{C}^{M}, \bm{z}_3\in\mathbb{C}^{N}} \mathcal{J}(\bm{z}_1, \bm{z}_2, \bm{z}_3),
\end{equation}
where
\begin{equation}
\mathcal{J}(\bm{z}_1, \bm{z}_2, \bm{z}_3)=\min_{\substack{\bm{x}\in\mathbb{C}^{N}, \bm{r}_1\in\mathbb{C}^{D}, \\ \bm{r}_2\in\mathbb{C}^{M}, \bm{r}_3\in\mathbb{C}^{N}}}\mathcal{L}_{\gamma}(\bm{x}, \bm{r}_1, \bm{r}_2, \bm{r}_3, \bm{z}_1, \bm{z}_2, \bm{z}_3)
\end{equation}
is the dual function. The main difference between SDMM and other primal-dual algorithms is that the optimization with respect to the primal variables is done in an alternating fashion by first minimizing $\mathcal{L}_{\gamma}$ with respect to $\bm{x}$ and then with respect to $\bm{r}_1$, $\bm{r}_2$, $\bm{r}_3$. The algorithm is shown to converge to a minimizer of \eqref{cvx3}. Convergence results of SDMM are based on convergence of the alternating-direction method of multipliers and can be found in \cite{boyd10}.

The minimizer of $\mathcal{L}_{\gamma}$ with respect to $\bm{x}$ with fixed variables $\bm{r}_i$, $\bm{z}_i$ is given by
\begin{equation}
\bm{x}^{*}=\arg\min_{\bm{x}\in\mathbb{C}^{N}}\sum_{i=1}^{3}\bm{z}_i^{\dagger}(\mathsf{L}_i\bm{x}-\bm{r}_i)+\frac{1}{2}\| \mathsf{L}_i\bm{x}-\bm{r}_i\|_2^2.
\end{equation}
Observe that the above problem is the minimization of a quadratic function, which is convex and differentiable. Therefore, necessary and sufficient optimality conditions are
\begin{equation}\label{qp1}
\nabla_{\bm{x}}\mathcal{L}_{\gamma}(\bm{x}^{*})=\sum_{i=1}^{3}\left [\mathsf{L}_i^{\dagger}\bm{z}_i + \mathsf{L}_i^{\dagger}(\mathsf{L}_i\bm{x}^{*}-\bm{r}_i) \right ]=0
\end{equation}
and the matrix $\mathsf{Q}=\sum_{i=1}^3\mathsf{L}_i^{\dagger}\mathsf{L}_i\in\mathbb{C}^{N\times N}$ should be invertible. For our particular problem $\mathsf{Q}=\mathsf{\Phi}^{\dagger}\mathsf{\Phi}+\mathsf{\Psi}\mathsf{\Psi}^{\dagger}+I$, which is positive-definite and invertible. Solving \eqref{qp1} for $\bm{x}^{*}$ yields
\begin{equation}\label{upd1}
\bm{x}^{*}=\mathsf{Q}^{-1}\sum_{i=1}^3\mathsf{L}_i^{\dagger}(\bm{r}_i-\bm{z}_i).
\end{equation}

The minimization over $\bm{r}_i$ can be carried out for all $i$ simultaneously since the problems are decoupled. Assume $i$ is fixed and also assume that $\bm{x}$ and $\bm{z}_i$ are fixed. Then the minimizer of $\mathcal{L}_{\gamma}$ with respect to $\bm{r}_i$ is 
\begin{equation}\label{upd21}
\bm{r}_i^{*}=\arg\min_{\bm{r}_i\in\mathbb{C}^N}f_i(\bm{r}_i)+\frac{1}{\gamma}\bm{z}_i^{\dagger}(\mathsf{L}_i\bm{x}-\bm{r}_i)+\frac{1}{2\gamma}\| \mathsf{L}_i\bm{x}-\bm{r}_i\|_2^2. 
\end{equation}
After some algebraic manipulations and adding the term $\frac{1}{2}\bm{z}_i^{H}\bm{z}_i$ to \eqref{upd21} we get
\begin{equation}\label{upd22}
\bm{r}_i^{*}=\arg\min_{\bm{r}_i\in\mathbb{C}^N} \gamma f_i(\bm{r}_i)+\frac{1}{2}\|\bm{r}_i - (\mathsf{L}_i\bm{x}+\bm{z}_i)\|_2^2,
\end{equation}
which is nothing but the proximity operator of $\gamma f_i$ applied to $\mathsf{L}_i\bm{x}+\bm{z}_i$. Thus, the minimizer with respect to $\bm{r}_i$ is computed as
\begin{equation}\label{upd2}
\bm{r}_i^{*}=\mathrm{prox}_{\gamma f_i}( \mathsf{L}_i\bm{x} + \bm{z}_i).
\end{equation}

The maximization over the dual variables is performed using a gradient ascend method. Again the optimization with respect to $\bm{z}_i$ can be carried out  simultaneously for all $i$ since the problems are decoupled. Thus, for a fixed $i$ the problem becomes
\begin{equation}
\bm{z}_i^{*}=\arg\max_{\bm{z}_i} \mathcal{J}=\arg\max_{\bm{z}_i}\bm{z}_i^{\dagger}(\mathsf{L}_i\bm{x}^{*} - \bm{r}_i^{*}).
\end{equation}
The gradient of $ \mathcal{J}$ with respect to $\bm{z}_i$ is given by $\mathsf{L}_i\bm{x}^{*} - \bm{r}_i^{*}$. Therefore, the dual ascend method yields updates of the form
\begin{equation}\label{upd3}
\bm{z}_{i}^{(t)}=\bm{z}_{i}^{(t-1)} + \mathsf{L}_i\bm{x}^{*} - \bm{r}_i^{*},
\end{equation}
for each iteration of the algorithm, where $t$ denotes the iteration variable. 

Note that the above described procedure can be easily extended for $S$ functions, thus providing a flexible framework for incorporating additional prior information either in the form of convex constraints or as additional convex penalty functions. The expressions in \eqref{upd1}, \eqref{upd2} and \eqref{upd3} constitutes the main iteration steps in our SDMM based solver, which is detailed in the next section.

\subsection{Implementation details}\label{ssec:opdet}
The resulting algorithm is summarized in Algorithm~\ref{alg1} where $S=3$. The algorithm is run for a fixed number of iterations, $T_{\rm{max}}$, or until a stopping criteria is met. The algorithm is stopped if the relative variation between the objective function evaluated at successive solutions, $\zeta=|f_1(\mathsf{L}_1\hat{\bm{x}}^{(t)})-f_1(\mathsf{L}_1\hat{\bm{x}}^{(t-1)}) |/| f_1(\mathsf{L}_1\hat{\bm{x}}^{(t-1)}) |$, is smaller than some bound $\xi\in(0,1)$ and if the normalized residual $\nu=\| \bm{y}-\mathsf{L}_2\hat{\bm{x}}^{(t)}\|_2/\epsilon$ is within the interval $[1 -\tau ,1 +\tau]$ for some tolerance $\tau\in(0,1)$, $\tau \ll 1$. In our implementation we fix $\xi=10^{-3}$ and $\tau=10^{-1}$. 
{\color{red}
\begin{algorithm}[h!]
\caption{SDMM }\label{alg1}
\begin{algorithmic}[1]
\STATE Initialize $\gamma>0$, $\hat{\bm{x}}^{(0)}$ and $\bm{z}_i^{(0)}=\bm{0}$, $i=1,\dots,S$.
\STATE $\bm{r}_i^{(0)}=\mathsf{L}_i\hat{\bm{x}}^{(0)}$, $i=1,\dots,S$.
\STATE $\bm{x}_i^{(0)}=\mathsf{L}_i^{\dagger}\bm{r}_i^{(0)}$, $i=1,\dots,S$.
\FOR{$t=1,\dots,T_{\rm{max}}$}
\STATE $\hat{\bm{x}}^{(t)}=\mathsf{Q}^{-1}\sum_{i=1}^S\bm{x}_i^{(t-1)}$.
\FORALL{$i=1,\dots,S$}
\STATE $\bm{r}_i^{(t)}=\mathrm{prox}_{\gamma f_i}( \mathsf{L}_i\hat{\bm{x}}^{(t)}+ \bm{z}_i^{(t-1)})$.
\STATE $\bm{z}_i^{(t)}=\bm{z}_i^{(t-1)} + \mathsf{L}_i\hat{\bm{x}}^{(t)} - \bm{r}_i^{(t)}$.
\STATE $\bm{x}_i^{(t)}=\mathsf{L}_i^{\dagger}(\bm{r}_i^{(t)}-\bm{z}_i^{(t)})$.
\ENDFOR
\IF{$\hat{\bm{x}}^{(t)}$ meets halting criteria}
\STATE Break.
\ENDIF
\ENDFOR
\RETURN $\hat{\bm{x}}^{(t)}$
\end{algorithmic}
\end{algorithm} 
}

In the following we detail the computation of the proximity operators used in Algorithm~\ref{alg1}. To compute the proximity operator of $f_1$, let us first define it entrywise as follows:
$f_1(\bm{r}_1)=\|\mathsf{W}\bm{r}_1\|_{1}=\sum_{j=1}^{D}\omega_j|r_{1,j}|$,
where $\omega_j=\mathsf{W}_{jj}$ (since $\mathsf{W}$ is a diagonal positive matrix) and $|\cdot|$ denotes the norm of a complex number. Since $f_1$ can be split as the sum of independent components of $\bm{r}_1$, the proximity operator of $\gamma f_1(\bm{r}_1)$ is given by 
\begin{equation}\label{prox1}
\mathrm{prox}_{\gamma f_1}(\bm{r}_1)= \mathsf{S}_{\gamma}(\bm{r}_1)=\{\mathrm{prox}_{\gamma \omega_j|\cdot|}(r_{1,j})\}_{1\leq j \leq D},
\end{equation}
where $\mathrm{prox}_{\lambda|\cdot|}$ is the entrywise soft-thresholding operator defined as
$\mathrm{prox}_{\lambda|\cdot|}(a)=\frac{a}{|a|}(|a|-\lambda)^{+}$,
with $(\cdot)^{+}=\max(0,\cdot)$. The proximity operator of $f_2(\bm{r}_2)=i_{B}(\bm{r}_2)$ is the projector onto the convex set $B=\{ \bm{r}_2\in\mathbb{C}^{M} : \| \bm{y}-\bm{r}_2\|_{2}\leq\epsilon \}$, and is computed as
\begin{equation}\label{prox2}
\mathrm{prox}_{\gamma f_2}(\bm{r}_2)=\min(1,\epsilon/\|\bm{r}_2\|_2)\bm{r}_2,
\end{equation}  
which is independent of $\gamma$. The proximity operator of $f_3(\bm{r}_3)$ is the projector onto the positive orthant and is given by 
\begin{equation}\label{prox3}
\mathrm{prox}_{\gamma f_3}(\bm{r}_3)=\left \{(r_{3,j})^{+}\right\}_{1\leq j \leq N},
\end{equation}
which is also independent of $\gamma$. See~\cite{combettes11} and references therein for derivation of these results. 

The bottleneck of Algorithm \ref{alg1}, in terms of computational resources, is the inversion of the matrix $\mathsf{Q}$. To invert this matrix we use the conjugate gradient algorithm \citep{saad03} to solve the system $\mathsf{Q}\hat{\bm{x}}^{(t)}=\sum_{i=1}^3\bm{x}_i^{(t-1)}$. The conjugate gradient algorithm is an iterative process that involves one matrix multiplication by $\mathsf{Q}$ at each iteration. Given that $\mathsf{Q}=\mathsf{\Phi}^{\dagger}\mathsf{\Phi}+\mathsf{\Psi}\mathsf{\Psi}^{\dagger}+I$, in general, each iteration requires one computation of the sensing operator $\mathsf{\Phi}$ and its adjoint, and, one computation of the sparsity operator $\mathsf{\Psi}$ and its adjoint. If we restrict the algorithm to use Parseval frames, i.e. $\mathsf{\Psi}\mathsf{\Psi}^{\dagger}=I$, the computation time can be considerably reduced since now $\mathsf{Q}=\mathsf{\Phi}^{\dagger}\mathsf{\Phi}+2I$. Examples of Parseval frames are orthogonal bases and the concatenation of orthogonal bases used in SARA.

Another important consideration in Algorithm \ref{alg1} is the choice of the penalty parameter $\gamma$. In theory any $\gamma >0$ guarantees convergence of the algorithm. However, in practice the convergence speed of the algorithm is severely affected by the value of this parameter. As it can be observed from the augmented Lagrangian function \eqref{cvx4}, small values of $\gamma$ place a large penalty on violations of primal feasibility, thus enforcing fast convergence of the dual variables $\bm{z}_i$. Conversely, large values of $\gamma$ place more weight on the original functions $f_i$, thus achieving a faster convergence rate on the objective function. Before discussing how to set the value of this parameter note that the proximity operators of $f_2$ and $f_3$, \eqref{prox2} and \eqref{prox3}, are independent of the value of $\gamma$ since $f_2$ and $f_3$ are indicator functions and the only effect of $\gamma$ in Algorithm \ref{alg1} is in the proximity operator of $f_1$. Therefore, $\gamma$ should scale with $\mathsf{\Psi}^{\dagger}\bm{x}^{*}$, where $\bm{x}^{*}$ denotes the true signal. Since $\bm{x}^{*}$ is unknown, we propose to set the penalty parameter as $\gamma=\beta\|\mathsf{\Psi}^{\dagger}\mathsf{\Phi}^{\dagger} \bm{y}\|_{\infty}$, i.e. a constant times the peak value of the dirty image in the sparsity domain. In our implementation we fix $\beta=10^{-3}$.

\subsection{Parallel and distributed optimization}
\label{ssec:parallel}
The SDMM structure offers several degrees of parallelization that can be further exploited. Firstly, the proximity operators can be implemented in parallel providing an acceleration factor of three. Secondly, as can be seen from \eqref{prox1}, \eqref{prox2} and \eqref{prox3}, the computation of the proximity operators is very simple and could support a high level of parallelization since it mostly involves simple entrywise operations. Finally, in the case of large-scale data problems, i.e. large number of visibilities $M\gg N$, the visibilities can no longer be processed on a single computer but rather in a computer cluster thus requiring a distributed processing of the data for the image reconstruction task. In this distributed scenario the data vector $\bm{y}$ and the measurement operator can be partitioned into $R$ blocks in the following manner:
\begin{equation}\label{blocks}
\bm{y}=
\begin{bmatrix}
\bm{y}_1\\
\vdots \\
\bm{y}_R
\end{bmatrix}
\textnormal{ and  }
\mathsf{\Phi}=
\begin{bmatrix}
\mathsf{\Phi}_1 \\
\vdots \\
\mathsf{\Phi}_R
\end{bmatrix},
\end{equation}
where $\bm{y}_i \in \mathbb{C}^{M_i}$, $\mathsf{\Phi}_i \in \mathbb{C}^{M_i \times N}$ and $M=\sum_{i=1}^R M_i$. Each $\bm{y}_i$ is modelled as $\bm{y}_i=\mathsf{\Phi}_i \bm{x} + \bm{n}_i$, where $\bm{n}_i\in \mathbb{C}^{M_i}$ denotes the noise vector. 

With this partition the optimization problem in \eqref{delta} can be rewritten as
\begin{equation}\label{dbp1}
\min_{\bar{\bm{x}}\in\mathbb{R}_{+}^{N}}\|\mathsf{W\Psi}^{\dagger}\bar{\bm{x}}\|_{1}
\textnormal{ subject to }\| \bm{y}_i-\mathsf{\Phi}_i\bar{\bm{x}}\|_{2}\leq\epsilon_i, i=1,\dots,R,
\end{equation}
where each $\epsilon_i$ is an appropriate bound for the  $\ell_2$ norm of the noise term $\bm{n}_i$. Observe that \eqref{dbp1} can be solved by SDMM (Algorithm \ref{alg1}) if we reformulate the problem as
\begin{equation}\label{dbp2}
\min_{\bm{x}\in\mathbb{C}^{N}} f_1(\mathsf{L}_1\bm{x})+\ldots +f_S(\mathsf{L}_S\bm{x}),
\end{equation}
with $S=R+2$. In this formulation $f_1$ and $f_2$ denote the $\ell_1$ sparsity term and the positivity constraint respectively, and $f_3$ to $f_S$ denote the $R$ data fidelity constraints. Thus $\mathsf{L}_1=\mathsf{\Psi}^{\dagger}$, $\mathsf{L}_2=\mathsf{I}$ and $\mathsf{L}_{i+2}=\mathsf{\Phi}_i$ for $i=1,\dots,S$. Note that steps 7 to 9 in Algorithm \ref{alg1} can be computed in parallel for each $i$. The advantages of this distributed optimization approach are: (i) the visibilities $\bm{y}_i$ and the measurement operators $\mathsf{\Phi}_i$ are local to each node in the cluster, therefore the memory requirements are distributed among  $R$ nodes, with a data dimensionality $M_i \ll M$; (ii) the measurement operators $\mathsf{\Phi}_i$, and their adjoint, are applied locally at each node thus distributing the processing load, for acceleration of the reconstruction process; (iii) the central processing node, where the global update $\hat{\bm{x}}^{(t)}=\mathsf{Q}^{-1}\sum_{i=1}^S\bm{x}_i^{(t-1)}$ is computed, and the parallel nodes, where the local updates $\bm{x}_i^{(t-1)}$ are computed, only need to exchange information of the size of the image vector at each iteration rather than of the size of the visibilities, thus alleviating the communication requirements to transfer information between nodes. Note that the composite operator $\mathsf{\Phi}^{\dagger}\mathsf{\Phi}$, needed in the conjugate gradient solver for the global update, can be applied in parallel by each node since $\mathsf{\Phi}^{\dagger}\mathsf{\Phi}=\sum_{i=1}^R\mathsf{\Phi}_i^{\dagger}\mathsf{\Phi}_i$. Although this approach would distribute the processing load of the conjugate gradient step into the parallel nodes, it would incur in a communication overhead since each parallel node needs to communicate its result at each iteration of the conjugate gradient algorithm. One approach that can be used to avoid this situation is to precompute and store the composite operator $\mathsf{\Phi}^{\dagger}\mathsf{\Phi}$ in the central processing node. The aforementioned distributed optimization approach could be very appealing for next-generation telescopes where massive amounts of data are acquired. These distributed optimization ideas are not implemented in the beta version of PURIFY, discussed in Section \ref{sec:purify}, and are the subject of ongoing work.

\section{The PURIFY package}
\label{sec:purify}
PURIFY\footnote{Package available at \url{http://basp-group.github.io/purify/}.} is a collection routines written in C that implements different tools for RI imaging including file handling (for both visibilities and fits images), implementation of the measurement operator and set-up of the different optimization problems used for image deconvolution. The code calls the generic Sparse OPTimization (SOPT\footnote{Package available at \url{http://basp-group.github.io/sopt/}.}) package, which is also written in C, to solve the imaging optimization problems. In the following we describe the different features included in PURIFY and SOPT. Note that the name PURIFY has no other meaning than that of a powerful alternative to CLEAN.

The optimization problems solved by SOPT within the SDMM structure are: (i) the weighted $\ell_1$ minimization problem in \eqref{delta} and (ii) the weighted TV minimization problem similar to \eqref{tvproblem} but with the TV norm replaced by a by a weighted TV norm defined as $\|\bar{\bm{x}}\|_{\rm{WTV}} = \| \mathsf{W}\nabla \bar{\bm{x}} \|_1$ where $\mathsf{W}$ is a matrix with positive weights applied to the image gradient. The non-reweighted problems can be solved just by setting the weight matrix to the identity matrix. In the case of the reweighted TV problem $f_1(\bm{x})=\|\bar{\bm{x}}\|_{\rm{WTV}}$, with the proximity operator computed using the fast first order iterative method described in \cite{beck09b}. For the $\ell_1$ problems a set of different dictionaries is supported, including: the Dirac basis, the Daubechies wavelets family and the concatenation of any of these bases. 

For the measurement operator, PURIFY implements a non-uniform FFT that maps a discrete image into continuous visibilities \citep{greengard04}. The operator is defined as
\begin{equation}\label{opmodel}
\mathsf{\Phi}=\mathsf{GFDZ}\mathsf{B}.
\end{equation}
The matrix $\mathsf{B}\in\mathbb{R}^{N\times N}$ is the diagonal matrix implementing the primary beam. The operator $\mathsf{Z}\in\mathbb{R}^{N'\times N}$ denotes the zero padding operator with $N'=kN$ and $k\geq 2$ needed to compute the discrete Fourier transform of $\bm{x}$ on an oversampled grid and achieve higher accuracy. The unitary matrix $\mathsf{F}\in\mathbb{C}^{N'\times N'}$ denotes the discrete Fourier transform. The matrix $\mathsf{G}\in\mathbb{R}^{M\times N'}$ represents a convolutional interpolation operator to model the map from a discrete frequency grid onto the continuous plane so that the FFT can be used to implement $\mathsf{F}$. PURIFY supports a Gaussian kernel in the frequency domain with a compact support, but support for other convolutional interpolation kernels can easily be included. Due to the kernel's compact support, the matrix $\mathsf{G}$ is highly sparse therefore allowing fast matrix-vector multiplications. The operator $\mathsf{D}\in\mathbb{R}^{N'\times N'}$ is a diagonal matrix that in practice implements a discrete version of the reciprocal of the inverse Fourier transform of the interpolation kernel, i.e. $d=1/\hat{g}$, where $\hat{g}$ denotes the inverse Fourier transform of the continuous interpolation kernel. The idea behind this procedure is to undo the effects of the convolution by the interpolation kernel in the frequency domain by dividing by the inverse Fourier transform of the interpolation kernel in the spatial domain. This operator and its adjoint are implemented in the package. Although the current version of PURIFY only supports the Gaussian kernel, other interpolation kernels, such as prolate spheroidal wave functions \citep{thompson01}, will be incorporated in future versions. 
 
Also note that our framework can easily incorporate DDEs, in particular the $w$-component effect, as additional convolution kernels in the frequency plane entering the matrix $\mathsf{G}$. Again, compact support of those kernels will ensure sparsity of $\mathsf{G}$, in turn ensuring its necessary fast implementation. This represents an alternative to the $w$-projection and the $\mathsf{A}$-projection algorithms \citep{bhatnagar08b,bhatnagar08}. See \cite{wolz13} for first steps in these directions.

Careful attention has been paid to the design of the interfaces of PURIFY. The solvers receive the measurement operators as pointers to functions implementing the forward and adjoint operators with a generic signature, thus other measurements operators can easily be used. Weighting matrices, such as complex antenna gains and natural or uniform weighting matrices, are not supported in the current implementation but their incorporation into the measurement operator is straightforward. The same philosophy is adopted for the sparsity operators allowing the incorporation of any sparsity dictionary. These interfaces will facilitate direct integration with standard packages for interferometric imaging such as CASA\footnote{\url{http://casa.nrao.edu/}.}.

The current version of SOPT does not exploit the parallel structure of SDMM. Firstly, the proximity operators are implemented in a serial manner rather than in parallel. Secondly, the computation of each proximity operators is implemented serially rather than in parallel thus not exploiting its separable structure. The only parallel structure that is exploited is the implementation of the sparsity averaging operator in SARA, i.e. each decomposition on the basis in the operator are computed in parallel. Therefore the highly redundant dictionary in SARA has an implementation as fast as a single orthonormal basis, which already represents a significant advantage. 
As discussed in  Section \ref{ssec:opdet}, the computation of the measurement operator $\mathsf{\Phi}$ is a major bottleneck for very high dimensional problems. In this case
the measurement operator $\mathsf{\Phi}$ can be parallelized by implementing a parallel matrix-vector product for the sparse matrix $\mathsf{G}$, e.g. partitioning $\mathsf{G}$ into several blocks $\mathsf{G}_i$ as done in \eqref{blocks} for $\mathsf{\Phi}$. Similar strategies might be adopted for the sparsity operator $\mathsf{\Psi}$. As discussed in Section \ref{ssec:opdet} the global update $\hat{\bm{x}}^{(t)}=\mathsf{Q}^{-1}\sum_{i=1}^S\bm{x}_i^{(t-1)}$ is the main bottleneck of the algorithm. One approach that could be implemented here is to precompute and store the sparse matrix $\mathsf{G}^{\dagger}\mathsf{G}=\sum_{i=1}^R\mathsf{G}_i^{\dagger}\mathsf{G}_i$ to accelerate the conjugate gradient solver\footnote{Note that \cite{sullivan12} also proposed to precompute $\mathsf{G}^{\dagger}\mathsf{G}$ to accelerate a CLEAN-based algorithm.}. These optimizations are the subject of ongoing work. 

\section{Simulations and results}
\label{sec:Simulations and results}

\begin{figure*}
    \centering
     
    \includegraphics[trim = 3.4cm 1.1cm 2.1cm 0.5cm, clip, keepaspectratio, width = 5.5cm]{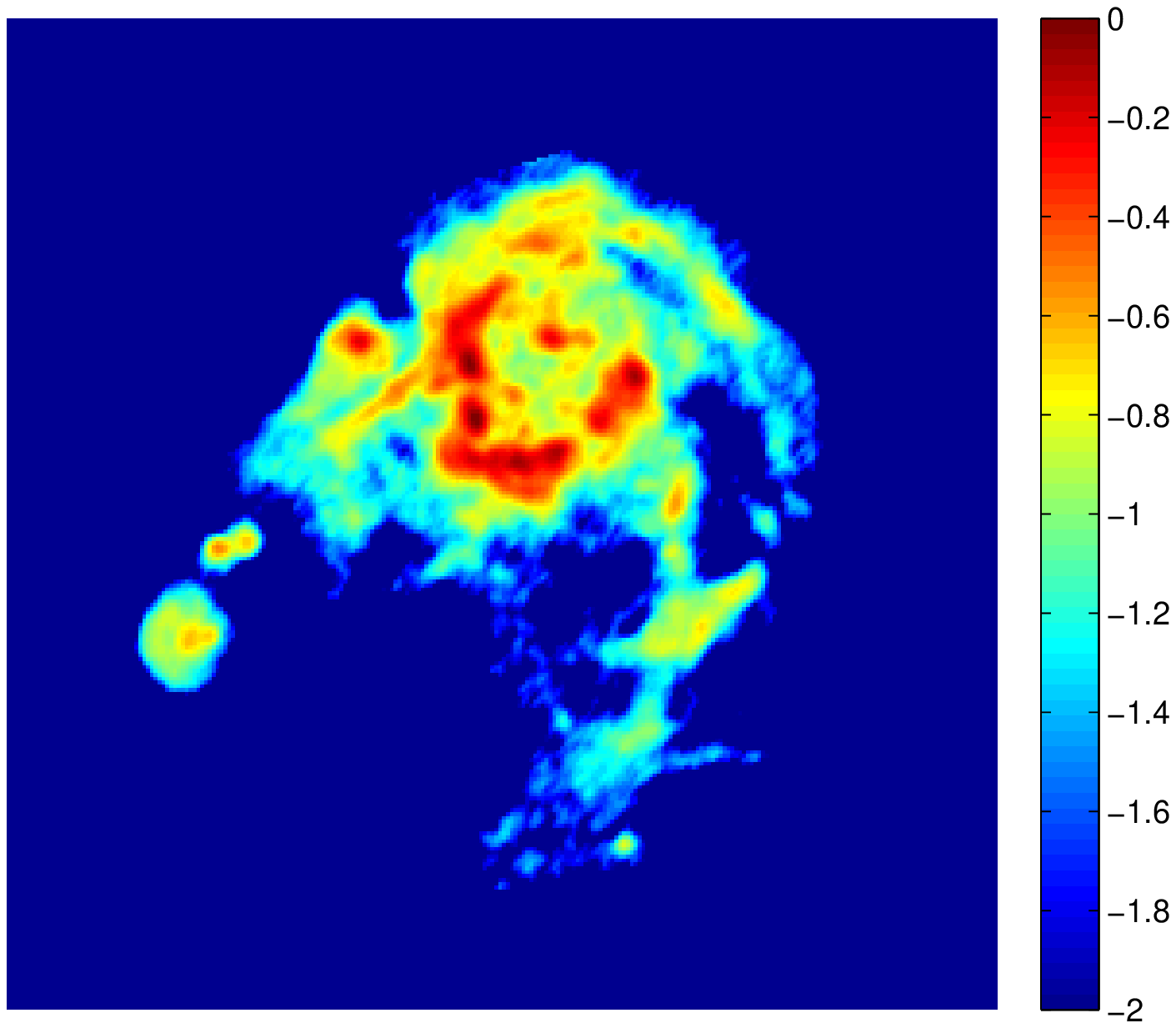}
    \includegraphics[trim = 3.4cm 1.1cm 2.1cm 0.5cm, clip, keepaspectratio, width = 5.5cm]{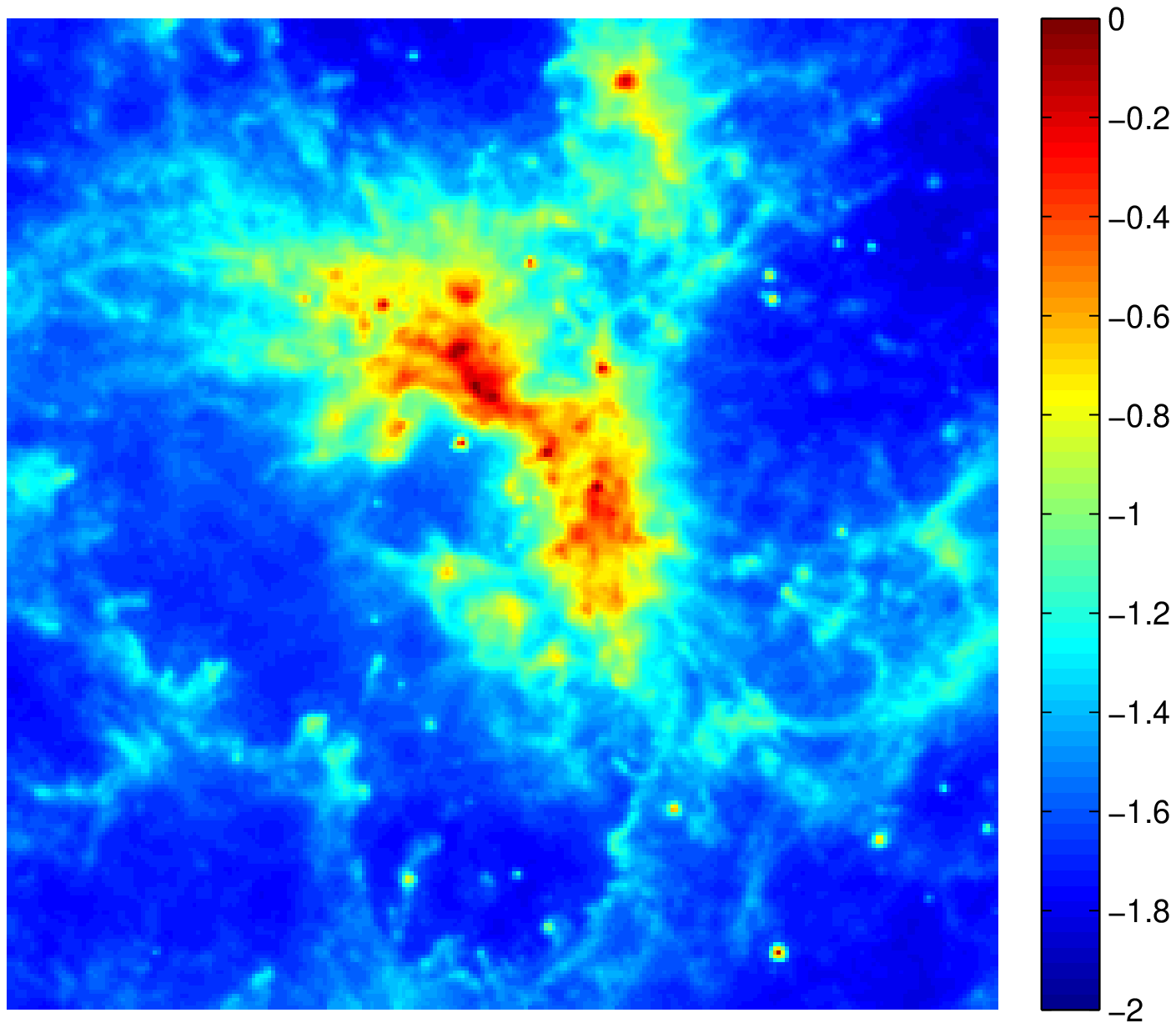}
    \includegraphics[trim = 3.4cm 1.1cm 2.1cm 0.5cm, clip, keepaspectratio, width = 5.5cm]{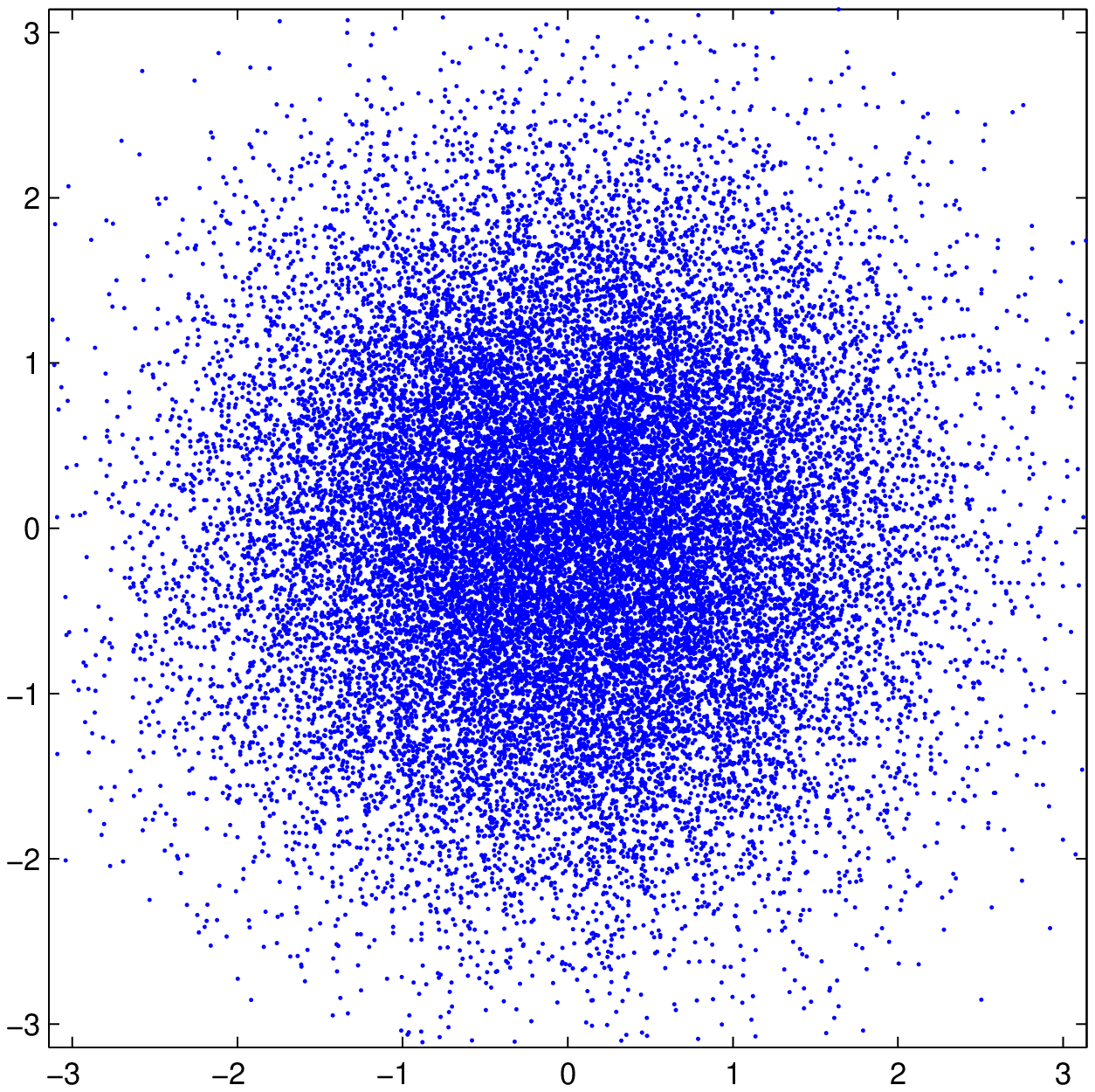}

\caption{Left and middle panels: original 256$\times$256 test images, M31 (left) and 30Dor (middle), shown in a $\log_{10}$ scale with brightness values in the interval $[0.01, 1]$. Right panel: Example of a simulated variable density coverage in the Fourier plane ($M=26374\approx 0.4N$).}
\label{fig:1}
\end{figure*}

In this section we illustrate the performance of the imaging algorithms implemented in PURIFY by recovering well known test images from simulated continuous frequency visibilities. The test images used in all simulations are M31, based on a HII region in the M31galaxy, and 30Dor, the 30 Doradus in the Large Magellanic Cloud. These images present different compact and extended structures thus being good candidates to evaluate different regularization priors. Figure~\ref{fig:1} shows the 256$\times$256 discrete models of M31 (left) and 30Dor (middle) used as ground truth images\footnote{Available at \url{http://casaguides.nrao.edu/index.php}.}. 

For our evaluation we compare constrained $\ell_1$ and TV minimization problems, as well as their reweighted versions, in terms of reconstruction quality and computation time. For the $\ell_1$ problems we study three different dictionaries $\mathsf{\Psi}$ in \eqref{delta}: the Dirac basis, the Daubechies 8 wavelet basis and the Dirac-Db1-Db8 concatenation highlited for the SARA algorithm in Section~\ref{ssec:RICS}. The associated algorithms are respectively denoted BP, BPDb8 and BPSA for the non-reweighted case. The reweighted versions are respectively denoted RWBP, RWBPDb8 and SARA. We also study the TV minimization problem in \eqref{tvproblem} with the additional constraint that $\bar{\bm{x}}\in\mathbb{R}_{+}^N$, denoted as TV, and its reweighted version, denoted as RWTV. Recall that $\ell_1$ minimization with a Dirac basis yields reconstruction qualities similar to CLEAN, thus we use BP as a proxy for CLEAN. Also, we use BPDb8 as a proxy for MS-CLEAN reconstruction quality since \cite{li11} reported that the isotropic undecimated wavelet transform outperformed MS-CLEAN and \cite{carrillo12} reported that BPDb8 outperformed the isotropic undecimated wavelet transform in the discrete setting.

We use as reconstruction quality metric the signal to noise ratio (SNR), which is defined as: 
\begin{equation}
\mathrm{SNR}=20\log_{10}\left( \frac{\|\bm{x}\|_2}{\|\bm{x}-\hat{\bm{x}}\|_2}\right)
\end{equation}
where $\bm{x}$ and $\hat{\bm{x}}$ denote the the original image and the estimated image respectively. The visibilities are corrupted by complex Gaussian noise with a fixed input SNR set to 30~dB. The input SNR is defined as $\mathrm{ISNR}=20\log_{10}(\| \bm{y}_0\|_2/\| \bm{n}\|_2)$, where $\bm{y}_0$ identifies the clean measurement vector. Assuming visibilities corrupted by i.i.d.~complex Gaussian noise with variance $\sigma_n$, the bound on the $\ell_{2}$ norm term in \eqref{delta}, $\epsilon$, is identical to a bound on a $\chi^{2}$ distribution with $2M$ degrees of freedom. Therefore, we set this bound as
 $\epsilon^2=(2M+4\sqrt{M})\sigma_n^2/2$, where $\sigma^2_n/2$ is the variance of both the real and imaginary parts of the noise. This choice provides a likely bound for $\|\bm{n}\|_2$ \citep{carrillo12}. We use the measurement operator described in \eqref{opmodel} with $\mathsf{B}=\mathsf{I}$ and an oversampling factor $k=2$.
 
\begin{figure*}

\centering
    \begin{tabular}{cc}
   
    \includegraphics[trim = 1.0cm 0.6cm 0.8cm 0.8cm, clip, keepaspectratio, width = 8.7cm]{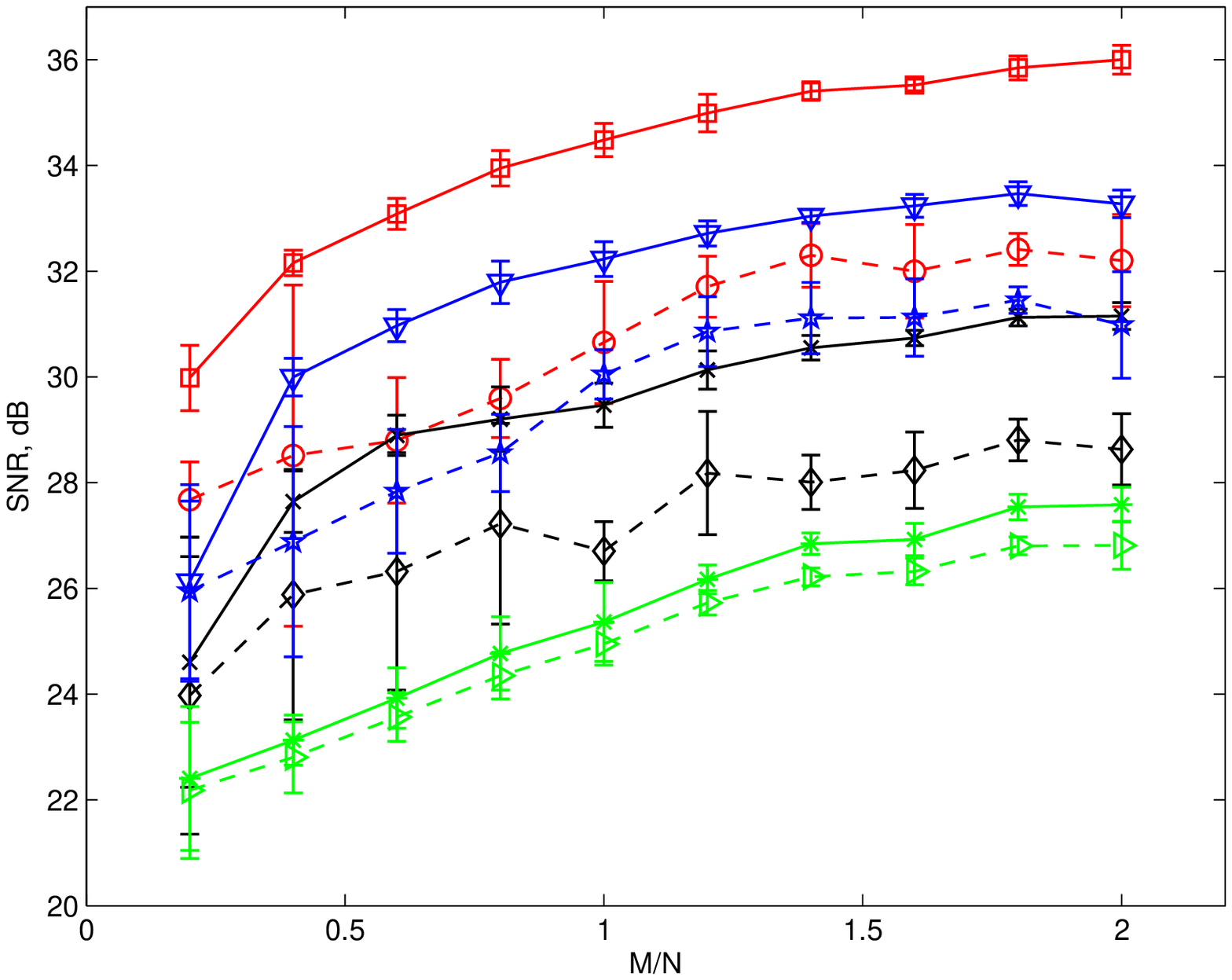}&
    \includegraphics[trim = 1.0cm 0.6cm 0.8cm 0.8cm, clip, keepaspectratio, width = 8.7cm]{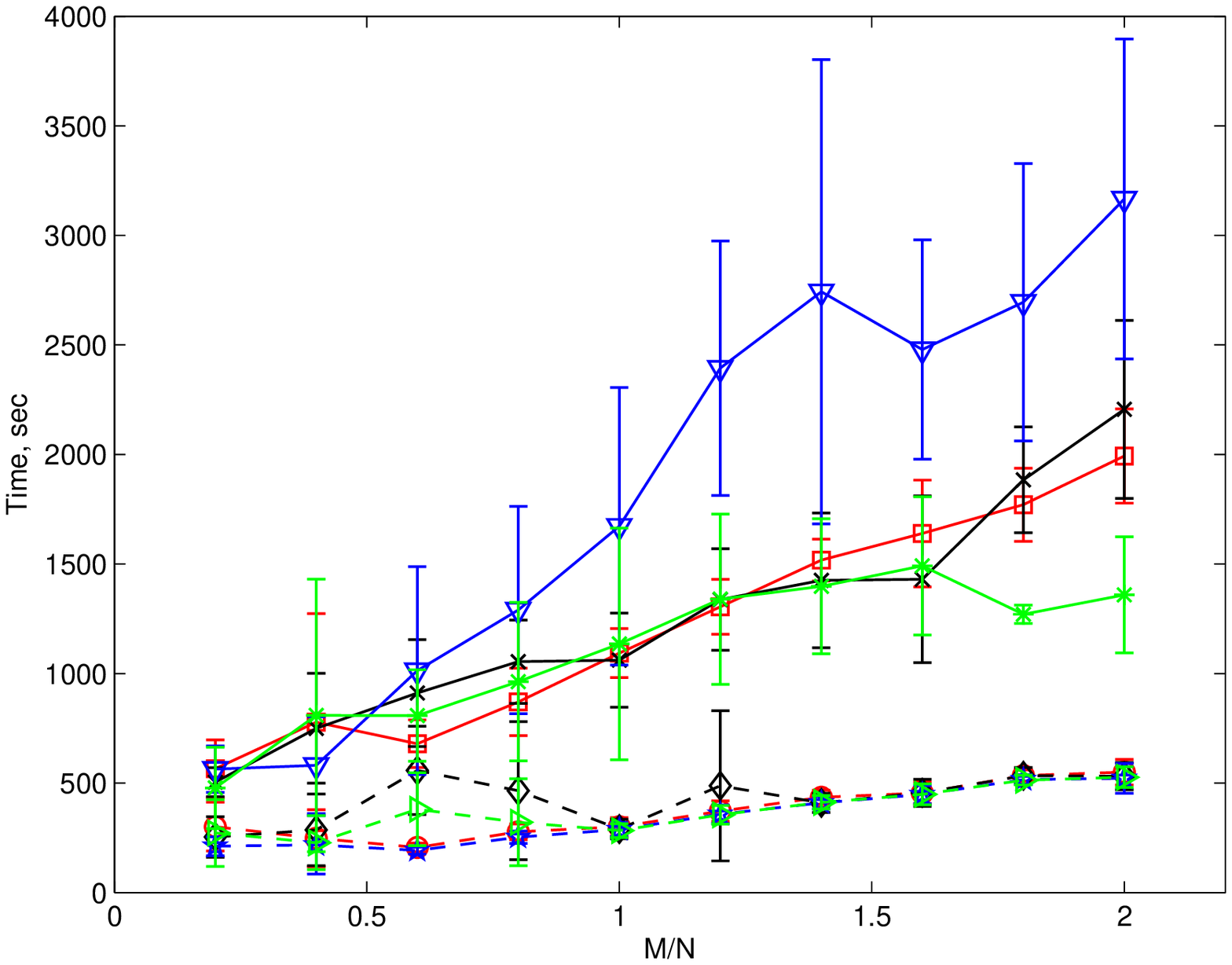}\\
    \includegraphics[trim = 1.0cm 0.6cm 0.8cm 0.8cm, clip, keepaspectratio, width = 8.7cm]{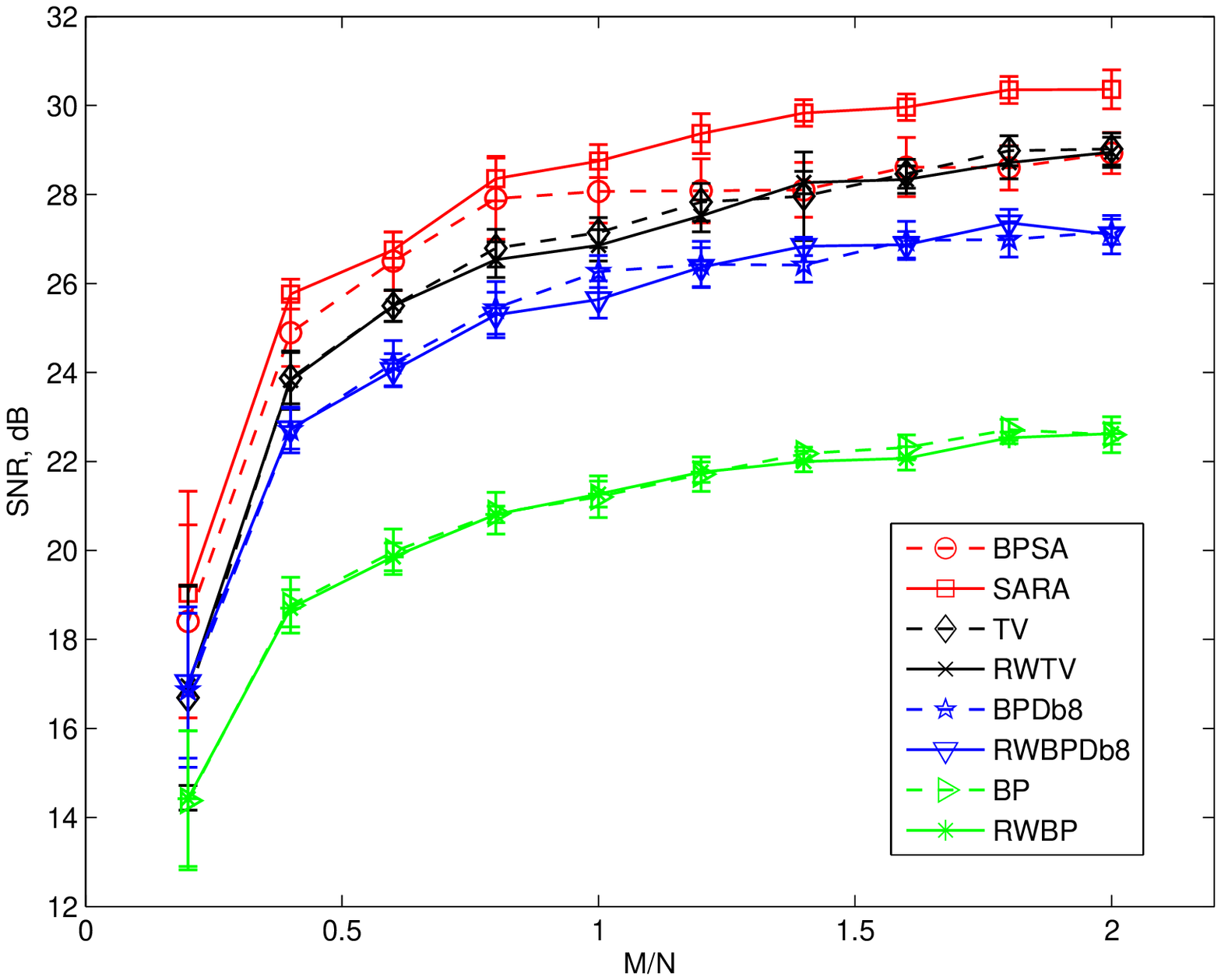}&
    \includegraphics[trim = 1.0cm 0.6cm 0.8cm 0.8cm, clip, keepaspectratio, width = 8.7cm]{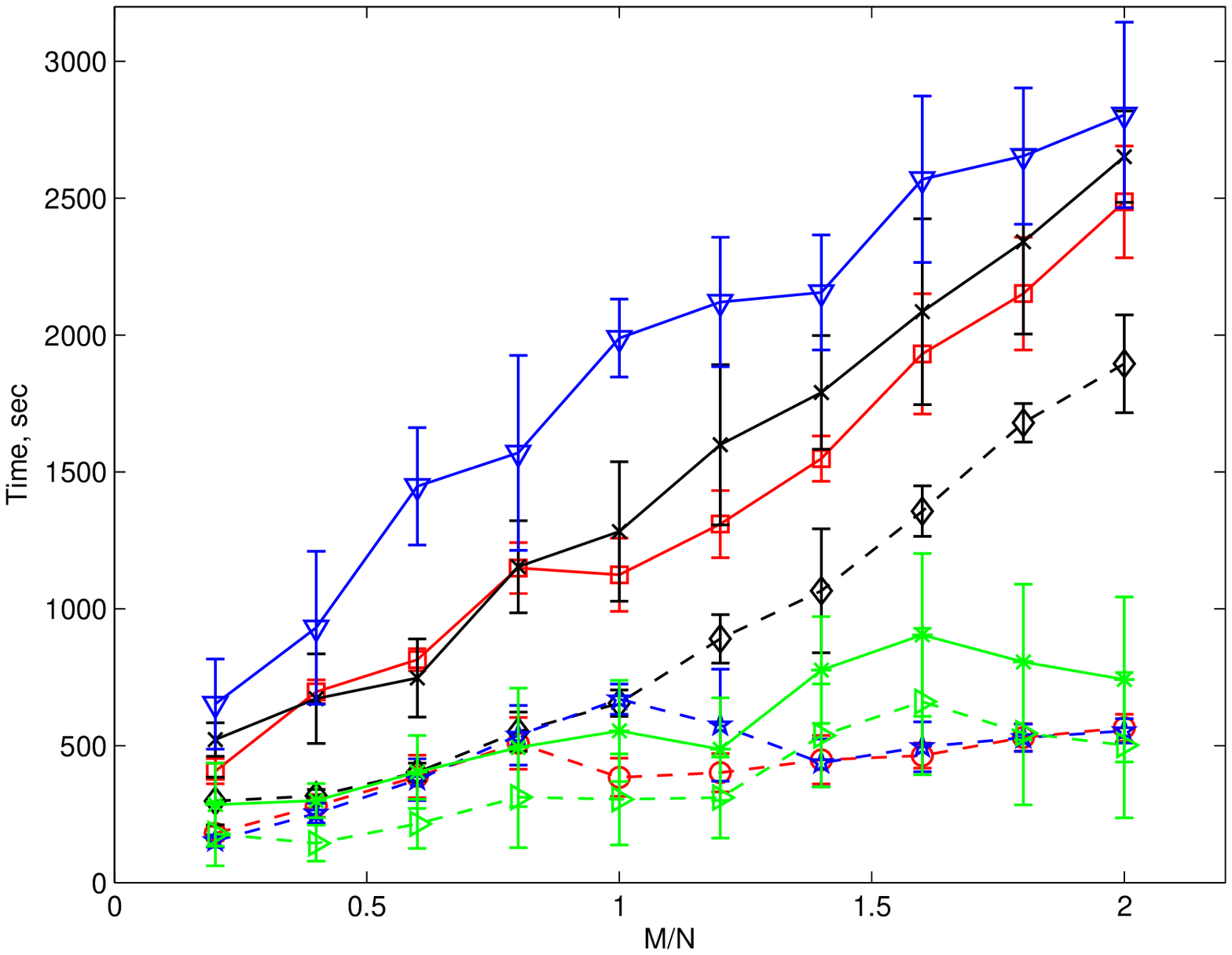}\\
    
    \end{tabular}

\caption{Reconstruction results for M31 (top row) and 30Dor (bottom row) 256$\times$256 test images. Left column: average reconstruction SNR against normalized number of visibilities $M/N$. Right column: average computation time. Vertical bars identify one standard deviation errors around the mean over 30 simulations. The input SNR is set to 30~dB. The results show that SARA outperforms all other methods in terms of reconstruction quality for both images.}
\label{fig:2}
\end{figure*}

The first experiment in this section considers incomplete visibility coverages generated by random variable density sampling profiles. Such profiles are characterized by denser sampling at low spatial frequencies than at high frequencies. This choice mimics common generic sampling patterns in radio interferometry. In order to make the simulated coverages more realistic we suppress the $(0,0)$ component of the Fourier plane from the measured visibilities. This generic profile approach allows us to make a thorough study of the reconstruction quality of the imaging algorithms with a large numbers of simulations for arbitrary number of visibilities and without concern for various telescope configurations. We vary the number of visibilities from $M=0.2N$ to $M=2N$. Reconstruction results for M31 and 30Dor are reported in the top and bottom rows of Figure~\ref{fig:2} respectively. Average values over 30 simulations and associated one standard deviation error bars are reported for all plots. 

The left panel of Figure~\ref{fig:2} shows SNR results for M31 (top) and 30Dor (bottom). The results show that SARA outperforms all other methods in reconstruction quality for both images. This confirms previous results reported by \cite{carrillo12} in the discrete case now for the more realistic continuous Fourier setting, including the case when $M>N$. Interestingly, BPSA shows the best reconstruction quality over all non-reweighted methods for both images. The results for M31, which exhibits a compact support with some extended structures, show that the second best method is RWBPDb8 having SNRs at most 4 dB below SARA. The results for 30Dor, which is a more complicated image with both extended structures and compact structures, show that TV and RWTV offer a good model for continuous extended structures achieving SNRs at most 2 dB below SARA. Note that BP and its reweighted version do not achieve good results for this image, as expected since the Dirac basis is not a good model for extended structures, achieving SNRs at least 4 dB below all other methods for coverages above $M=0.2N$.

Computation times, on a 2.4 GHz Xeon quad core and using the current non-optimized software version, are reported in the right panel of Figure~\ref{fig:2} for M31 (top) and 30Dor (bottom). As expected the reweighted methods are most costly having reconstruction times ranging from tens of minutes for $M=0.2N$ to one hour for $M=2N$. Even though the concatenation of bases in SARA makes the algorithm structure more costly in theory, the parallel implementation of the bases in SARA yields a competitive algorithm in terms of computation time. In fact, the results show that RWBPDb8, with a single wavelet basis, is the slowest method and the most unstable with respect to convergence rate, as can be observed from the large error bars. This result indicates that RWBPDb8 might need more iterations to achieve convergence than other methods. RWTV reports similar reconstruction times to SARA. The results also show that the non-reweighted methods are fast, achieving reconstruction times below 10 minutes for all coverages, except for TV in 30Dor which has a similar behaviour as the reweighted methods. An interesting observation is that the reconstruction times scale linearly with the number of visibilities for the reweighted methods. This is due to the fact that the complexity of the SDMM algorithm is dominated by the cost of solving the linear system at step 5 of Algorithm~\ref{alg1}, which needs to apply the sensing operator $\mathsf{\Phi}$ and its adjoint at every iteration of the conjugate gradient algorithm. Therefore beyond having a fast implementation of $\mathsf{\Phi}$, alternative strategies to accelerate the solution of the linear system should be explored such as the use of preconditioned conjugate gradient solvers and faster implementations of the Gram matrix $\mathsf{\Phi}^{\dagger}\mathsf{\Phi}$.

\begin{figure*}
    \centering
     
    \includegraphics[trim = 3.4cm 1.1cm 2.1cm 0.5cm, clip, keepaspectratio, width = 5.5cm]{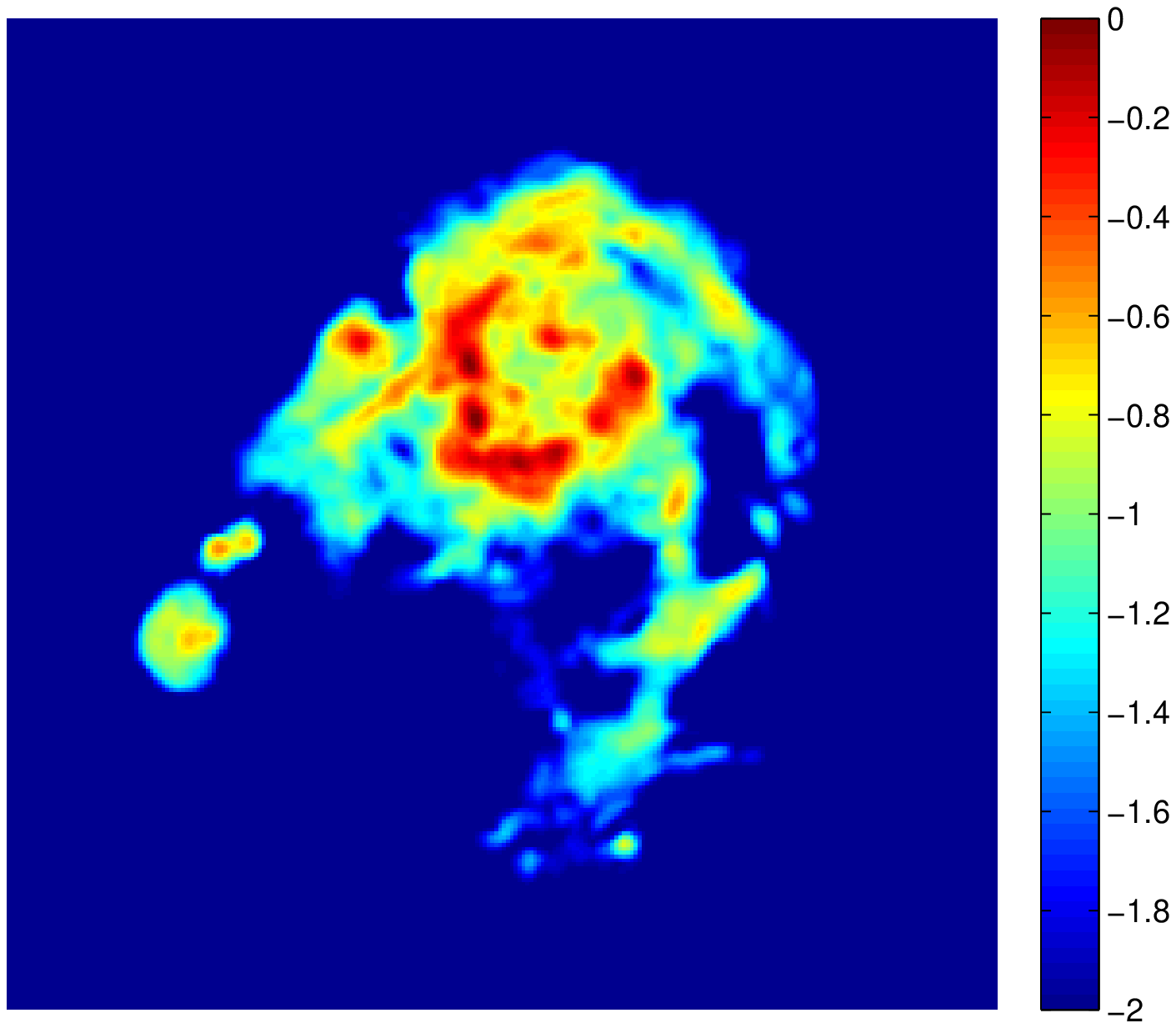}
    \includegraphics[trim = 3.4cm 1.1cm 2.1cm 0.5cm, clip, keepaspectratio, width = 5.5cm]{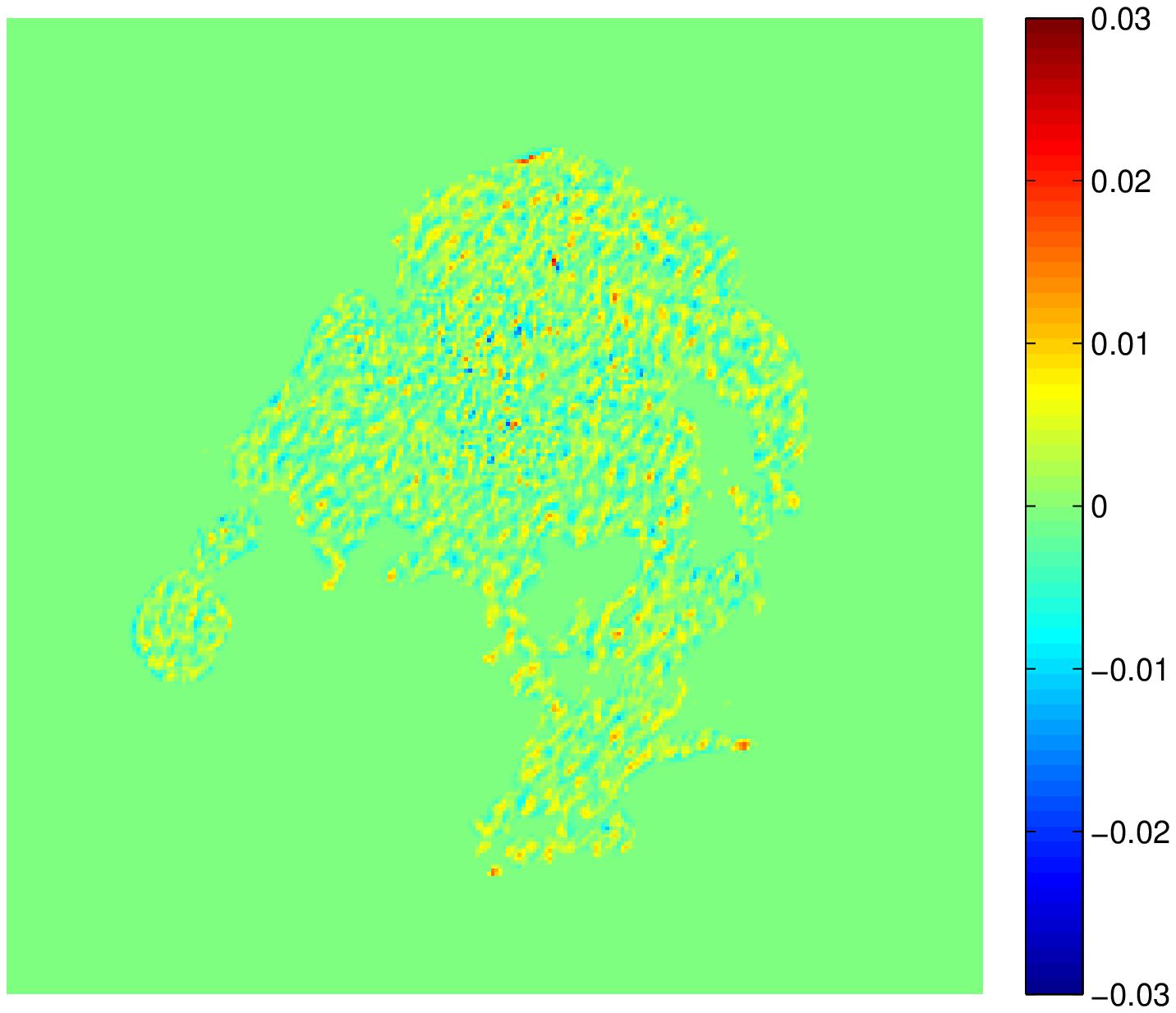}
    \includegraphics[trim = 3.4cm 1.1cm 2.1cm 0.5cm, clip, keepaspectratio, width = 5.5cm]{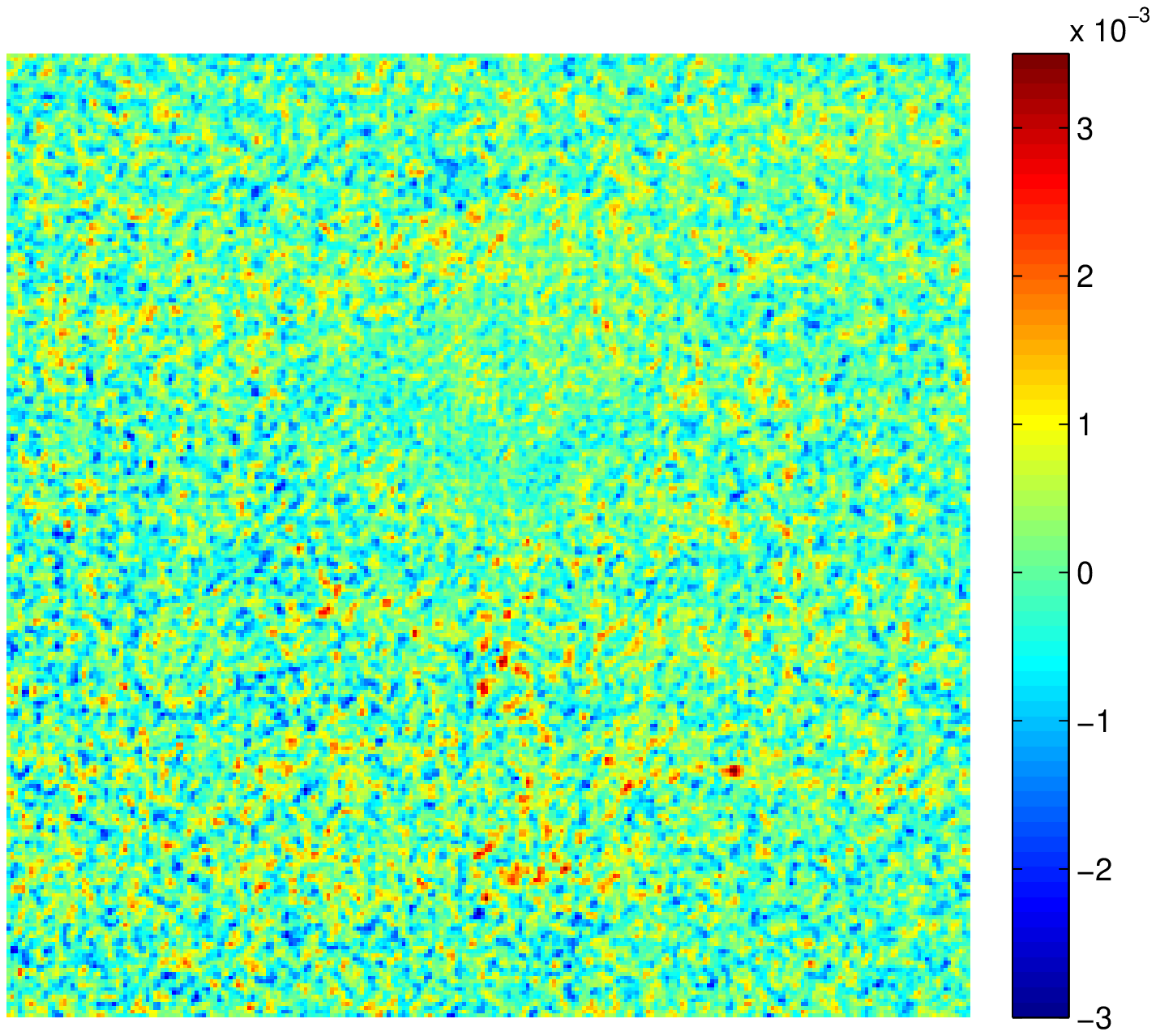}
    \includegraphics[trim = 3.4cm 1.1cm 2.1cm 0.5cm, clip, keepaspectratio, width = 5.5cm]{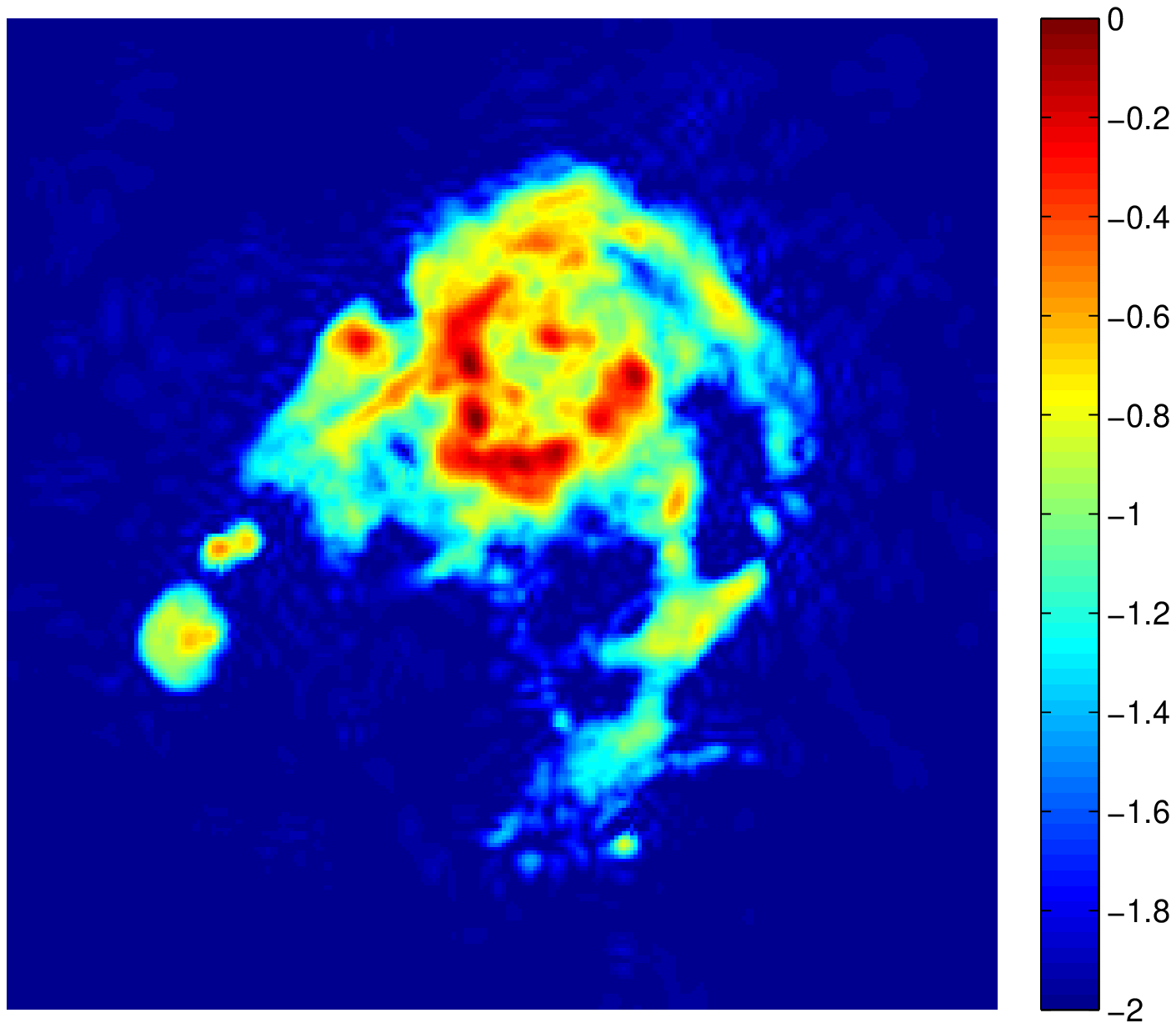}
    \includegraphics[trim = 3.4cm 1.1cm 2.1cm 0.5cm, clip, keepaspectratio, width = 5.5cm]{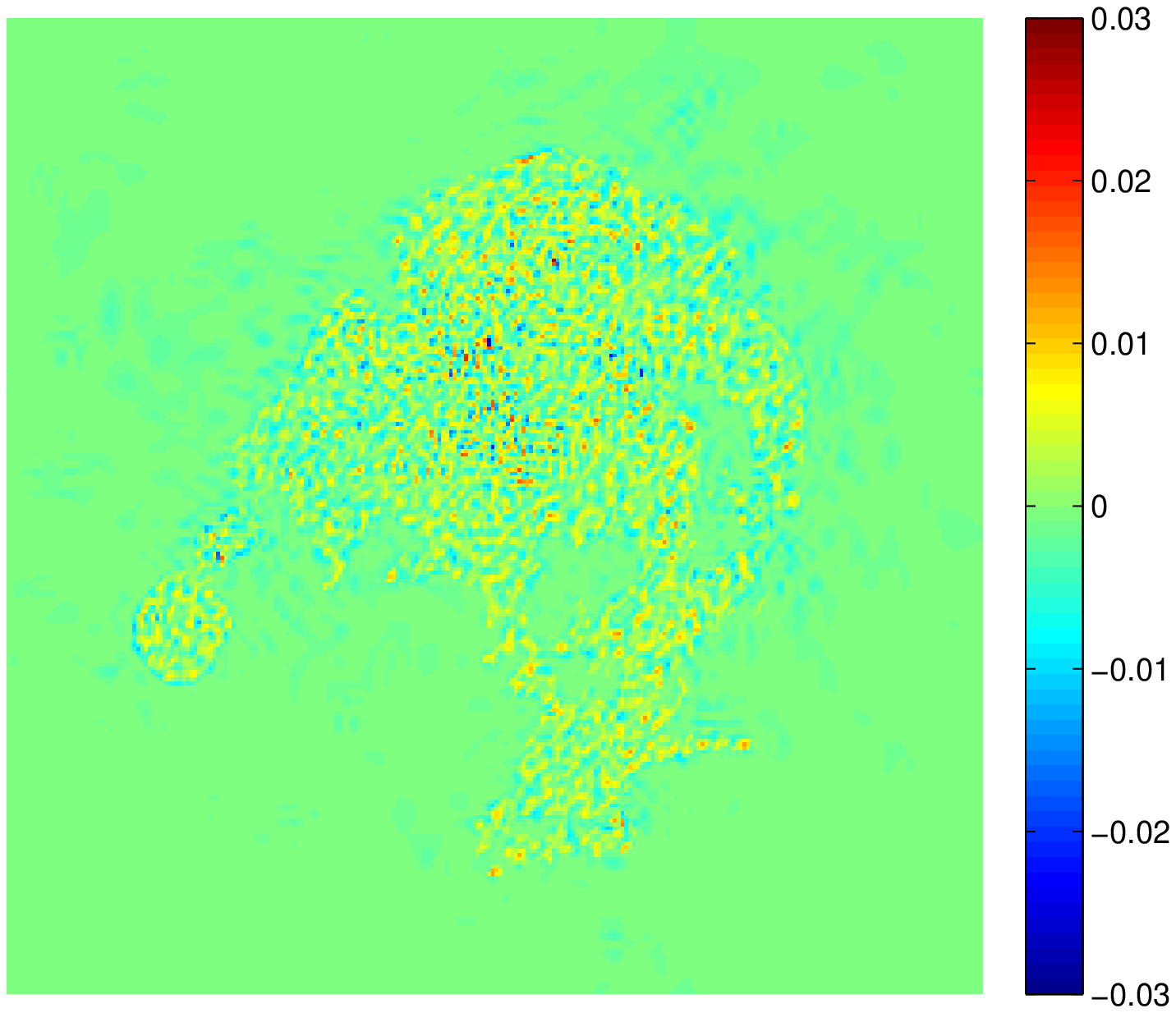}
    \includegraphics[trim = 3.4cm 1.1cm 2.1cm 0.5cm, clip, keepaspectratio, width = 5.5cm]{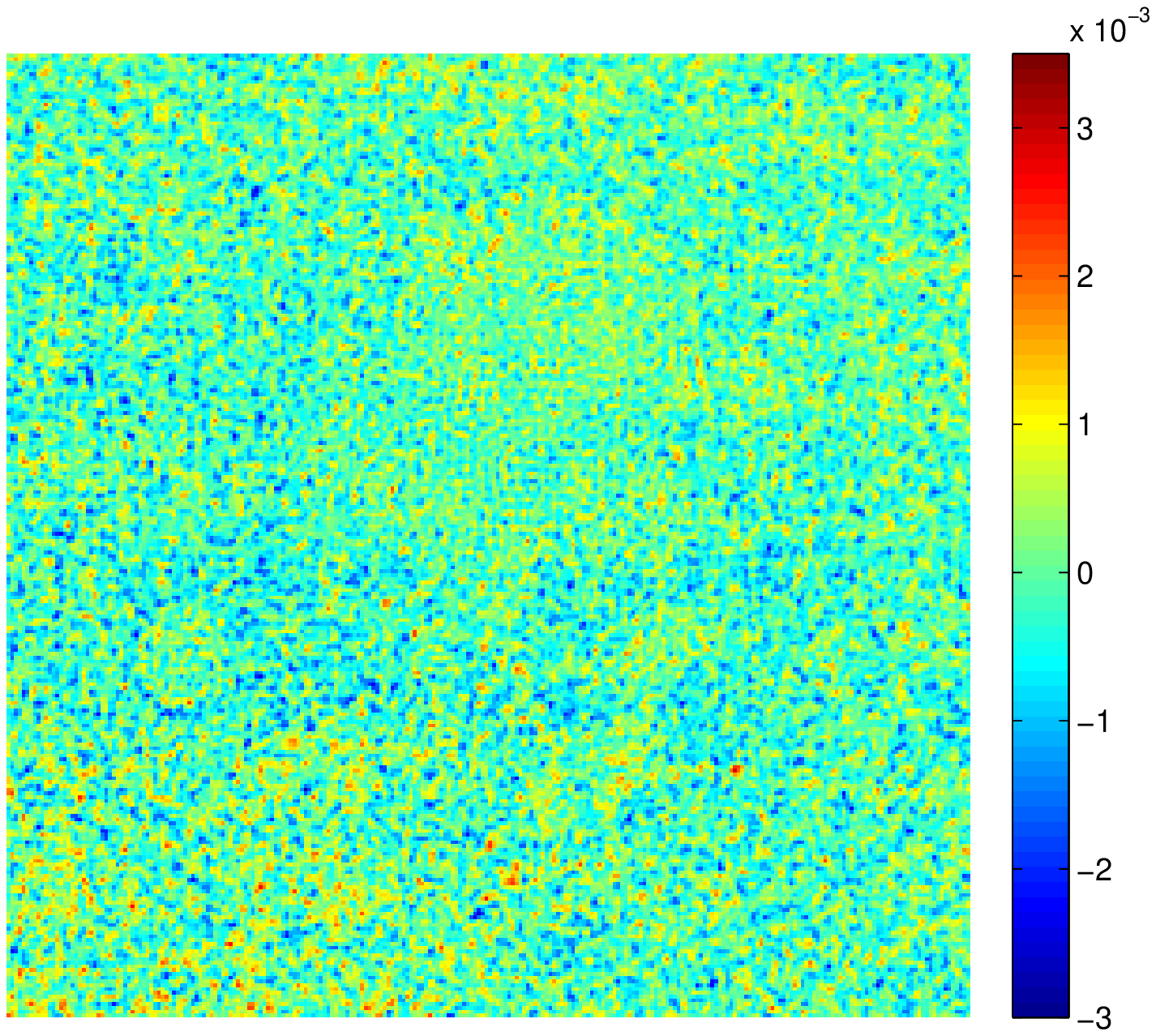}
    \includegraphics[trim = 3.4cm 1.1cm 2.1cm 0.5cm, clip, keepaspectratio, width = 5.5cm]{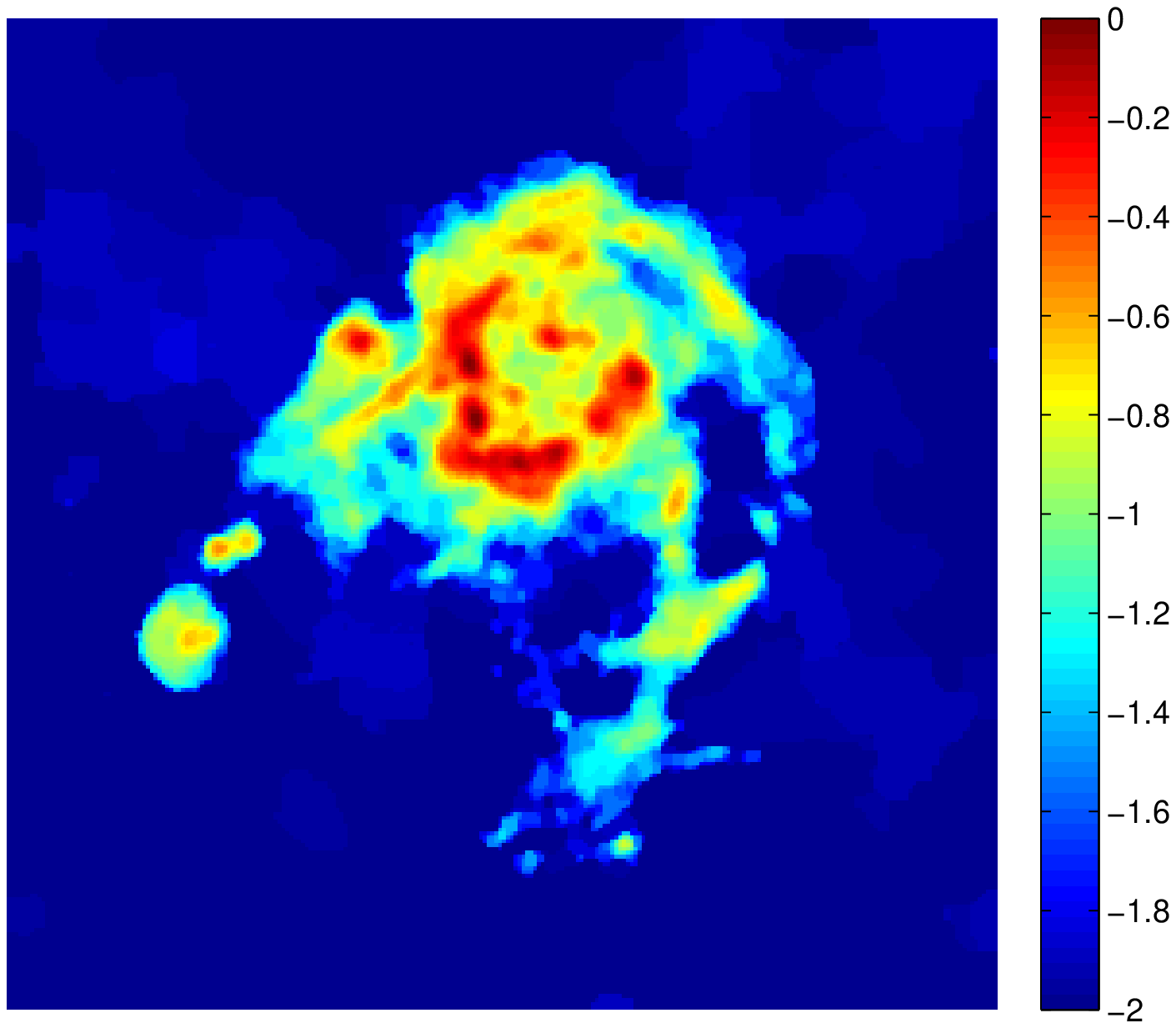}
    \includegraphics[trim = 3.4cm 1.1cm 2.1cm 0.5cm, clip, keepaspectratio, width = 5.5cm]{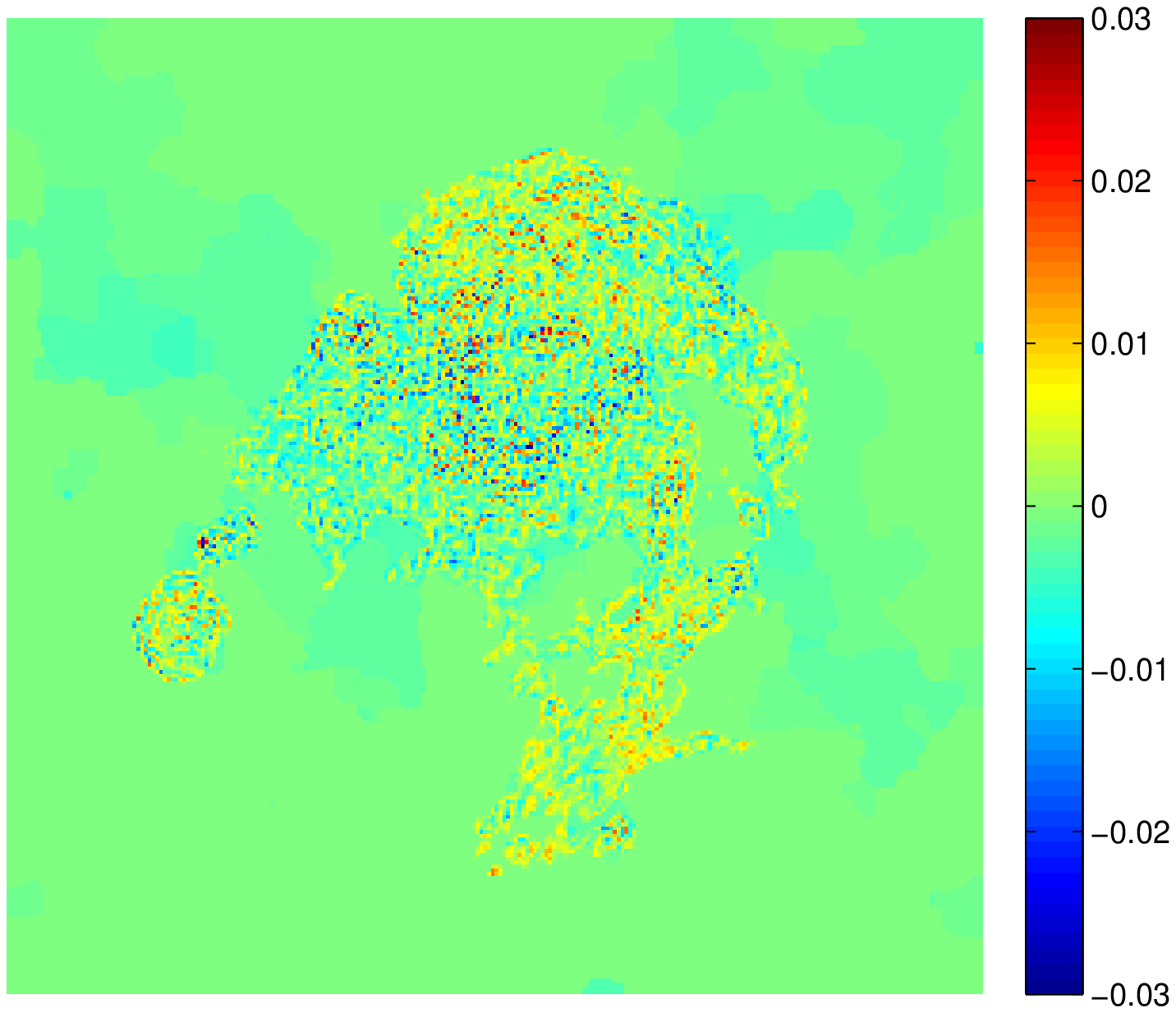}
    \includegraphics[trim = 3.4cm 1.1cm 2.1cm 0.5cm, clip, keepaspectratio, width = 5.5cm]{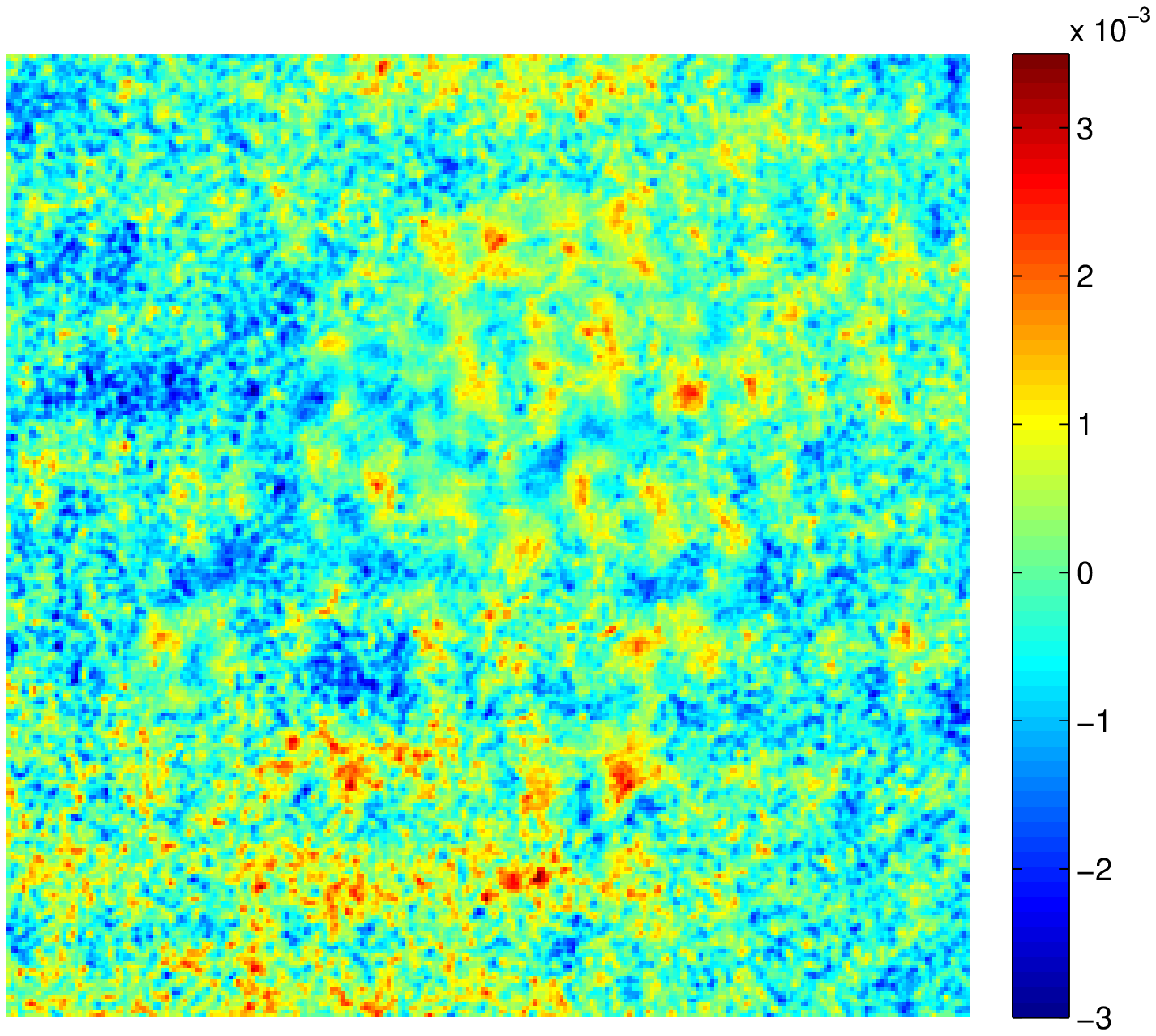}
    \includegraphics[trim = 3.4cm 1.1cm 2.1cm 0.5cm, clip, keepaspectratio, width = 5.5cm]{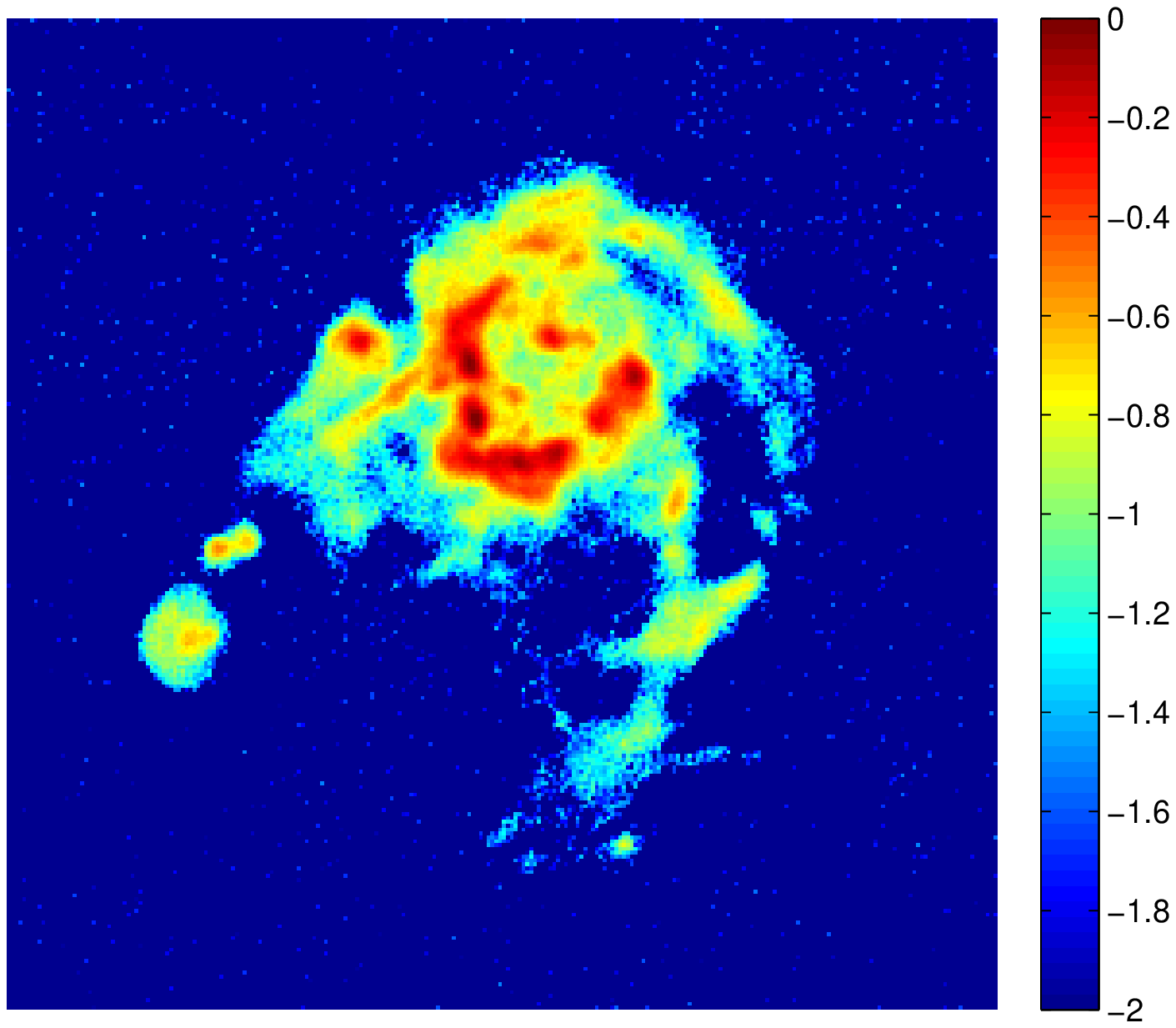}
    \includegraphics[trim = 3.4cm 1.1cm 2.1cm 0.5cm, clip, keepaspectratio, width = 5.5cm]{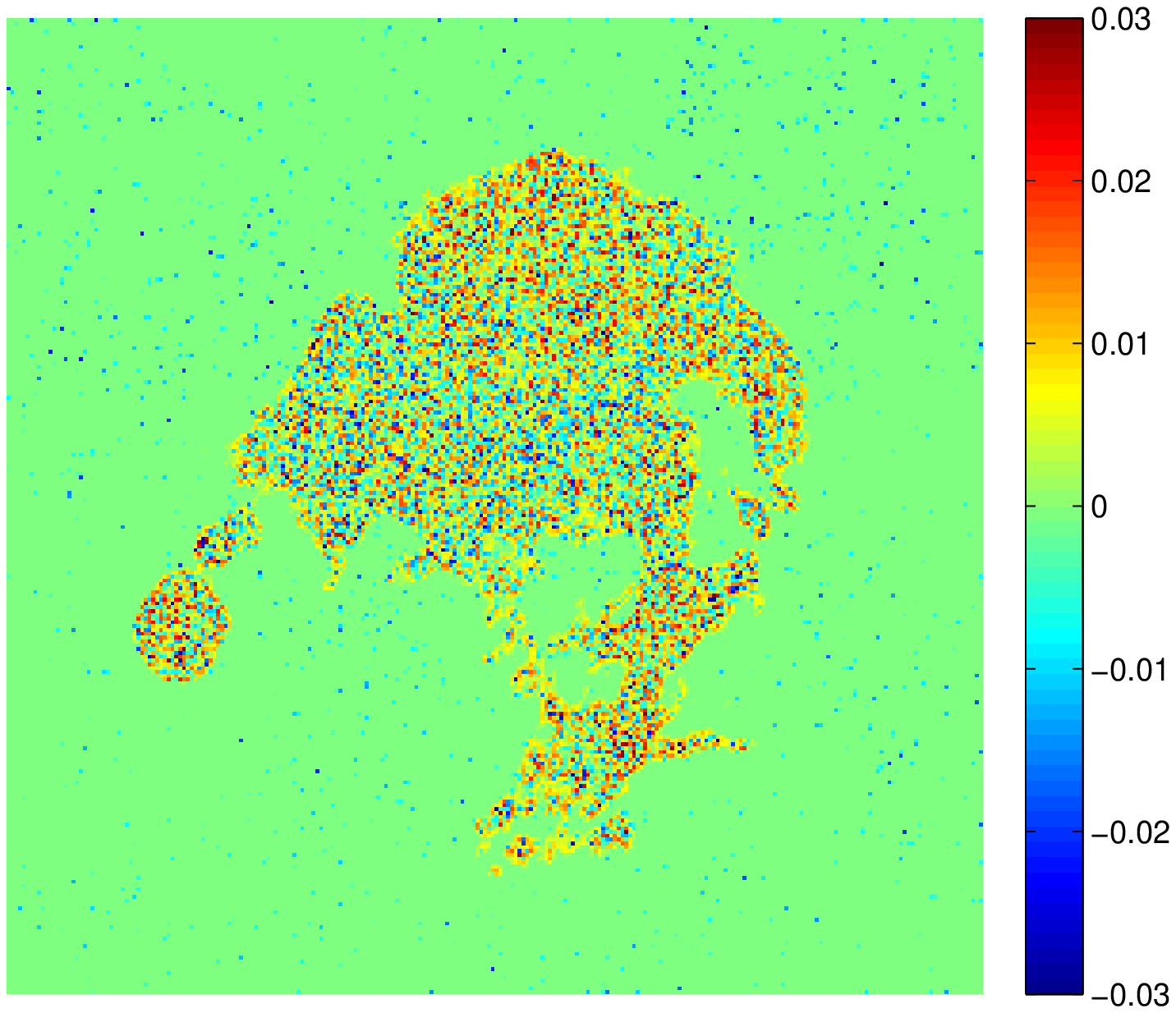}
    \includegraphics[trim = 3.4cm 1.1cm 2.1cm 0.5cm, clip, keepaspectratio, width = 5.5cm]{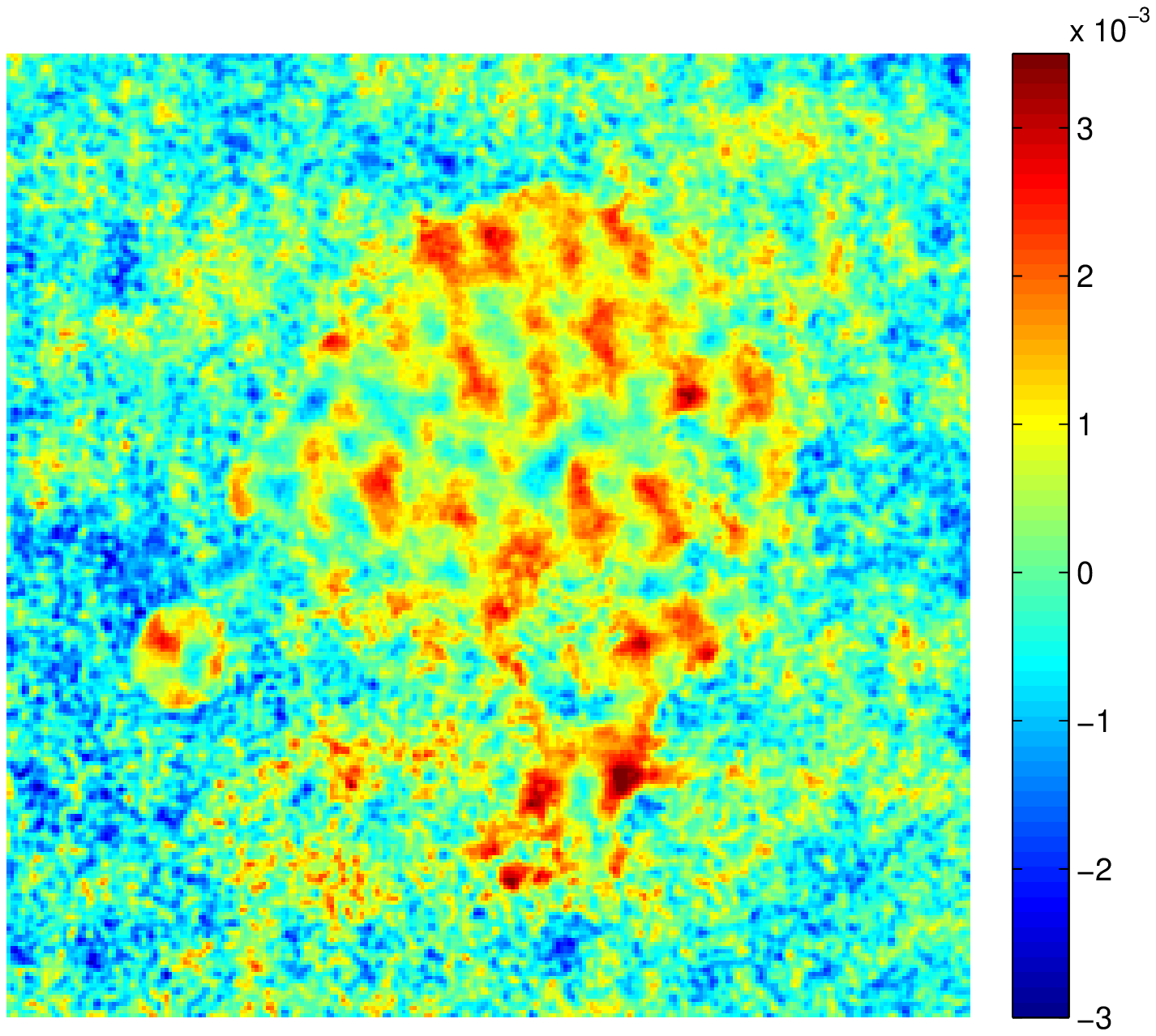}

\caption{Reconstruction example of M31 (256$\times$256) for a $u$-$v$ coverage with $M=0.4N$ sampling frequencies. The results are shown from top to bottom for SARA (SNR=32.4~dB), RWBPDb8 (SNR=30.6~dB), RWTV (SNR=28.6~dB) and RWBP (SNR=23.4~dB) respectively. The first column shows the reconstructed images in a $\log_{10}$ scale, the second column shows the error images in linear scale, and the third column shows the residual dirty images also in linear scale.}
\label{fig:3}
\end{figure*}

\begin{figure*}
    \centering
     
    \includegraphics[trim = 3.4cm 1.1cm 2.1cm 0.5cm, clip, keepaspectratio, width = 5.5cm]{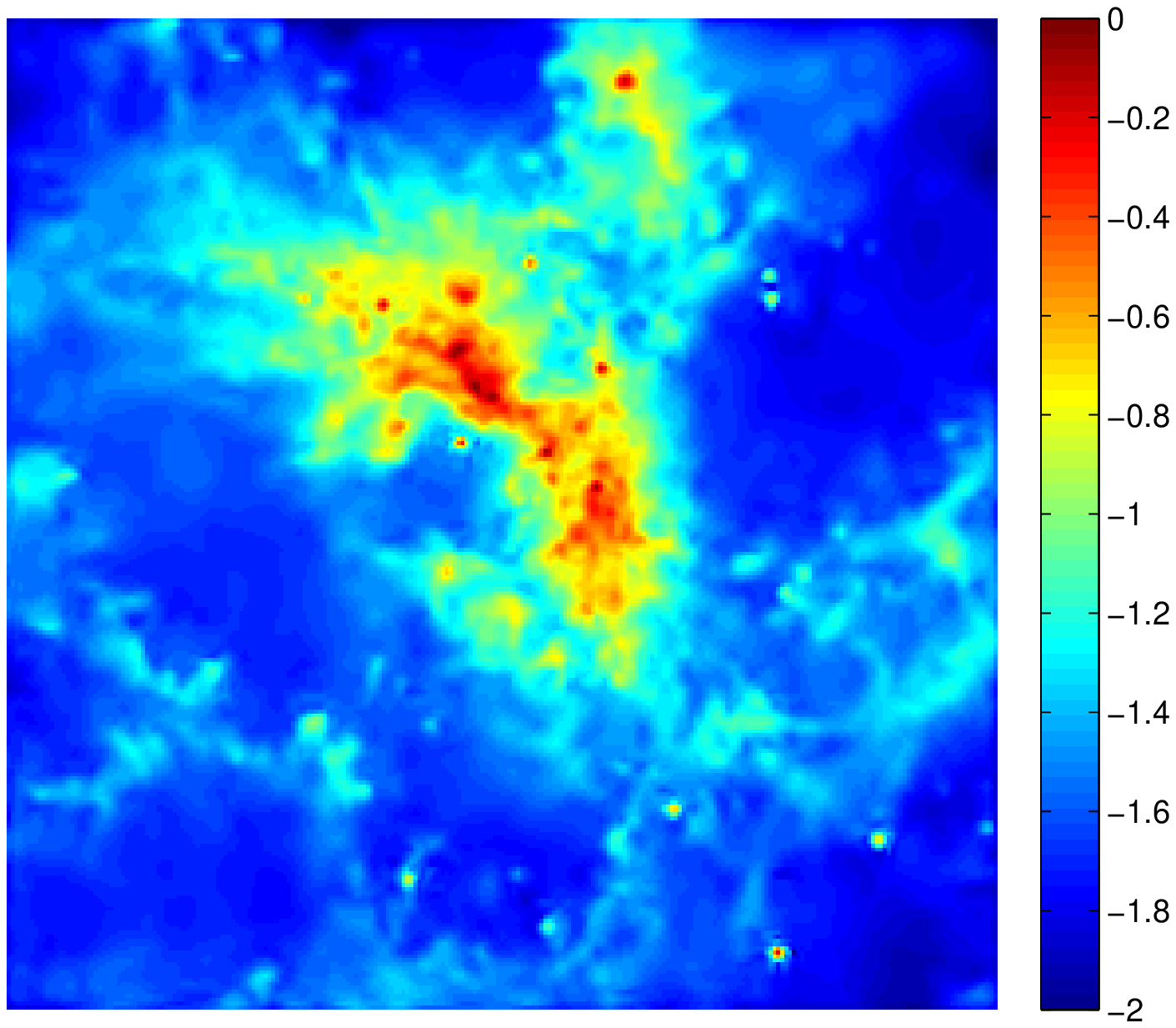}
    \includegraphics[trim = 3.4cm 1.1cm 2.1cm 0.5cm, clip, keepaspectratio, width = 5.5cm]{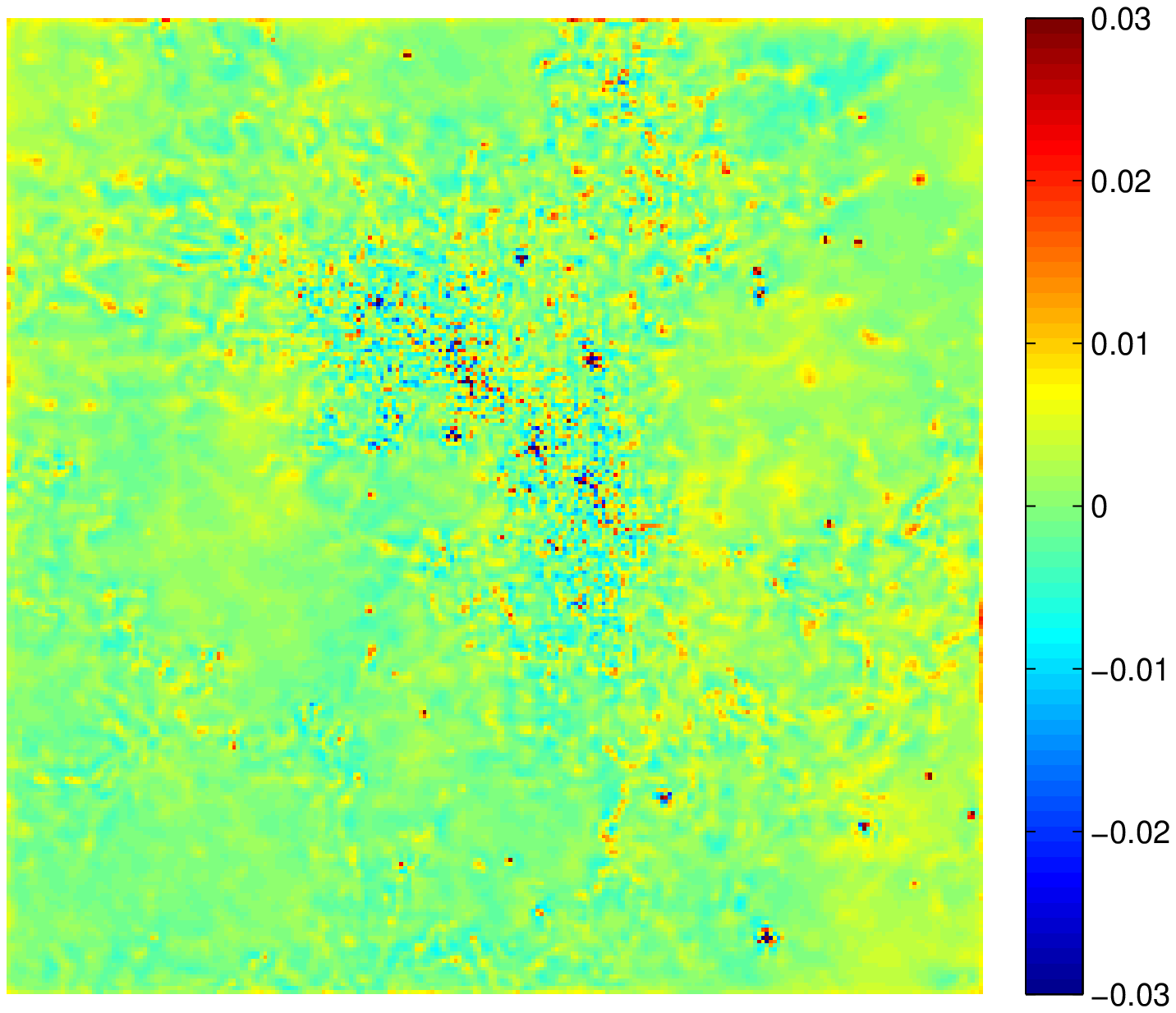}
    \includegraphics[trim = 3.4cm 1.1cm 2.1cm 0.5cm, clip, keepaspectratio, width = 5.5cm]{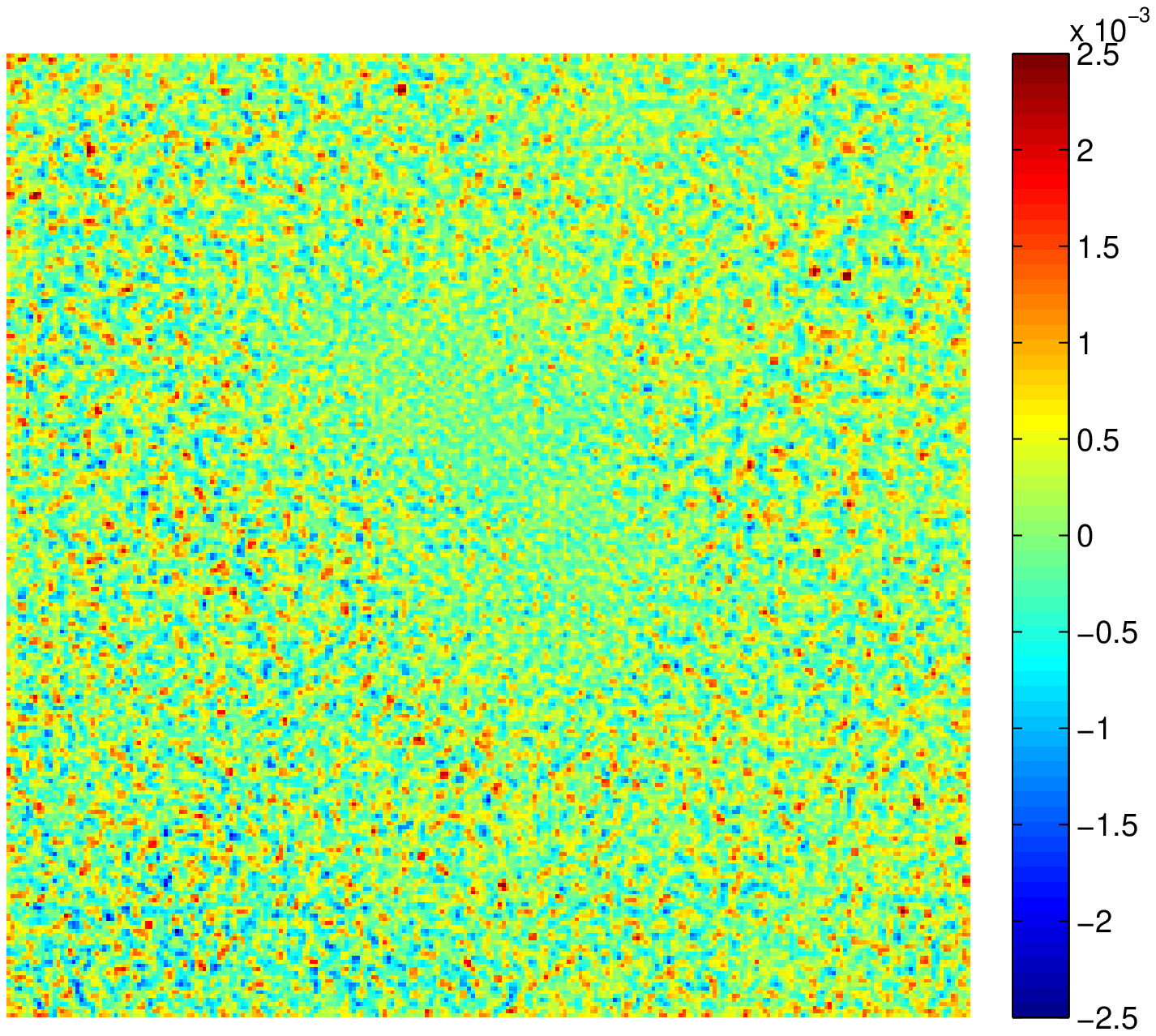}
    \includegraphics[trim = 3.4cm 1.1cm 2.1cm 0.5cm, clip, keepaspectratio, width = 5.5cm]{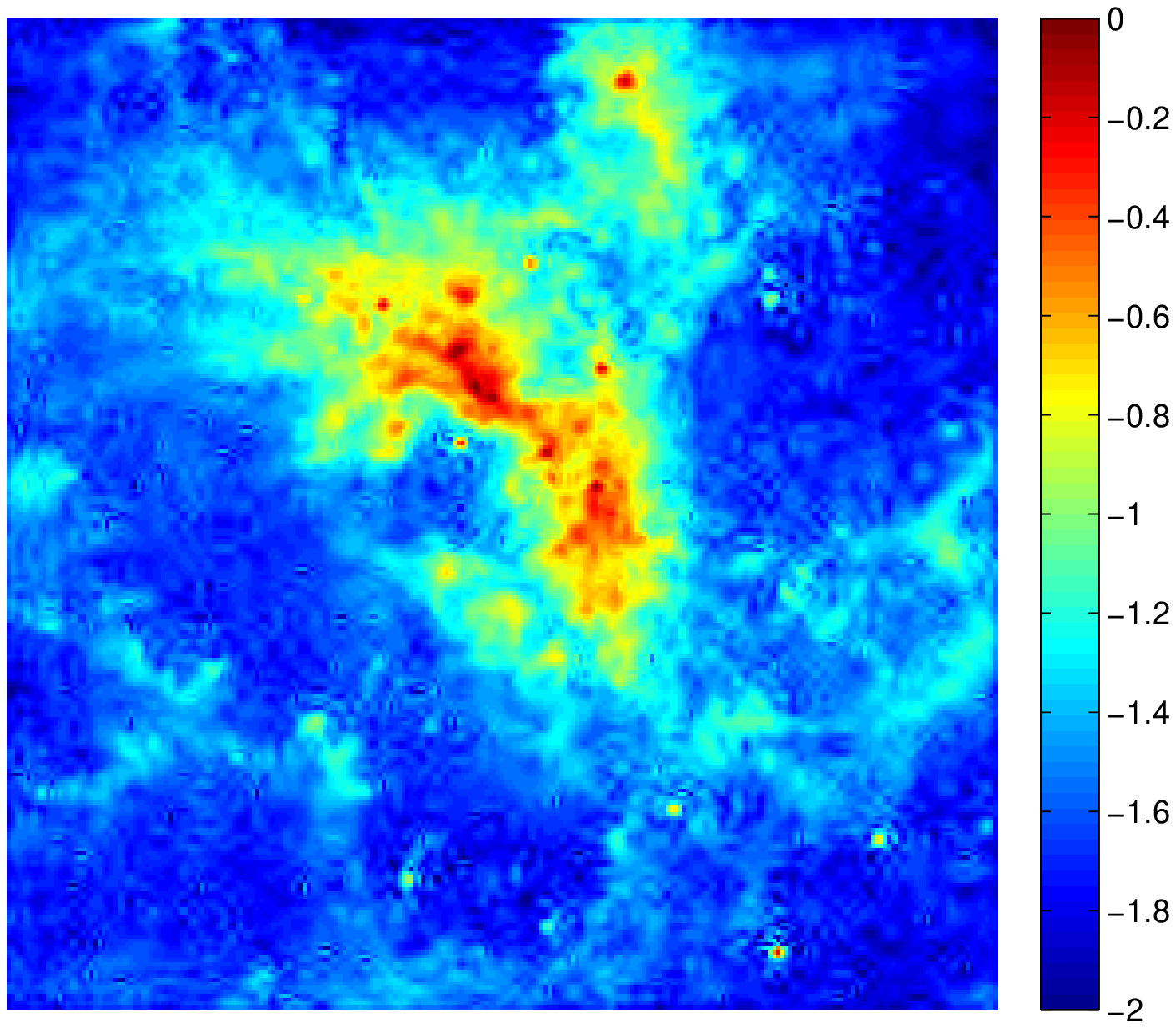}
    \includegraphics[trim = 3.4cm 1.1cm 2.1cm 0.5cm, clip, keepaspectratio, width = 5.5cm]{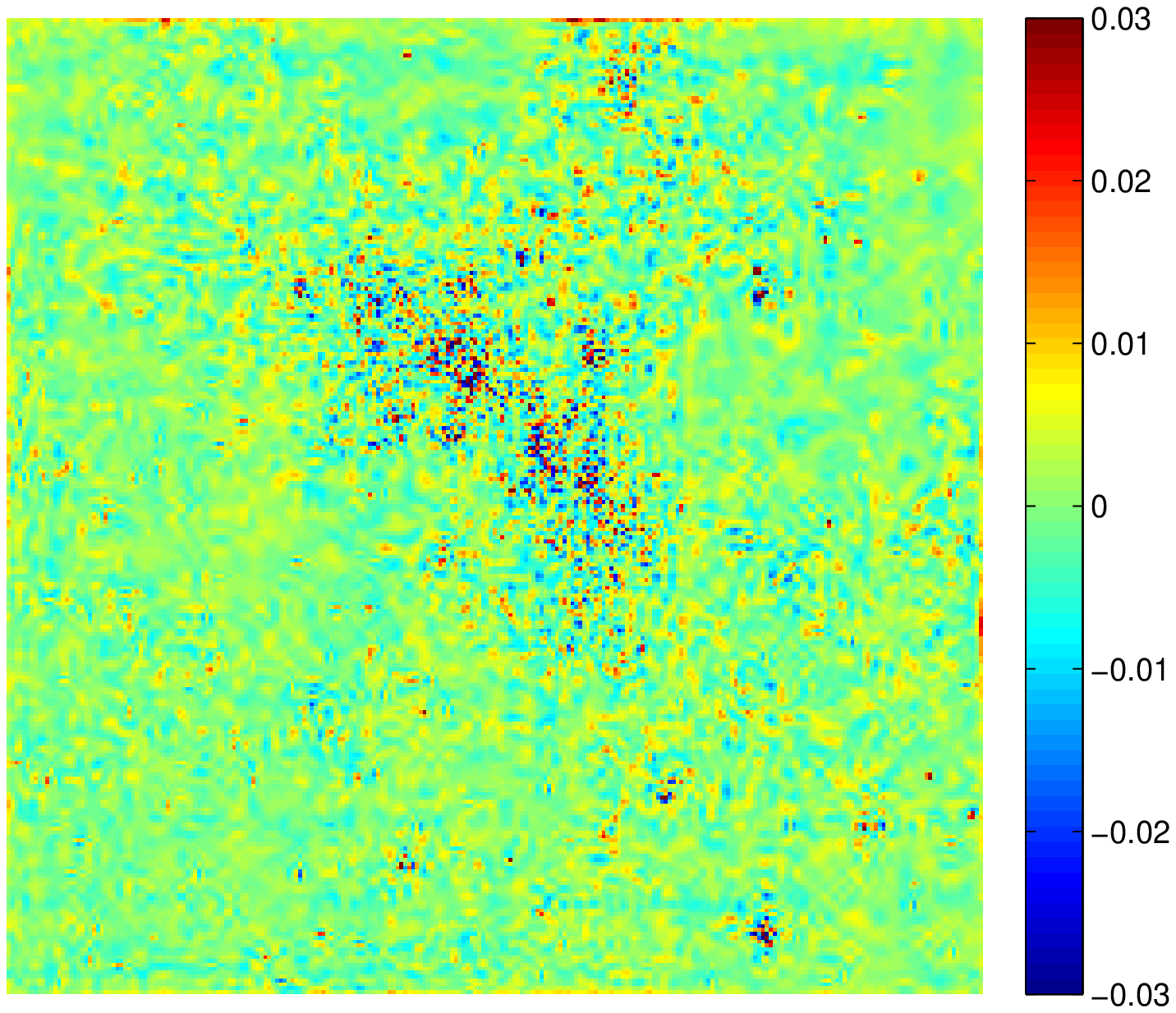}
    \includegraphics[trim = 3.4cm 1.1cm 2.1cm 0.5cm, clip, keepaspectratio, width = 5.5cm]{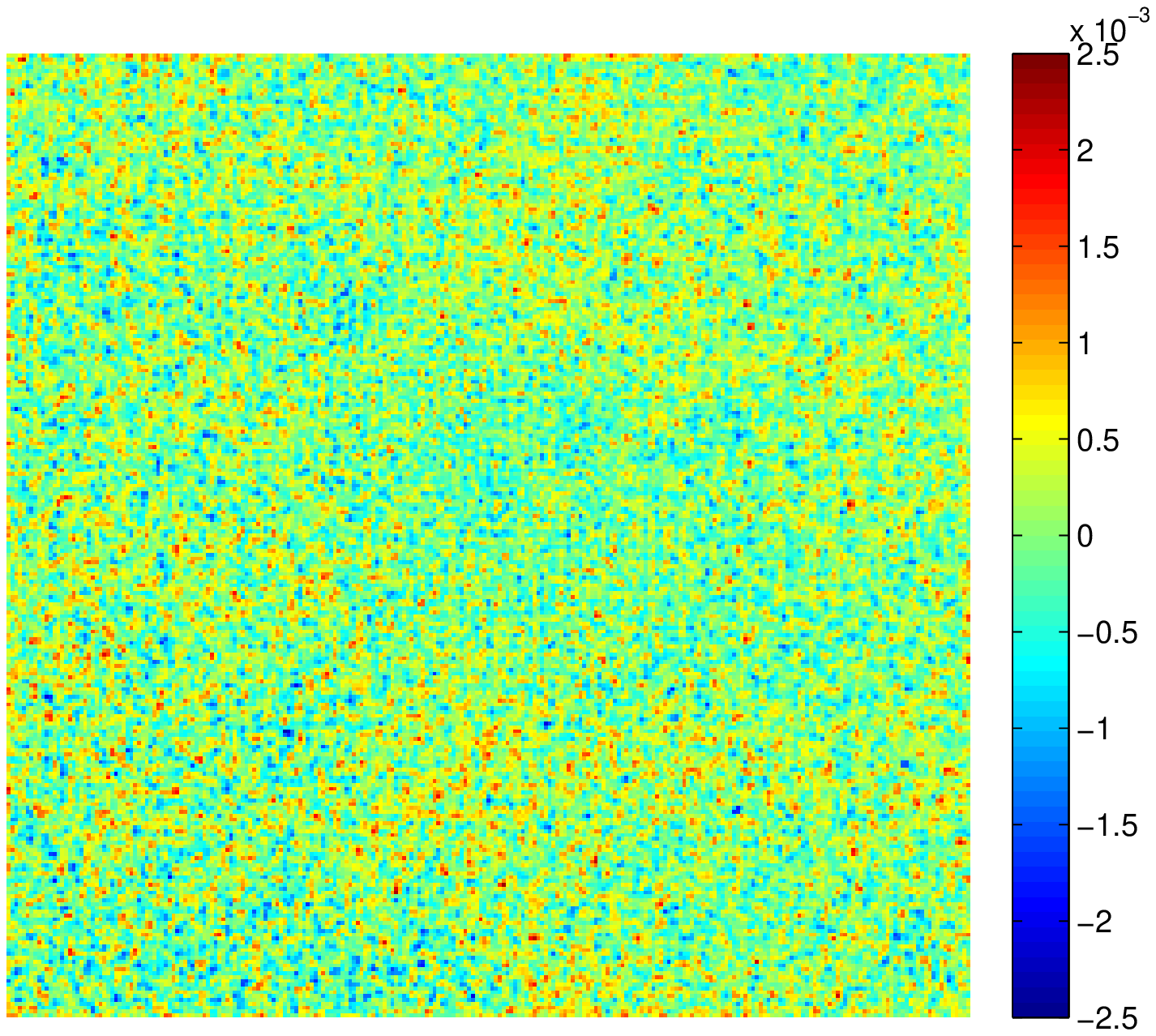}
    \includegraphics[trim = 3.4cm 1.1cm 2.1cm 0.5cm, clip, keepaspectratio, width = 5.5cm]{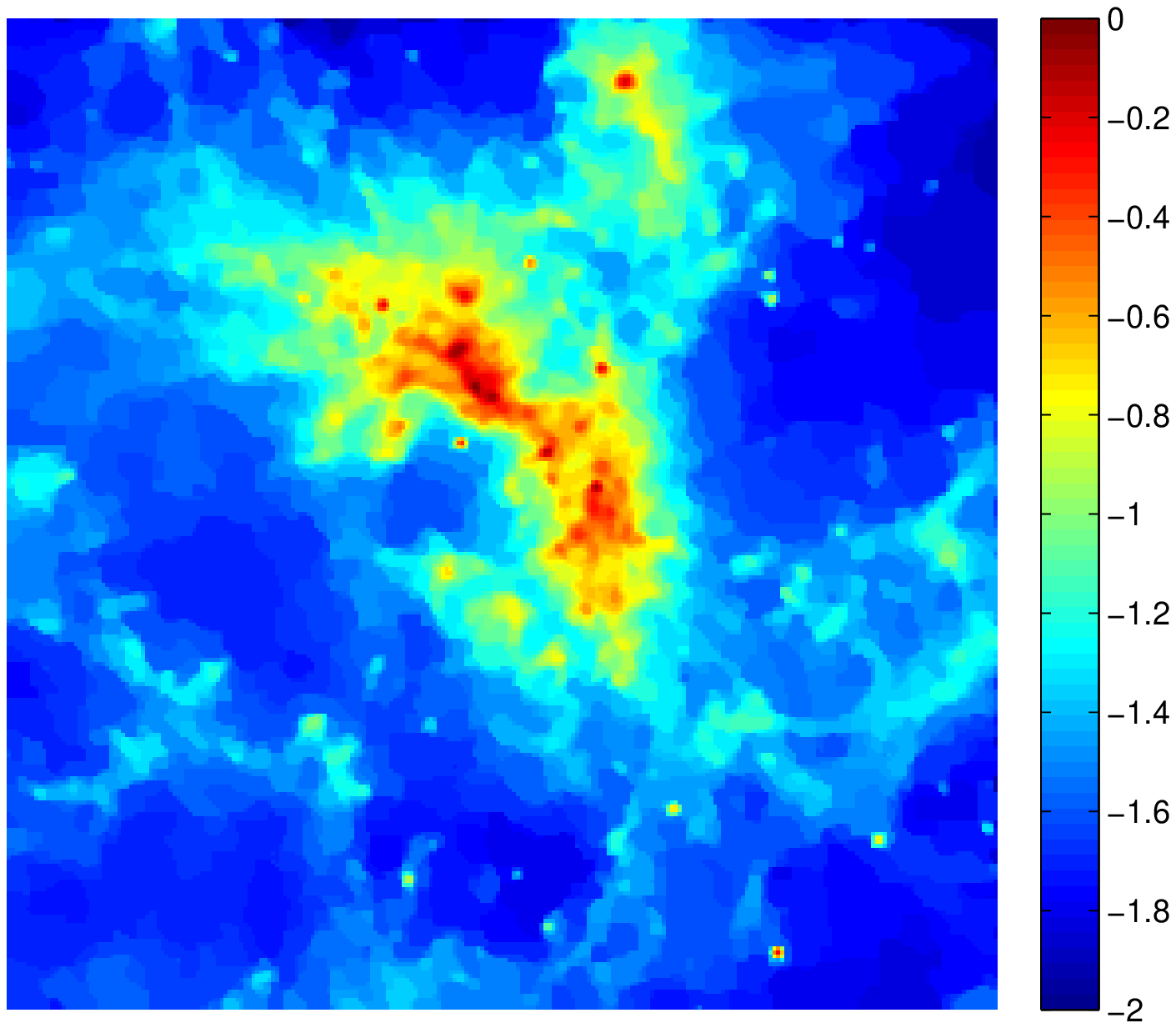}
    \includegraphics[trim = 3.4cm 1.1cm 2.1cm 0.5cm, clip, keepaspectratio, width = 5.5cm]{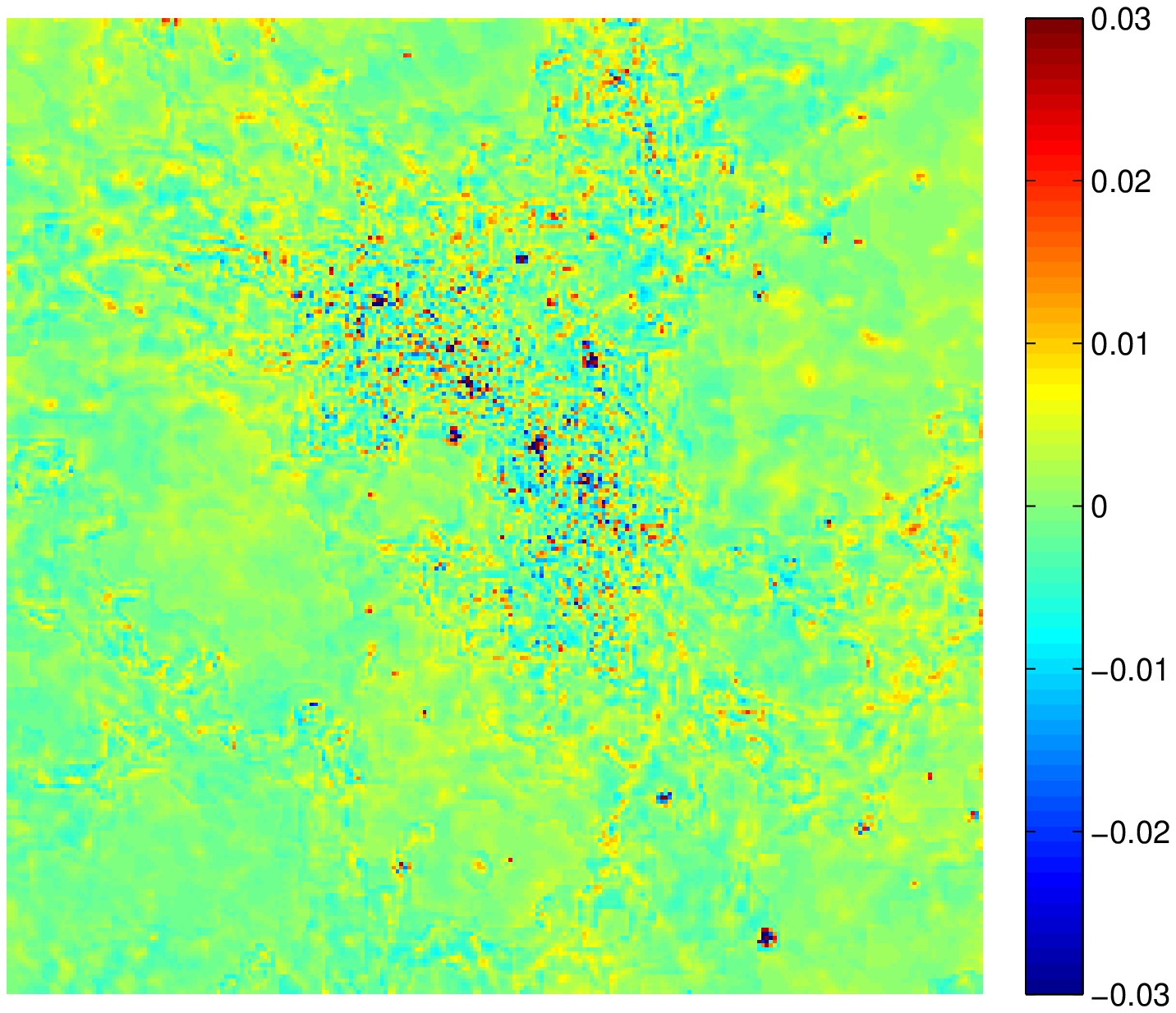}
    \includegraphics[trim = 3.4cm 1.1cm 2.1cm 0.5cm, clip, keepaspectratio, width = 5.5cm]{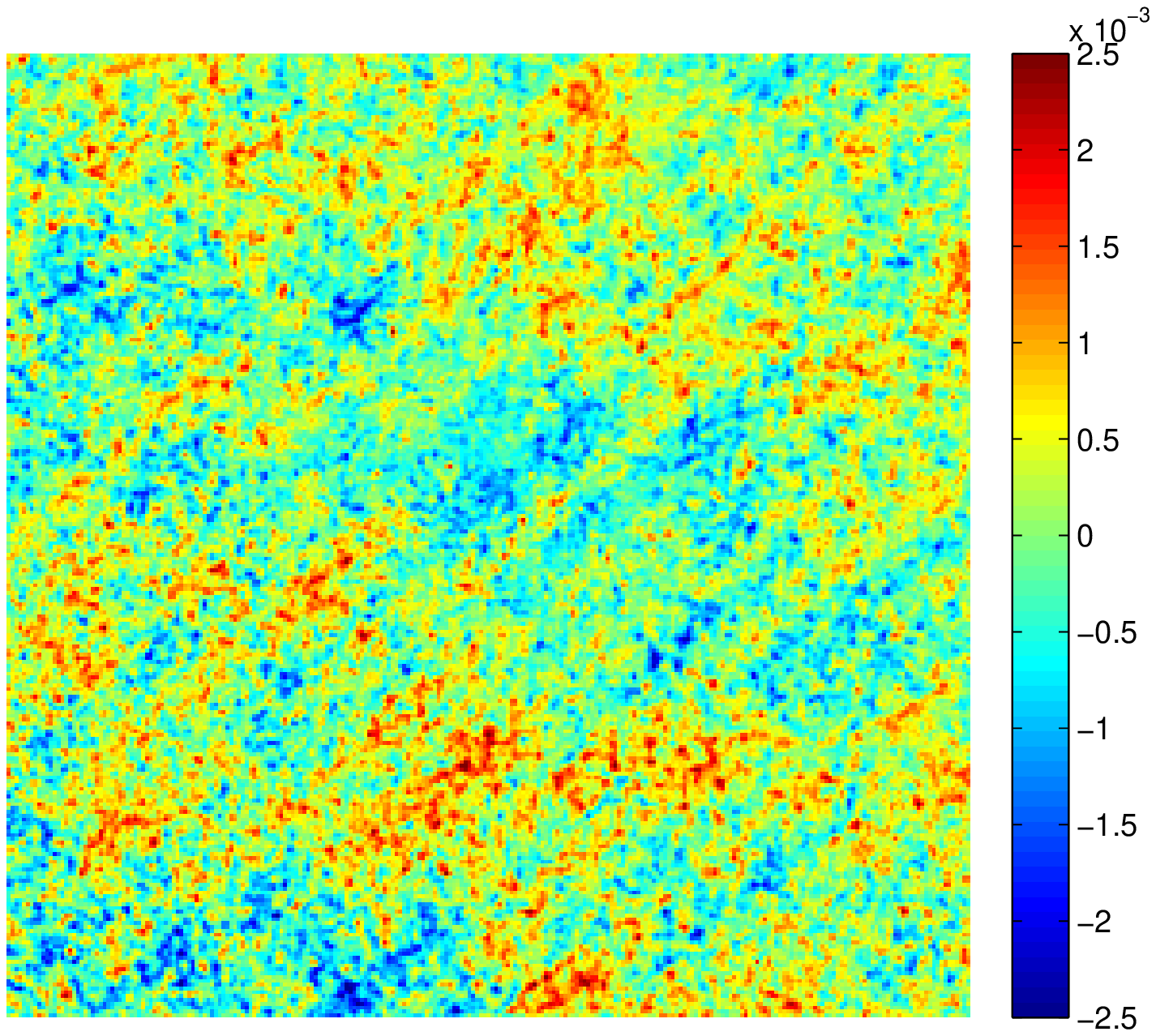}
    \includegraphics[trim = 3.4cm 1.1cm 2.1cm 0.5cm, clip, keepaspectratio, width = 5.5cm]{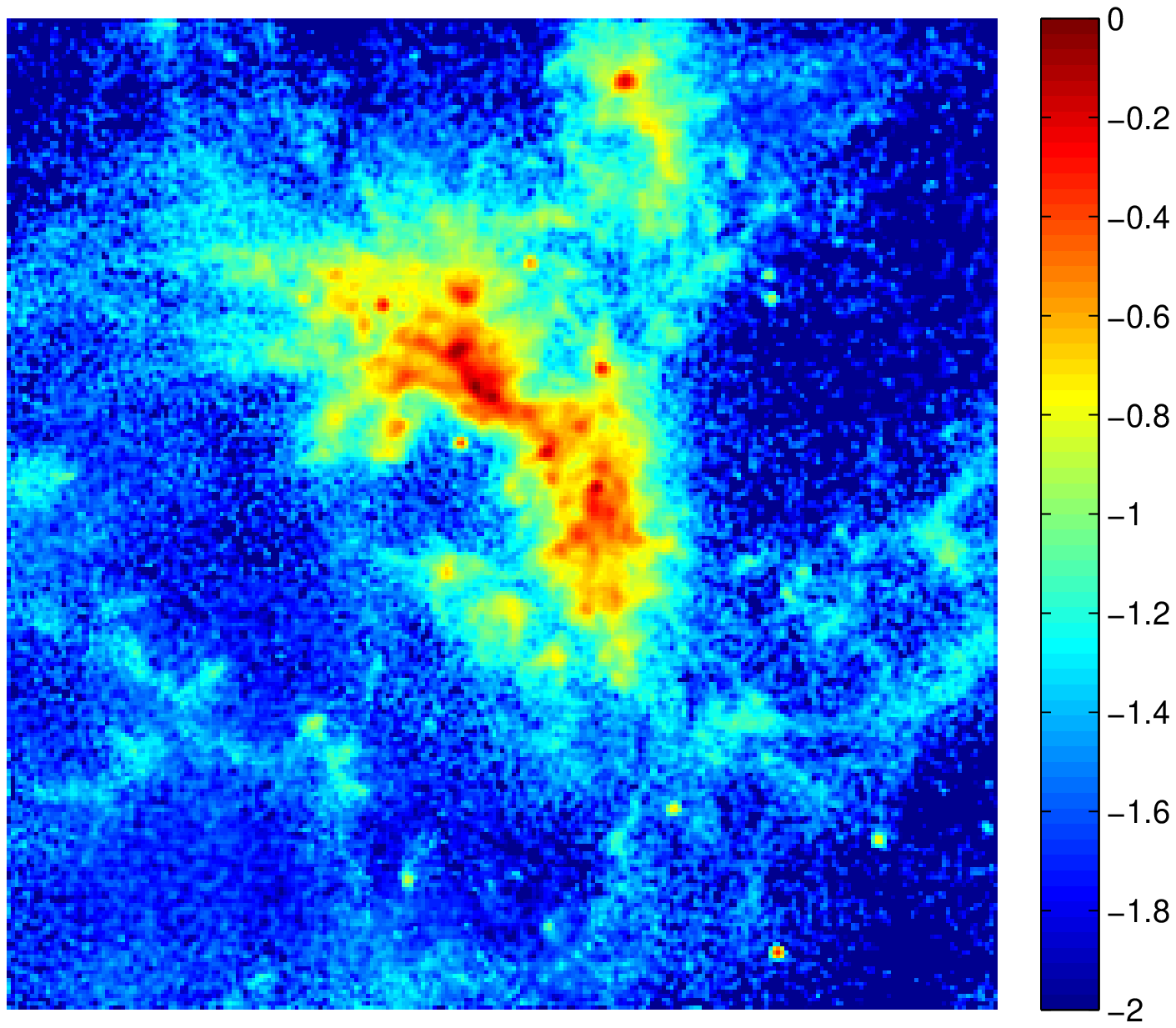}
    \includegraphics[trim = 3.4cm 1.1cm 2.1cm 0.5cm, clip, keepaspectratio, width = 5.5cm]{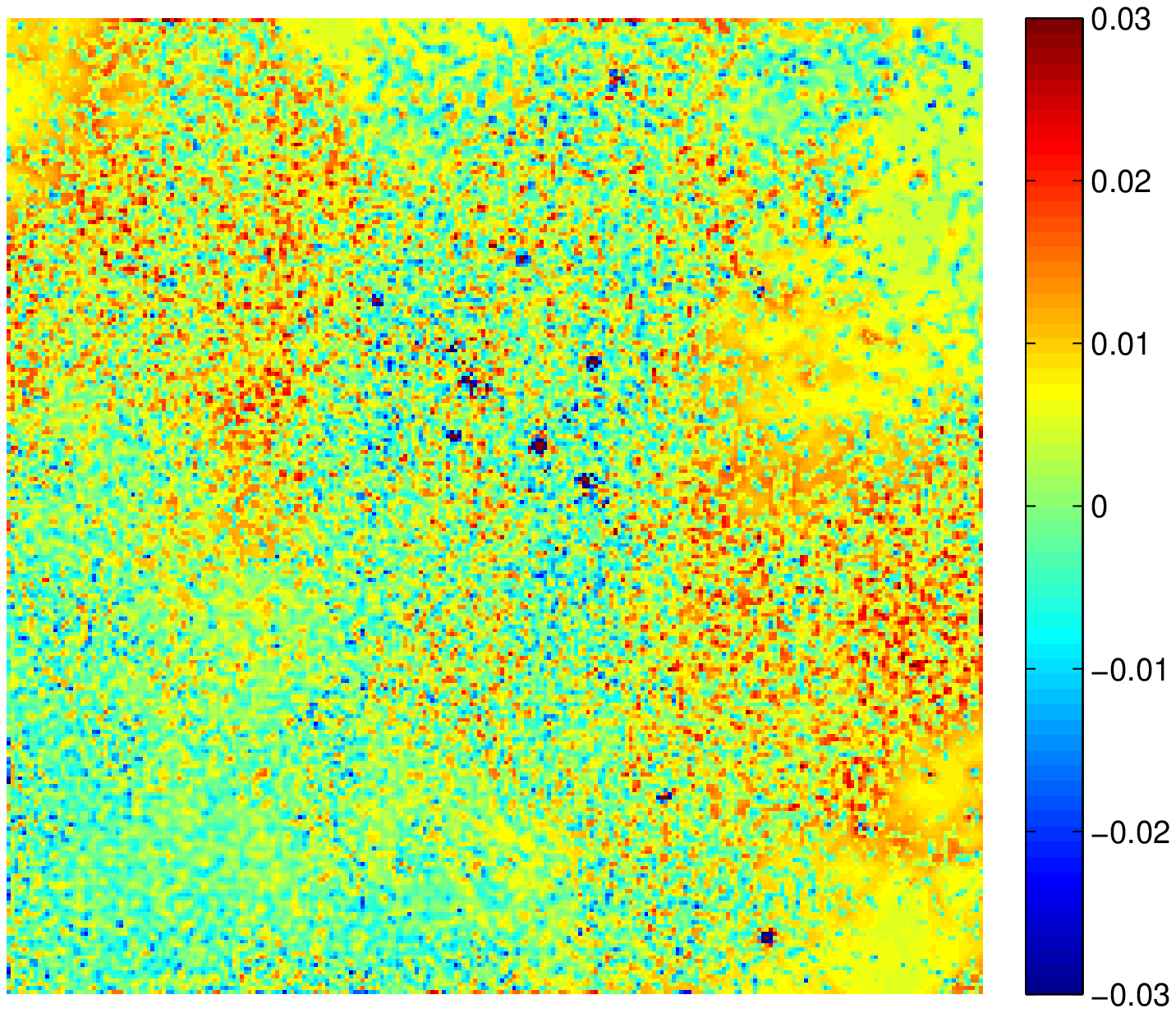}
    \includegraphics[trim = 3.4cm 1.1cm 2.1cm 0.5cm, clip, keepaspectratio, width = 5.5cm]{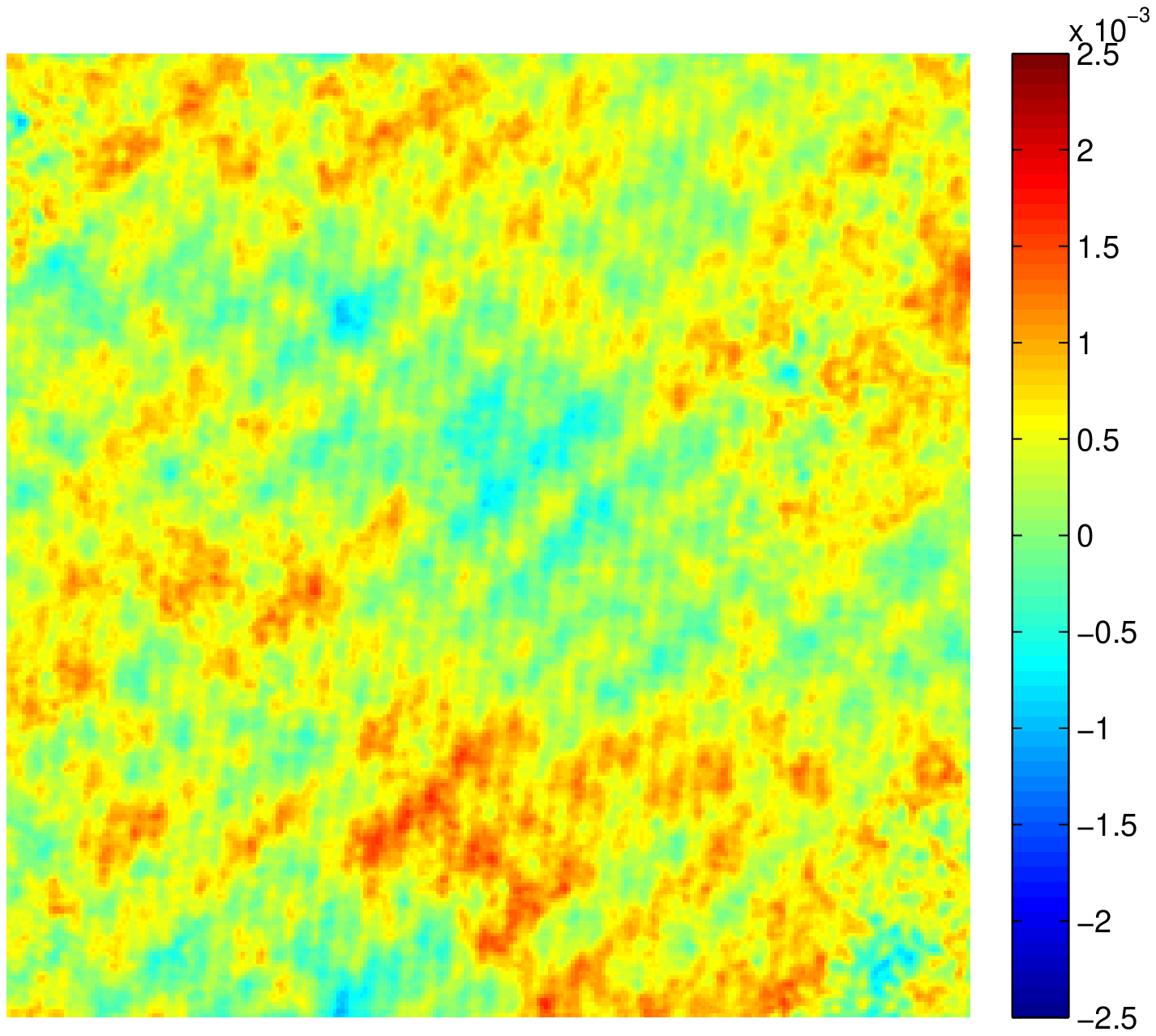}

\caption{Reconstruction example of 30dor (256$\times$256) for a $u$-$v$ coverage with $M=0.4N$ sampling frequencies. The results are shown from top to bottom for SARA (SNR=25.3~dB), RWBPDb8 (SNR=22.6~dB), RWTV (SNR=24.1~dB) and RWBP (SNR=18.8~dB) respectively. The first column shows the reconstructed images in a $\log_{10}$ scale, the second column shows the error images in linear scale, and the third column shows the residual dirty images also in linear scale.}
\label{fig:4}
\end{figure*}

Next we present a visual assessment of the reconstruction quality of the different algorithms. Figure~\ref{fig:3} and Figure~\ref{fig:4} show the results from M31 and 30Dor respectively for a $u$-$v$ coverage of $M=26374\approx 0.4N$ visibilities. The results are shown from top to bottom for SARA, RWBPDb8, RWTV and RWBP respectively. The first column shows the reconstructed images in a $\log_{10}$ scale, the second column shows the error images, defined as $\bm{x}-\hat{\bm{x}}$, in linear scale, and, the third column shows the real part of the residual dirty images, defined as the difference between dirty images and dirty images constructed from recovered images, i.e. $\bm{r}=\mathsf{\Phi}^{\dagger}\bm{y}-\mathsf{\Phi}^{\dagger}\mathsf{\Phi}\hat{\bm{x}}$, also in linear scale. These images confirm the previous results found by examining recovered SNR levels; SARA yields reconstructed images with fewer artifacts in the background regions and smaller errors in the structured inner regions than the other methods. Interestingly RWBPDb8 yields a nearly flat residual map for 30Dor. However, this does not necessarily translate into a better reconstruction quality as can be observed in the error image. This phenomenon can also be seen in the reconstructed image by RWTV of 30Dor, which shows a small error image compared to RWBPDb8 but showing a residual map with a lot of structures. This highlights the fact that the common criterion of flatness of residual image is not an optimal measure of reconstruction fidelity as emphasized in our previous work \citep{carrillo12}.


\begin{figure*}
    \centering
     
    \includegraphics[trim = 3.4cm 1.1cm 2.1cm 0.5cm, clip, keepaspectratio, width = 5.5cm]{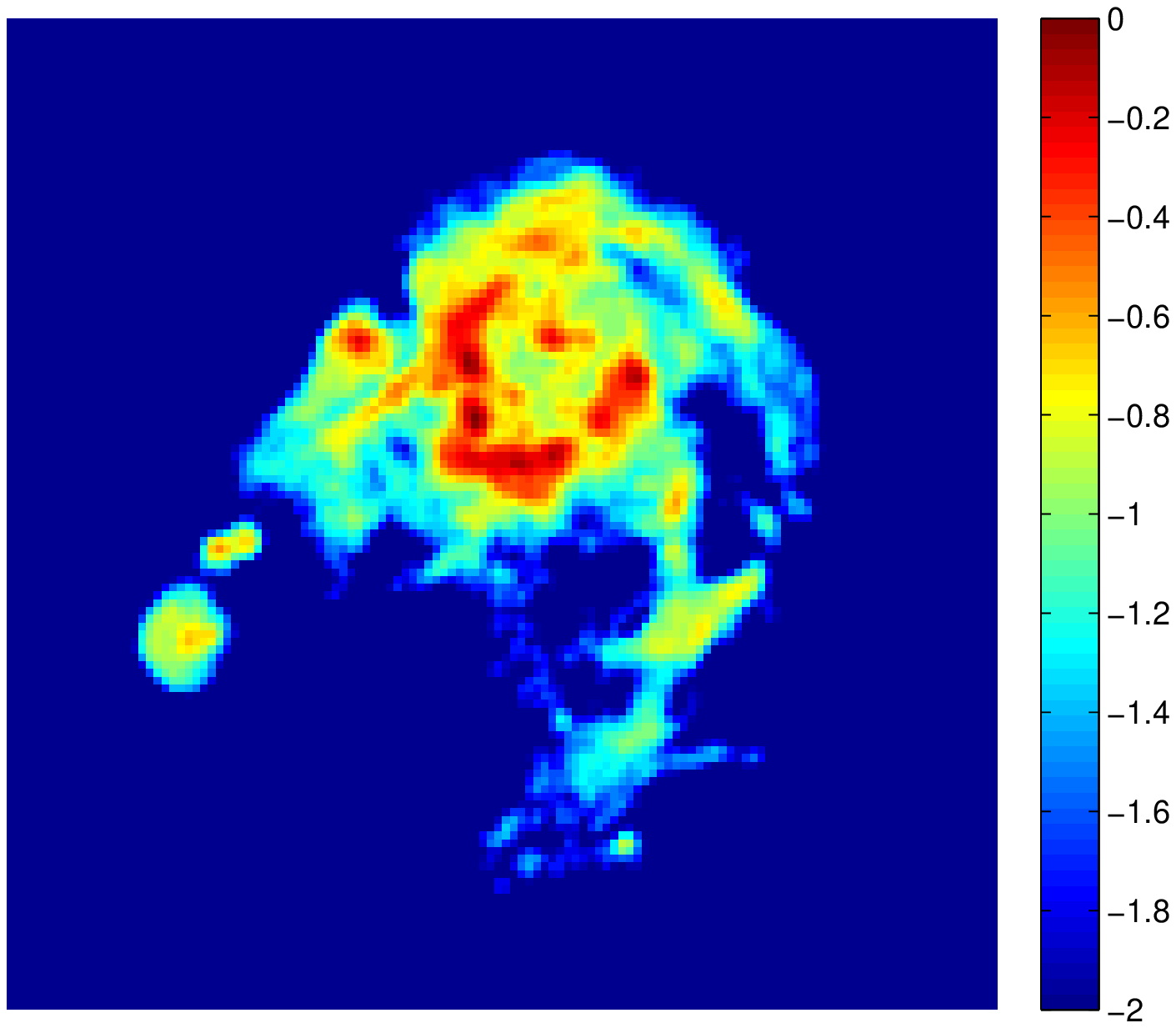}
    \includegraphics[trim = 3.4cm 1.1cm 2.1cm 0.5cm, clip, keepaspectratio, width = 5.5cm]{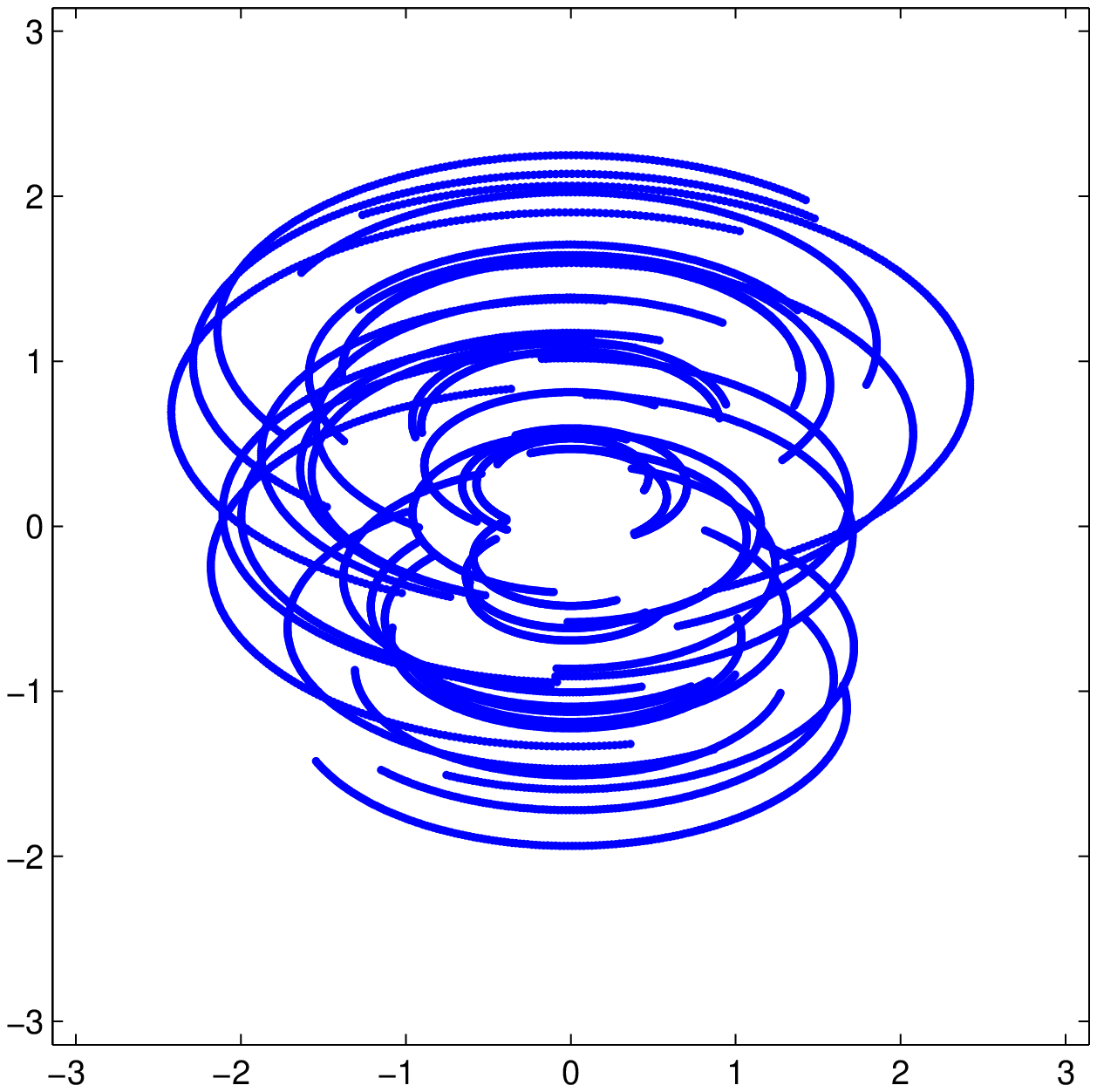}
    \includegraphics[trim = 3.4cm 1.1cm 2.1cm 0.5cm, clip, keepaspectratio, width = 5.5cm]{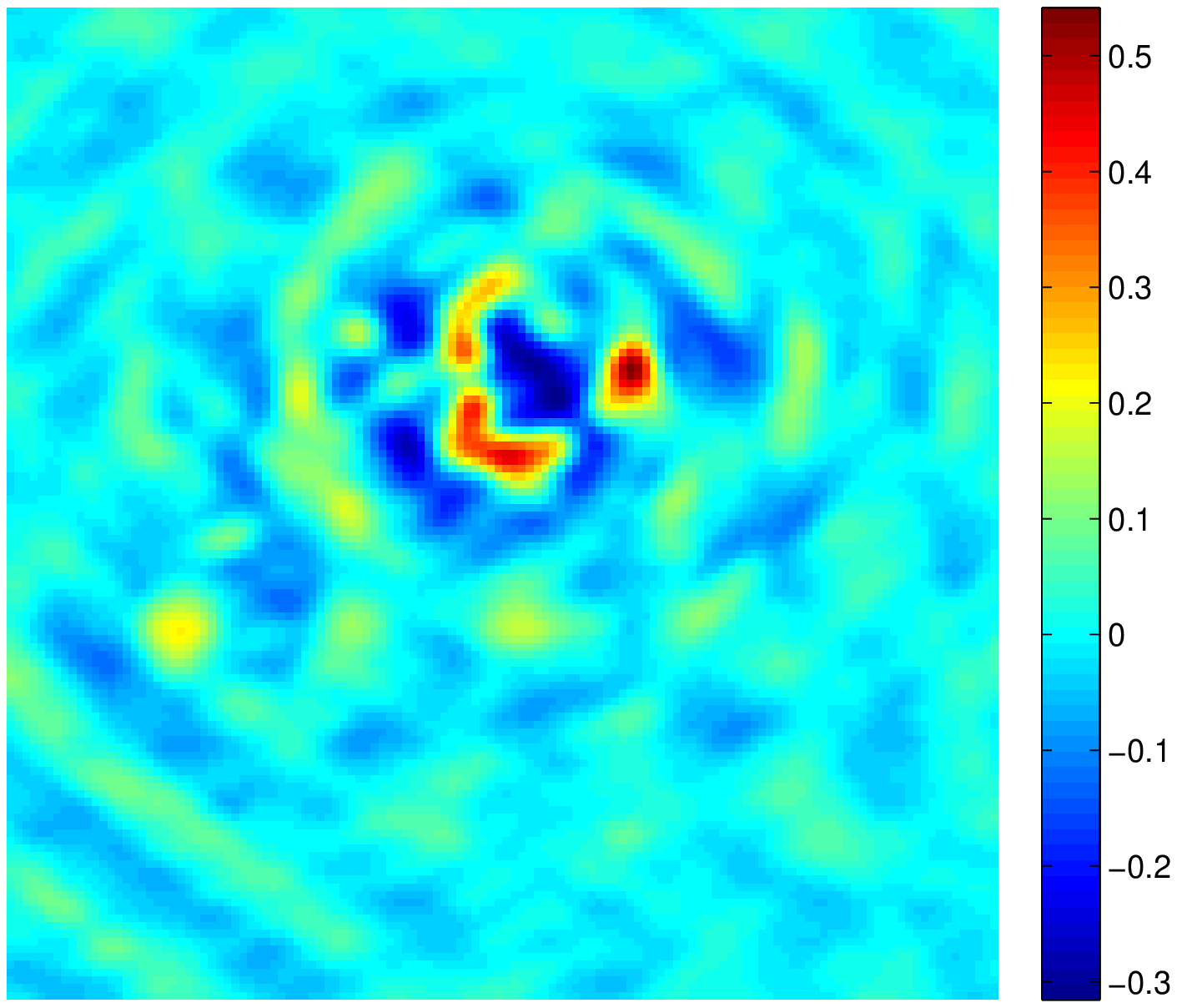}
    \includegraphics[trim = 3.4cm 1.1cm 2.1cm 0.5cm, clip, keepaspectratio, width = 5.5cm]{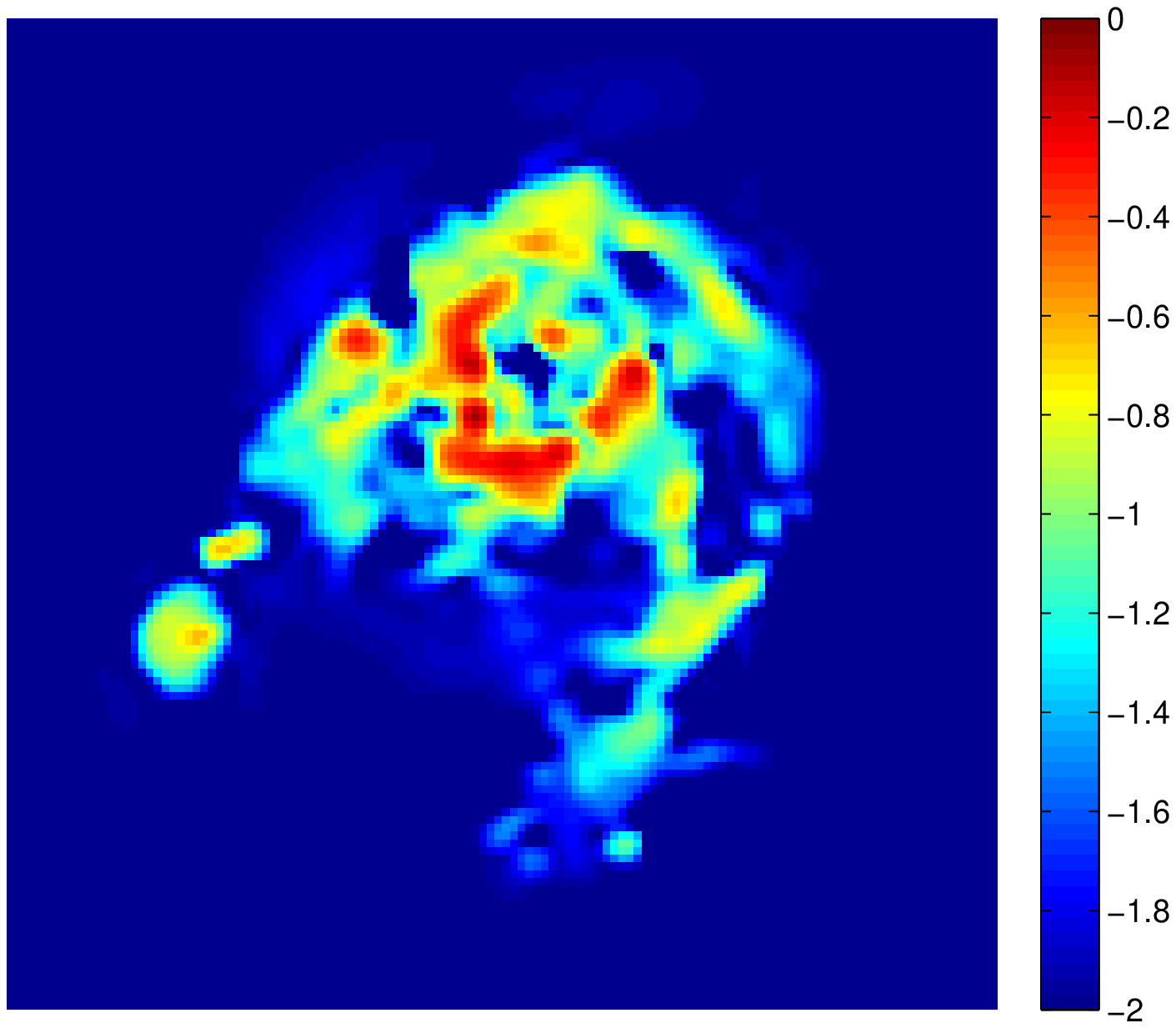}
    \includegraphics[trim = 3.4cm 1.1cm 2.1cm 0.5cm, clip, keepaspectratio, width = 5.5cm]{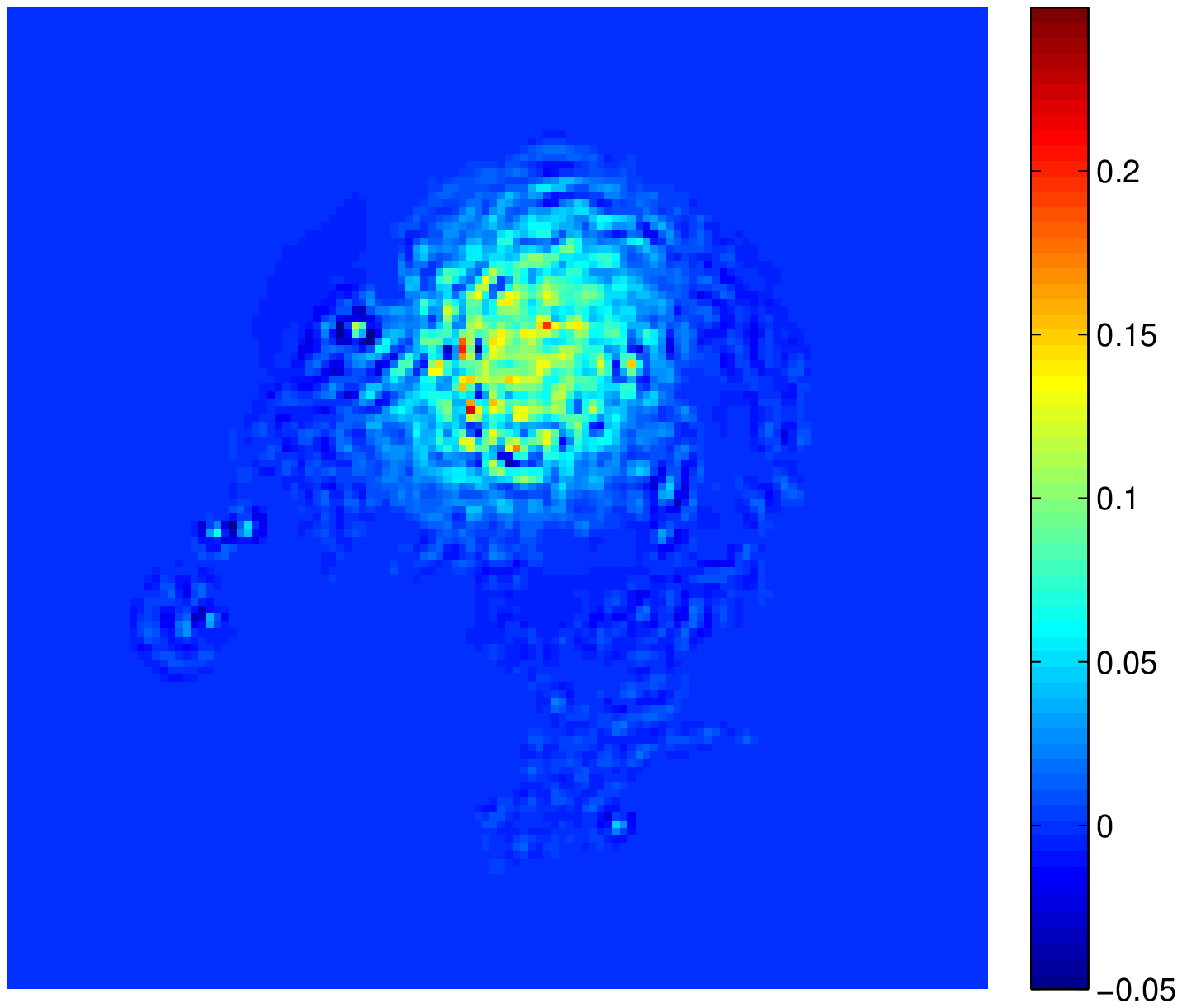}
    \includegraphics[trim = 3.4cm 1.1cm 2.1cm 0.5cm, clip, keepaspectratio, width = 5.5cm]{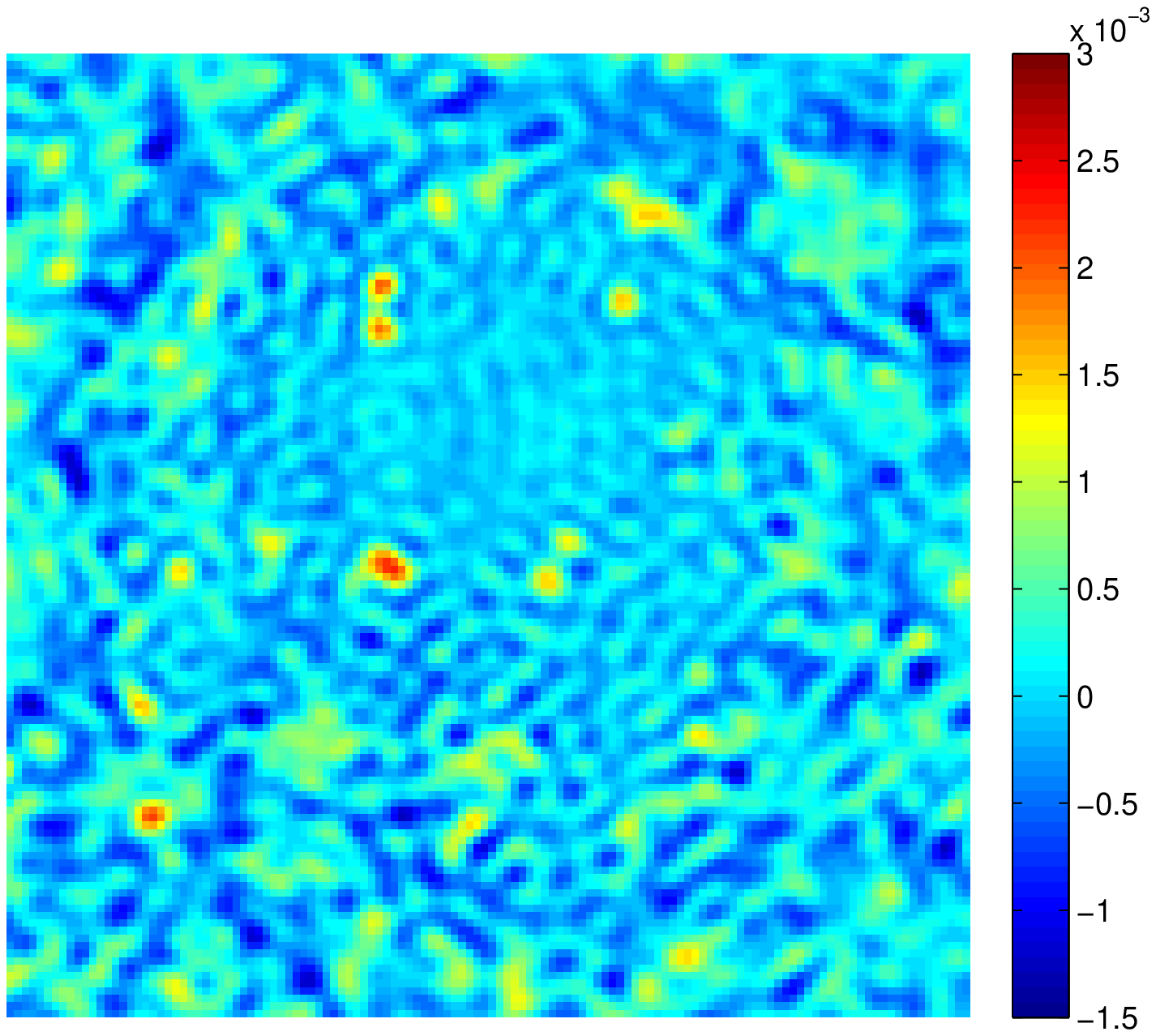}
    \includegraphics[trim = 3.4cm 1.1cm 2.1cm 0.5cm, clip, keepaspectratio, width = 5.5cm]{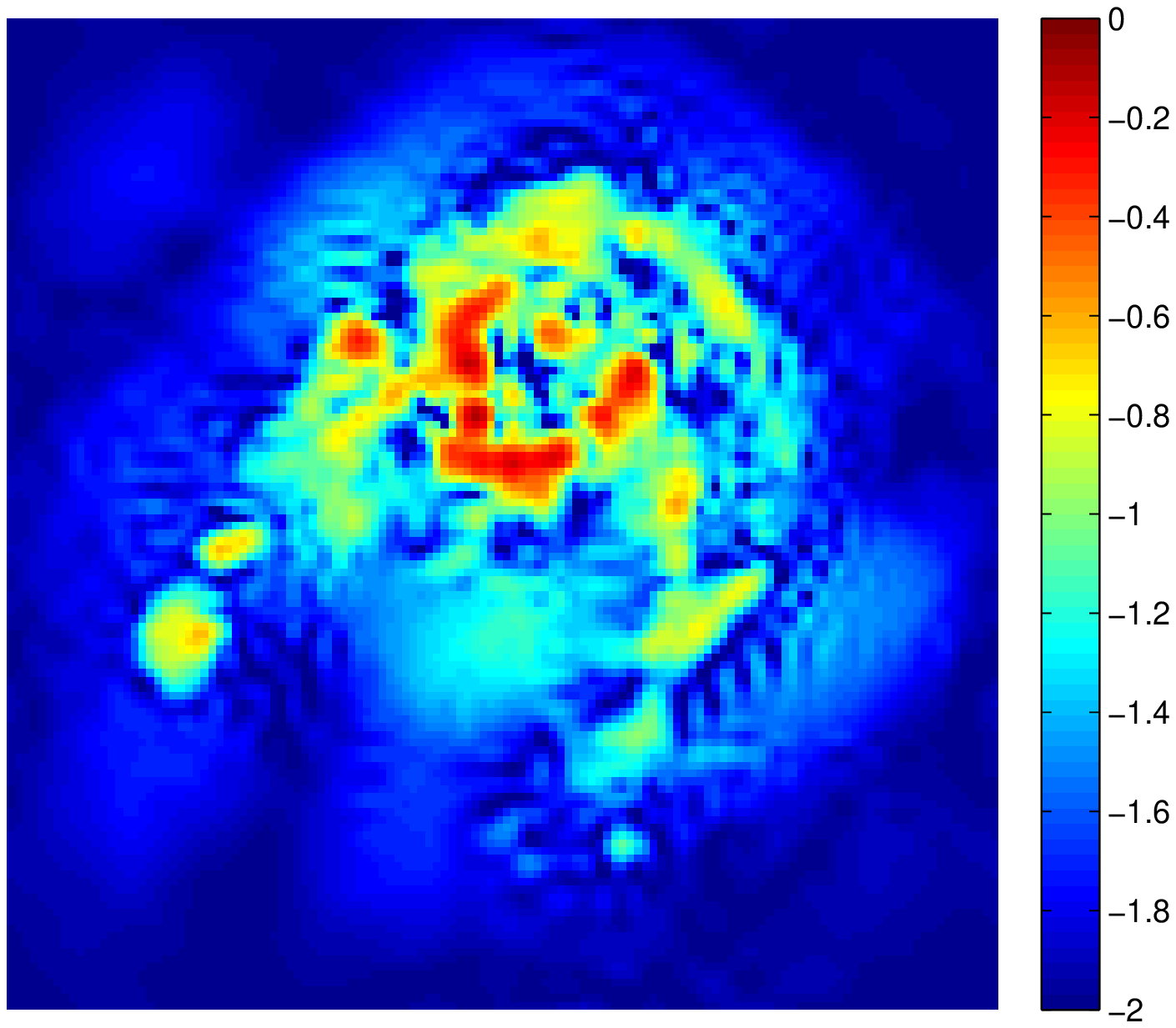}
    \includegraphics[trim = 3.4cm 1.1cm 2.1cm 0.5cm, clip, keepaspectratio, width = 5.5cm]{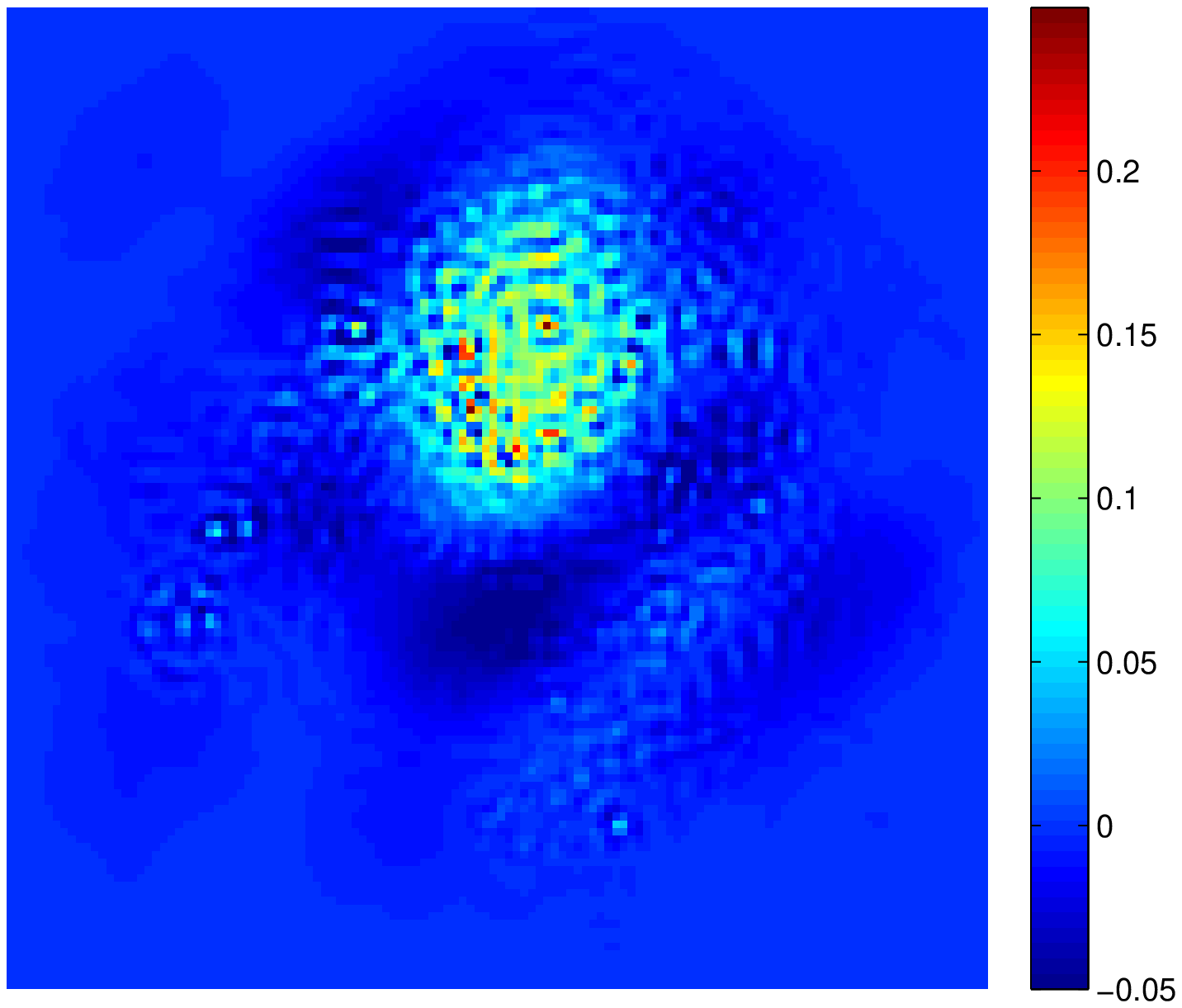}
    \includegraphics[trim = 3.4cm 1.1cm 2.1cm 0.5cm, clip, keepaspectratio, width = 5.5cm]{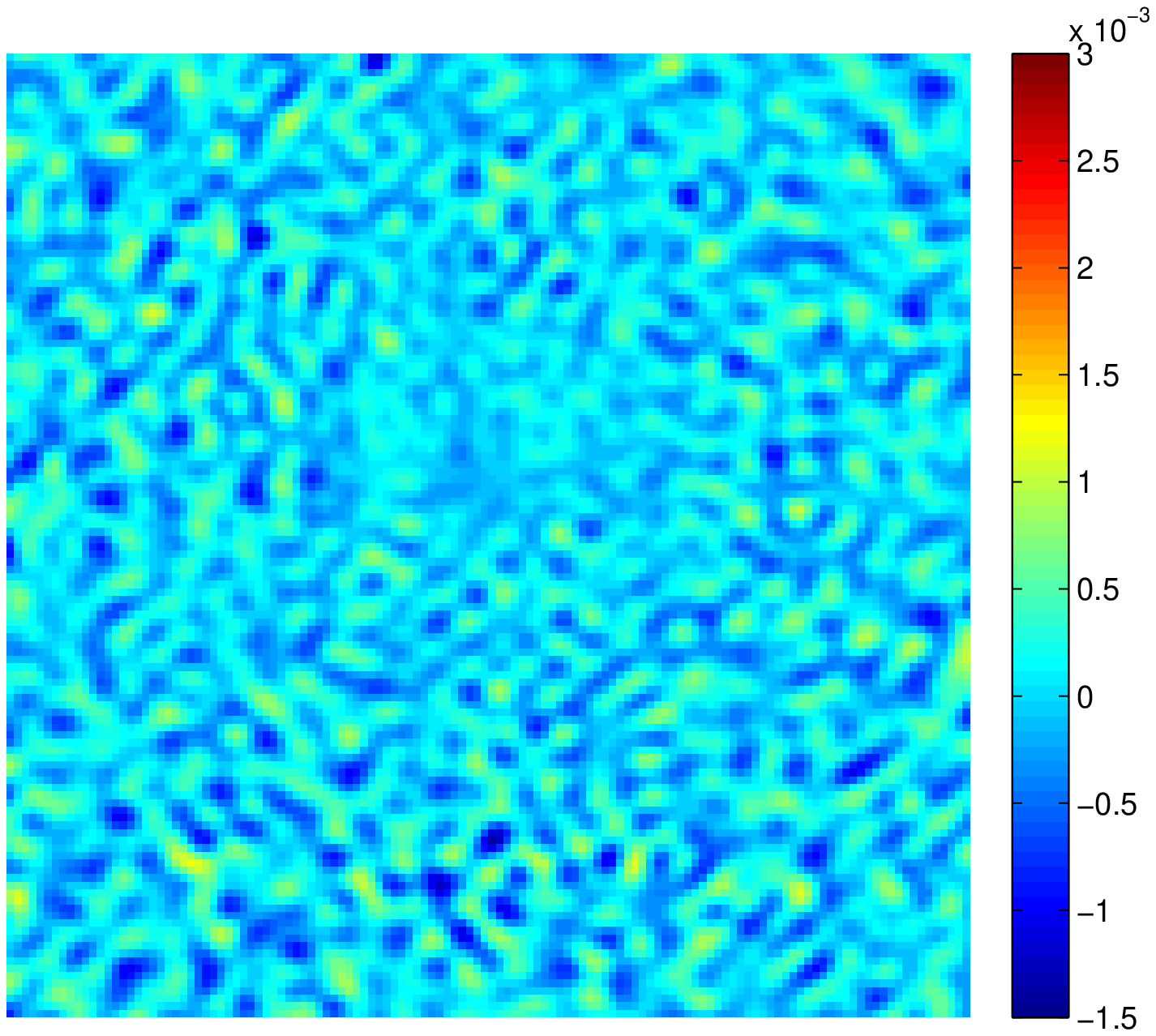}
    \includegraphics[trim = 3.4cm 1.1cm 2.1cm 0.5cm, clip, keepaspectratio, width = 5.5cm]{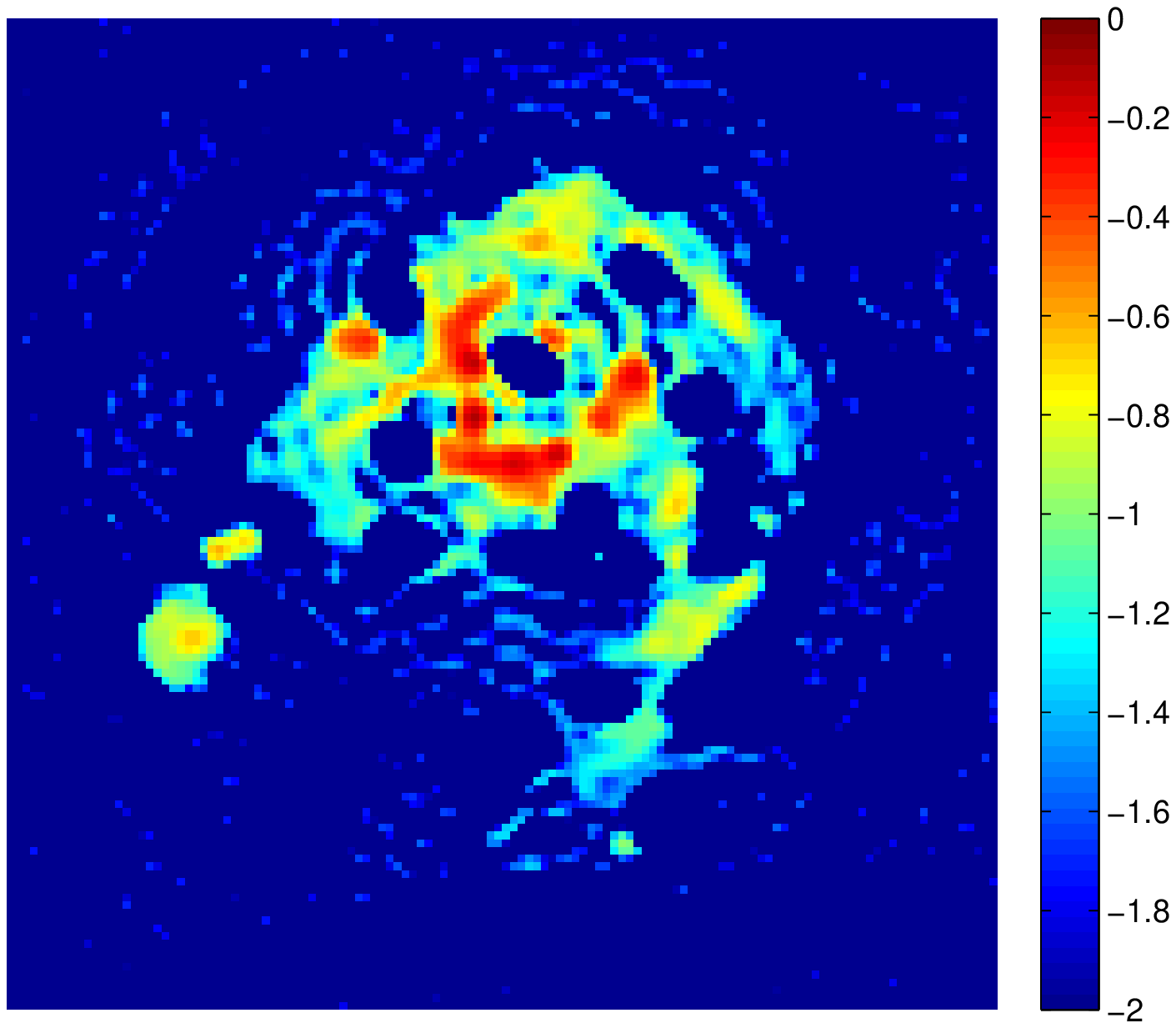}
    \includegraphics[trim = 3.4cm 1.1cm 2.1cm 0.5cm, clip, keepaspectratio, width = 5.5cm]{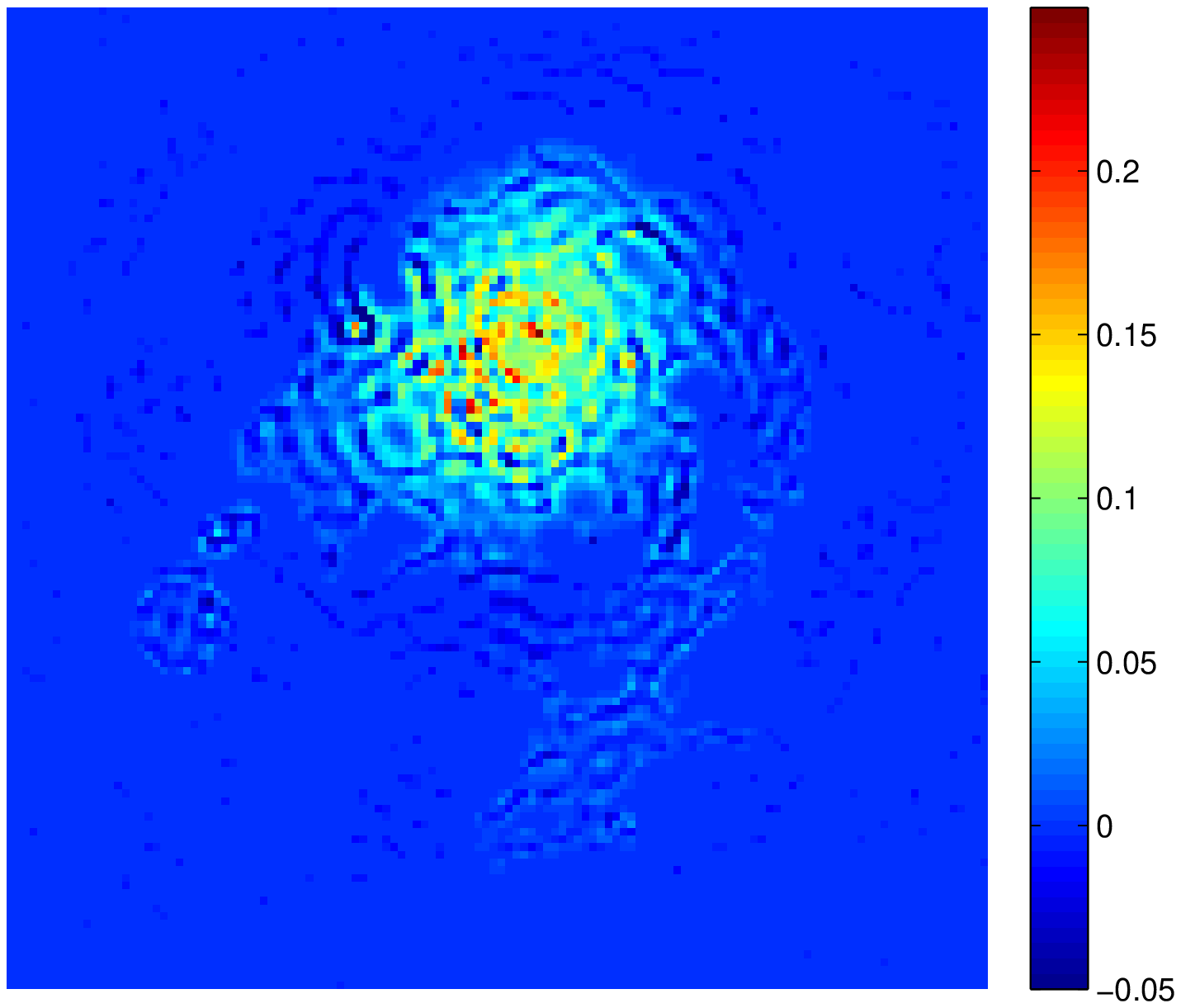}
    \includegraphics[trim = 3.4cm 1.1cm 2.1cm 0.5cm, clip, keepaspectratio, width = 5.5cm]{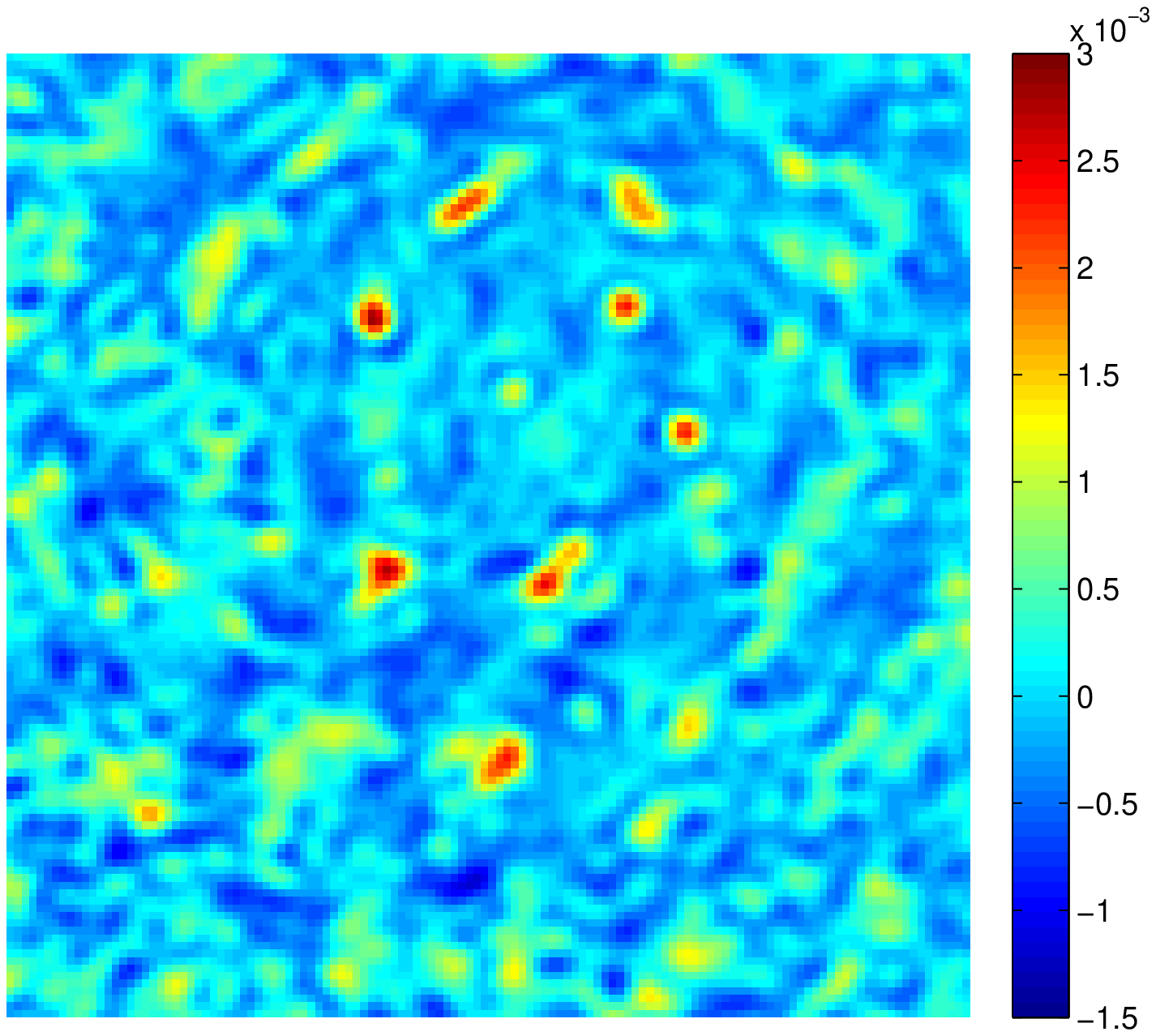}

\caption{AMI coverage example. First row from left to right: original M31 128$\times$128 test image in $\log_{10}$ scale, $u$-$v$ coverage in normalized angular frequency units ($M=9413$) and corresponding dirty image in linear scale. Second to last rows: reconstruction results for SARA (SNR=14.3~dB), RWBPDb8 (SNR=12.3~dB) and RWBP (SNR=10.9~dB). The first column shows the reconstructed images in a $\log_{10}$ scale, the second column shows the error images in linear scale, and the third column shows the residual dirty images also in linear scale.}
\label{fig:5}
\end{figure*}

The last experiment presents an illustration with a realistic radio telescope coverage. We use a simulation of the Arcminute Microkelvin Imager (AMI) \citep{zwart08} array to obtain a $u$-$v$ coverage with $M=9413$ points. For this experiment we use a low resolution 128$\times$128 version of M31. The top row in Figure~\ref{fig:5} shows the original test image in $\log_{10}$ scale, the $u$-$v$ coverage and the corresponding dirty image in linear scale. The SNR of the recovered image for each algorithm is as follows: BP (10.7dB), RWBP (SNR=10.9~dB), BPDb8 (11.6~dB), RWBPDb8 (SNR=12.3~dB), TV (10.6~dB), RWTV (10.5~dB), BPSA (12.4~dB) SARA (14.3~dB). The second and third rows in Figure~\ref{fig:5} show the reconstructed images along with the corresponding error and residual dirty images images for SARA, RWBPDb8 and RWBP. SARA provides not only a SNR increase but also a significant reduction of visual artifacts relative to all other methods.

\section{Concluding Remarks}
\label{sec:Conclusion}

In this paper we have proposed an algorithmic framework based on the simultaneous-direction method of multipliers to solve sparse imaging problems in RI imaging. The new algorithm provides a parallel implementation structure, therefore offering an attractive framework to handle continuous visibilities and associated high dimensional problems. A variety of state-of-the-art sparsity regularization priors, including our recent average sparsity approach SARA, as well as discrete and continuous measurement operators are available in the new PURIFY software. Source code for PURIFY is publicly available. Experimental results confirm both the superiority of SARA for continuous Fourier measurements and the fact that the new algorithmic structure offers a promising path to handle large-scale problems. 

In future work we will extend the current PURIFY implementation to take full advantage of the parallel and distributed structure of SDMM as discussed in Section \ref{ssec:parallel}. We expect that parallel and hardware implementations of the measurement and sparsity operators as well as the proximity operators could achieve drastic accelerations of the algorithms. Also, different strategies will be explored to accelerate the convergence of the conjugate gradient solver, e.g.\ using preconditioners for the operator $\mathsf{Q}$ and precomputing the sparse matrix $\mathsf{G}^{\dagger}\mathsf{G}$ to avoid multiplications by $\mathsf{G}$ and $\mathsf{G}^{\dagger}$ separately, which involve an intermediate high dimensional vector of length $M>N$, at each iteration of the conjugate gradient solver. Finally, DDEs will be incorporated into PURIFY. Recall that DDEs can easily be included in the matrix $\mathsf{G}$ as additional convolution kernels in the frequency plane. Compact support kernels will ensure sparsity of $\mathsf{G}$ and a fast matrix-vector multiplication. Integration with standard packages for interferometric imaging, such as CASA, will allow to take advantage of their built-in real data handling and also to have a full comparison with standard algorithms such as MS-CLEAN and ASP-CLEAN.

\section*{Acknowledgments}
We thank Keith Grainge for providing the visibility coverage corresponding to an example observation made by the AMI telescope. We thank Pierre Vandergheynst and Jean-Philippe Thiran for providing the infrastructure to support our research. REC is supported by the Swiss National Science Foundation (SNSF) under grant 200020-140861. JDM is supported in part by a Newton International Fellowship from the Royal Society and the British Academy. YW is supported in part by the Center for Biomedical Imaging (CIBM) of the Geneva and Lausanne Universities, EPFL and the Leenaards and Louis-Jeantet foundations.  

\bibliographystyle{mymnras_eprint}
\bibliography{abrev,sara}

\label{lastpage}

\end{document}